\documentclass[conference]{IEEEtran}
\ifCLASSINFOpdf
\else
\fi
\usepackage{amsbsy}

\newtheorem{teigi}{Definition}
\newtheorem{rei}{Example}
\newtheorem{teiri}{Theorem}

\begin{document}
%
\title{Construction of Parallel RIO Codes
  using Coset Coding with Hamming Codes}

\author{\IEEEauthorblockN{Akira Yamawaki, Hiroshi Kamabe
and Shan Lu}
\IEEEauthorblockA{Graduate School of Engineering\\
Gifu University\\
1--1, Yanagido, Gifu, 501--1193\\
Email: yamawaki@kmb.info.gifu-u.ac.jp, kamabe@ieee.org,
shan.lu.jp@ieee.org}
}


%


\maketitle

\begin{abstract}
Random input/output (RIO) code is a coding scheme
that enables reading of one logical page
using a single read threshold
in multilevel flash memory.
The construction of RIO codes is
equivalent to the construction of WOM codes.
Parallel RIO (P-RIO) code is an RIO code
that encodes all pages in parallel.
In this paper,
we utilize coset coding with Hamming codes
in order to construct P-RIO codes.
Coset coding is
a technique that
constructs WOM codes
using linear binary codes.
We leverage
the information on the data
of all pages to
encode each page.
Our constructed codes store
more pages
than RIO codes
constructed via coset coding.
\end{abstract}


%
\IEEEpeerreviewmaketitle

\section{Introduction}
Flash memory is the prevalent type of
non-volatile memory in use today,
and
flash devices are employed in
universal serial bus
(USB) memory technology,
solid state drives
(SSD), and mobile applications.
A flash memory consists of an array of cells,
in which 
information bits are stored
in the form of the amount of charge.
In conventional flash memory,
one cell stores a single bit
and the information is read
using a single read threshold.
Recently, multilevel flash memory
technology has been
introduced.
In multilevel flash memory,
each cell
can
represent
one of more than two levels,
and
these levels are distinguished by
multiple read thresholds.
We consider
the
triple-level cell (TLC)
that is currently being utilized
in multilevel flash memory.
A TLC can
represent one of eight levels,
that is,
it stores three bits.
Each level corresponds to a three-bit sequence
as shown in Table \ref{table:cell},
and each bit represents
one logical page.
\begin{table}[h]
\caption{Triple-level cell (TLC)}
\label{table:cell}
\begin{center}
\begin{tabular}{cccc}
Level & Page 1 & Page 2 & Page 3 \\ \hline
7 & 1 & 0 & 0 \\ \hline
6 & 1 & 0 & 1 \\ \hline
5 & 1 & 1 & 1 \\ \hline
4 & 1 & 1 & 0 \\ \hline
3 & 0 & 1 & 0 \\ \hline
2 & 0 & 1 & 1 \\ \hline
1 & 0 & 0 & 1 \\ \hline
0 & 0 & 0 & 0 \\ \hline
\end{tabular}%
\end{center}
\end{table}
Then, a group of such cells
stores three logical pages,
referred to as pages 1--3.
To read pages 1--3,
the required numbers of read thresholds are 1, 2, and 4,
respectively.
Therefore, the average number of read thresholds
is 2.33.
As stated above,
multiple read thresholds are
required in order to read a single logical page
in multilevel flash memory.
However,
use of a large number of read thresholds degrades
the read performance of the flash memory device.

In order to solve this problem,
the random input/output (RIO) code has been proposed by
Sharon and Alrod \cite{riocode}.
This is a coding scheme
in which one logical page
can be read using a single read threshold.
Further,
Sharon and Alrod showed that
the construction of the RIO code
is equivalent to the construction of the WOM code
\cite{riocode}.

However,
a distinction exists between the RIO code
and the WOM code.
In the WOM code,
the data are stored sequentially.
Therefore,
when the data is encoded,
the subsequent data are unknown.
On the other hand,
in the RIO code,
the data of all logical pages are stored
simultaneously;
thus,
all the data are known in advance.
Previously,
Yaakobi and Motwani proposed
the parallel RIO (P-RIO) code,
in which
the encoding of each page is
performed in parallel \cite{parallel}.
These researchers demonstrated P-RIO codes
having parameters for which
WOM codes or RIO codes
do not exist.
In addition,
they proposed an algorithm to
construct a P-RIO code via a computer search.
However, the complexity of this algorithm
increases exponentially
with the code length.

In this paper,
P-RIO codes
are constructed
using coset coding.
Coset coding,
introduced by Cohen, Godlewski, and Merkx,
is a technique that constructs WOM codes
using linear binary codes \cite{cosetcoding}.
When Hamming codes are used
as the linear codes,
we leverage the information on the data of
all logical pages to construct the P-RIO codes.
Our constructed codes
store more pages than RIO codes
constructed via coset coding
with Hamming codes.
The remainder of the paper is
structured as follows.
In section \ref{section:preliminary},
some preliminary notation and definitions
are presented,
whereas the construction of P-RIO codes
using coset coding is demonstrated in section
\ref{section:construction}.
Section \ref{section:conclusion} contains
a brief concluding section.

\section{Preliminaries}
\label{section:preliminary}

We first present some preliminary definitions and notation.
For a positive integer $n$,
we define 
$\mathcal{I}_n = \{1, \ldots, n\}$.
In addition,
for two vectors
$\boldsymbol{x} = (x_1, \ldots, x_n)$ and
$\boldsymbol{y} = (y_1, \ldots, y_n)$,
we denote $\boldsymbol{x} \leq \boldsymbol{y}$
if $x_i \leq y_i$ for any $i \in \mathcal{I}_n$.
For a binary vector
$\boldsymbol{x} = (x_1, \ldots, x_n)$,
we define
$I(\boldsymbol{x}) = \{ i \mid i \in \mathcal{I}_n,
x_i = 1 \}$.

\subsection{WOM Code}

In a WOM, a cell stores a single bit and
its content can be
changed from $0$ to $1$;
however,
it
cannot be changed from $1$ to $0$.
Rivest and Shamir have proposed a WOM code
that allows multiple writings into WOM
\cite{womcode}.
\begin{teigi}
\label{teigi:womcode}
An $[n,l,t]$ WOM code is a coding scheme
that permits the writing of $l$ data bits into
$n$ cells $t$ times.
This scheme 
is defined by
the
encoding map $\mathcal{E}$
and the decoding map $\mathcal{D}$.
The former is defined by
\begin{equation}
  \mathcal{E} : \{0,1\}^l \times Im(\mathcal{E}^{i-1})
  \to \{0,1\}^n, \nonumber
\end{equation}
where $Im(\mathcal{E}^0) = \{(0,\ldots,0)\}$.
For all $(\boldsymbol{d}, \boldsymbol{c})
\in \{0,1\}^l \times Im(\mathcal{E}^{i-1})$
with
$i \in \mathcal{I}_t$,
$\boldsymbol{c} \leq \mathcal{E}(\boldsymbol{d}, \boldsymbol{c})$
is satisfied.
Further,
$\mathcal{D}$ is defined by
\begin{equation}
  \mathcal{D} : Im(\mathcal{E}^{i})
  \to \{0,1\}^l, \nonumber
\end{equation}
such that
$\mathcal{D}(\mathcal{E}(\boldsymbol{d}, \boldsymbol{c}))
= \boldsymbol{d}$ for all
$(\boldsymbol{d}, \boldsymbol{c})
\in \{0,1\}^l \times Im(\mathcal{E}^{i-1})$
with 
$i \in \mathcal{I}_t$.
The sum-rate is defined as $lt/n$.
\end{teigi}
\begin{rei}
Previously,
Rivest and Shamir presented the
$[3,2,2]$ WOM code \cite{womcode},
which 
is shown in Table \ref{table:womcode}.
\begin{table}[tb]
\caption{$[3,2,2]$ WOM code}
\label{table:womcode}
\begin{center}
\begin{tabular}{ccc}
Data bits & First write & Second write \\ \hline
00 & 000 & 111 \\ \hline
01 & 100 & 011 \\ \hline
10 & 010 & 101 \\ \hline
11 & 001 & 110 \\ \hline
\end{tabular}%
\end{center}
\end{table}
\end{rei}

\subsection{Coset Coding}

In this paper,
a linear binary code of length $n$ and dimension $k$
is referred to as an $(n,k)$ code.
Coset coding is
used to construct an $[n,n-k,t]$ WOM code using
$(n,k)$ code $C$,
where $t$ depends on $C$ \cite{cosetcoding}.
Let $H$ be the parity check matrix of $C$.
For all $(\boldsymbol{d}, \boldsymbol{c}) \in
\{0,1\}^{n-k} \times Im(\mathcal{E}^{i-1})$
with
$i \in \mathcal{I}_t$,
the encoding map $\mathcal{E}$ is as follows.
\begin{equation}
  \mathcal{E}(\boldsymbol{d}, \boldsymbol{c})
  = \boldsymbol{c} + \boldsymbol{x},
  \nonumber
\end{equation}
where
\begin{equation}
\boldsymbol{x} \in
\{ \boldsymbol{v} \mid \boldsymbol{v} \in \{0,1\}^n,
\boldsymbol{v} H^T = \boldsymbol{d} - \boldsymbol{c} H^T,
I(\boldsymbol{c}) \cap I(\boldsymbol{v}) = \emptyset \}.
\nonumber
\end{equation}
Further,
for all $\boldsymbol{c} \in Im(\mathcal{E}^i)$
with
$i \in \mathcal{I}_t$,
the decoding map $\mathcal{D}$ is
\begin{equation}
  \mathcal{D}(\boldsymbol{c}) = \boldsymbol{c} H^T.
  \nonumber
\end{equation}

The following theorem
has been 
proven in \cite{cosetcoding}.
\begin{teiri}
\label{teiri:hamming}
When the linear code is
the
$(7,4)$ Hamming code,
coset coding can be used to
construct the $[7,3,3]$ WOM code.
Let $r$ be an integer greater than $3$.
Then a $[2^r-1,r,2^{r-2}+2]$ WOM code can be constructed
via coset coding with a $(2^r-1,2^r-r-1)$ Hamming code.
\end{teiri}

\subsection{RIO Code}

In this paper,
it is assumed that
each cell of a flash memory represents
one of
$q$ levels ($0,1,\ldots,q-1$).
These levels are distinguished by
$(q-1)$ different read thresholds.
For each $i \in \mathcal{I}_{q-1}$,
we denote the threshold between
levels $(i-1)$ and $i$ by $i$.
For the state of $n$ cells
$\boldsymbol{c} = (c_1, \ldots, c_n)
\in \{0, \ldots, q-1\}^n$,
the operation for read threshold $i$
is denoted by $RT_i(\boldsymbol{c})$.
We define
$RT_i(\boldsymbol{c})
= (r_1, \ldots, r_n) \in \{0,1\}^n$
where, for each $j \in \mathcal{I}_n$,
$r_j = 1$ if $c_j \geq i$;
otherwise, $r_j = 0$.
For any $\boldsymbol{c}_1, \ldots, \boldsymbol{c}_{q-1}
\in \{0,1\}^n$
such that $\boldsymbol{c}_1 \leq \boldsymbol{c}_2
\leq \ldots \leq \boldsymbol{c}_{q-1}$,
the following property holds:
For each $i \in \mathcal{I}_{q-1}$,
\begin{eqnarray}
  RT_{q-i} \left(\sum_{j=1}^{q-1} \boldsymbol{c}_j \right)
  = \boldsymbol{c}_i.
  \label{siki:property3}
\end{eqnarray}
RIO code is a coding scheme that
stores $t$ pages in a $(t+1)$-level flash memory,
such that
page $i$ can be read using read threshold $(t+1-i)$
for each $i \in \mathcal{I}_{t}$ \cite{riocode}.
From (\ref{siki:property3}),
pages $1, \ldots, t$ are encoded into
$\boldsymbol{c}_1, \ldots, \boldsymbol{c}_{t}$,
respectively,
such that
$\boldsymbol{c}_1 \leq \boldsymbol{c}_2 \leq \cdots
\leq \boldsymbol{c}_{t}$,
and the cell state is set to
$\sum_{i=1}^{t} \boldsymbol{c}_i$,
where $\boldsymbol{c}_i$ is a binary vector
for each $i \in \mathcal{I}_{t}$.
\begin{teigi}
\label{teigi:riocode}
An $[n,l,t]$ RIO code is a coding scheme
that enables storage of $t$ pages
of $l$ data bits in
$n$ $(t+1)$-level cells,
such that
each page can be read using a single read threshold.
For each $i \in \mathcal{I}_{t}$,
the encoding map of page $i$
$\mathcal{E}_i$ is defined by
\begin{equation}
  \mathcal{E}_i : \{0,1\}^l \times Im(\mathcal{E}_{i-1})
  \to \{0,1\}^n, \nonumber
\end{equation}
where $Im(\mathcal{E}_0) = \{(0,\ldots,0)\}$.
For all $(\boldsymbol{d}, \boldsymbol{c})
\in \{0,1\}^l \times Im(\mathcal{E}_{i-1})$,
$\boldsymbol{c} \leq
\mathcal{E}_i(\boldsymbol{d}, \boldsymbol{c})$
is satisfied.
The cell state is the sum of the $\mathcal{E}_i$ results
for all $i \in \mathcal{I}_{t}$.
Further,
for each $i \in \mathcal{I}_{t}$,
the decoding map of page $i$ $\mathcal{D}_i$
is defined by
\begin{equation}
  \mathcal{D}_i : Im(\mathcal{E}_i)
  \to \{0,1\}^l, \nonumber
\end{equation}
such that
$\mathcal{D}_i(\mathcal{E}_i(\boldsymbol{d}, \boldsymbol{c}))
= \boldsymbol{d}$
for all $(\boldsymbol{d}, \boldsymbol{c})
\in \{0,1\}^l \times Im(\mathcal{E}_{i-1})$.
If the cell state is
$\boldsymbol{c} \in \{0,\ldots,t\}^n$,
the argument of $\mathcal{D}_i$ is
$RT_{t+1-i}(\boldsymbol{c})$ for each $i \in \mathcal{I}_{t}$.
\end{teigi}

From Definitions \ref{teigi:womcode} and
\ref{teigi:riocode},
it is apparent that
the construction of an $[n,l,t]$ RIO code is
equivalent to that of an $[n,l,t]$ WOM code
\cite{riocode}.
\begin{rei}
The $[3,2,2]$ RIO code based on
the $[3,2,2]$ WOM code
is shown in Table \ref{table:riocode}.
\begin{table}[tb]
\caption{$[3,2,2]$ RIO code}
\label{table:riocode}
\begin{center}
\begin{tabular}{c|cccc}
\multicolumn{1}{c|}{Data of page 2} &
\multicolumn{4}{c}{Data of page 1} \\ \cline{2-5}
 & 00 & 01 & 10 & 11 \\ \hline
00 & 000 & 211 & 121 & 112 \\ \hline
01 & 100 & 200 & 021 & 012 \\ \hline
10 & 010 & 201 & 020 & 102 \\ \hline
11 & 001 & 210 & 120 & 002 \\ \hline
\end{tabular}%
\end{center}
\end{table}
As an example,
let the data of pages $1$ and $2$
be $10$ and $01$, respectively.
Then, from Table \ref{table:womcode},
$\mathcal{E}_1(10,000) = 010$ and
$\mathcal{E}_2(01,010) = 011$.
Therefore,
the cell state is $021$.
\end{rei}

\subsection{P-RIO Code}

In the RIO code,
each page encoding
depends on
the page data only.
In contrast,
for P-RIO code,
information on the data of
all pages is leveraged
during
encoding of each page
\cite{parallel}.
\begin{teigi}
An $[n,l,t]$ P-RIO code is
an $[n,l,t]$ RIO code
for which
all pages are encoded
in parallel.
The encoding map $\mathcal{E}$ is
defined by
\begin{equation}
  \mathcal{E} : \prod_{i=1}^{t} \{0,1\}^l \to
  \prod_{i=1}^{t} \{0,1\}^n. \nonumber
\end{equation}
For all $(\boldsymbol{d}_1, \ldots, \boldsymbol{d}_{t})
\in \prod_{i=1}^{t} \{0,1\}^l$,
$\boldsymbol{c}_1 \leq \boldsymbol{c}_2
\leq \cdots \leq \boldsymbol{c}_{t}$ is
satisfied, where
$(\boldsymbol{c}_1, \ldots, \boldsymbol{c}_{t})
= \mathcal{E}(\boldsymbol{d}_1, \ldots, \boldsymbol{d}_{t})$.
The cell state is the sum of
the components of the $\mathcal{E}$ result.
For each $i \in \mathcal{I}_{t}$,
the decoding map $\mathcal{D}_i$ of page $i$ is
defined by
\begin{equation}
  \mathcal{D}_i : \{0,1\}^n \to \{0,1\}^l, \nonumber
\end{equation}
such that $\mathcal{D}_i(\boldsymbol{c}_i) =
\boldsymbol{d}_i$
for all $(\boldsymbol{d}_1, \ldots, \boldsymbol{d}_{t})
\in \prod_{i=1}^{t} \{0,1\}^l$,
where
$(\boldsymbol{c}_1, \ldots, \boldsymbol{c}_{t})
= \mathcal{E}(\boldsymbol{d}_1, \ldots, \boldsymbol{d}_{t})$.
If the cell state is
$\boldsymbol{c} \in \{0,\ldots,t\}^n$,
the argument of $\mathcal{D}_i$ is
$RT_{t+1-i}(\boldsymbol{c})$ for each $i \in \mathcal{I}_{t}$.
\end{teigi}

An algorithm to construct P-RIO code has been proposed
in a previous study
\cite{parallel},
and run to
yield P-RIO codes that
store two pages for moderate code lengths.
These codes have parameters for which RIO codes do not
exist \cite{parallel}.

In this study,
we use coset coding to
construct P-RIO codes.
When Hamming codes
are used as the linear codes,
from Theorem \ref{teiri:hamming},
$[7,3,3]$ RIO code and
$[15,4,6]$ RIO code can
be constructed.
Then,
we leverage the information
on the data of all pages
to
construct P-RIO codes
that store more pages than
these RIO codes.

\section{Construction of P-RIO Code using Coset Coding}
\label{section:construction}

Prior to
construction of P-RIO codes using coset coding,
we first discuss several properties.

\subsection{Properties}

We have the following theorem.
\begin{teiri}
\label{teiri:sufficientcondition}
Let $H$ be the parity check matrix of
$(n,k)$ code $C$.
The sufficient condition
that uses coset coding with $C$
to construct an $[n,n-k,t]$ P-RIO code
is as follows.
For any
$\boldsymbol{s}_1, \ldots, \boldsymbol{s}_{t} \in \{0,1\}^{n-k}$,
there exist $\boldsymbol{x}_1, \ldots, \boldsymbol{x}_{t}
\in \{0,1\}^n$ that satisfy the following conditions:
\begin{enumerate}
 \item For all $i \in \mathcal{I}_{t}$,
 $\boldsymbol{x}_i H^T = \boldsymbol{s}_i$;
 \item For all $i,i' \in \mathcal{I}_{t}, i \not= i'$,
 $I(\boldsymbol{x}_i) \cap I(\boldsymbol{x}_{i'})
 = \emptyset$.
\end{enumerate}
\end{teiri}
\begin{IEEEproof}
For any $\boldsymbol{d}_1, \ldots, \boldsymbol{d}_{t}
\in \{0,1\}^{n-k}$,
we define $\boldsymbol{s}_i = \boldsymbol{d}_{i} -
\boldsymbol{d}_{i-1}$ for each $i \in \mathcal{I}_{t}$,
where $\boldsymbol{d}_0 = (0, \ldots, 0)$.
Suppose that there exist
$\boldsymbol{x}_1, \ldots, \boldsymbol{x}_{t}
\in \{0,1\}^n$ that satisfy
the above conditions for $\boldsymbol{s}_1,
\ldots, \boldsymbol{s}_{t}$.
We define
$\boldsymbol{c}_i = \sum_{i'=1}^i \boldsymbol{x}_{i'}$
for each $i \in \mathcal{I}_{t}$.
Then, we have
$\boldsymbol{c}_i H^T = \boldsymbol{d}_i$
for each $i \in \mathcal{I}_{t}$
and $\boldsymbol{c}_1 \leq \boldsymbol{c}_2
\leq \cdots \leq \boldsymbol{c}_{t}$.
Therefore,
an $[n,n-k,t]$ P-RIO code can be constructed,
where 
$\mathcal{E}(\boldsymbol{d}_1, \ldots,
\boldsymbol{d}_{t}) =
(\boldsymbol{c}_1, \ldots, \boldsymbol{c}_{t})$
and
$\mathcal{D}_i(\boldsymbol{c}_i) = \boldsymbol{c}_i H^T$.
\end{IEEEproof} 

For $r \geq 3$,
we denote the parity check matrix of
the $(2^r-1,2^r-r-1)$ Hamming code
by
\begin{equation}
H_r = \left(
\begin{array}{cccc}
\boldsymbol{h}_1^T & \boldsymbol{h}_2^T & \cdots & \boldsymbol{h}_{2^r-1}^T
\end{array}
\right), \nonumber
\end{equation}
where
$\{\boldsymbol{h}_1, \boldsymbol{h}_2, \ldots, \boldsymbol{h}_{2^r-1}\}
= \{0,1\}^r \setminus \{(0,\ldots,0)\}$.
For $\boldsymbol{s} \in \{0,1\}^r$,
we define
$V(\boldsymbol{s}) = \{ \{ j \} \mid j \in \mathcal{I}_{2^r-1},
\boldsymbol{h}_{j} = \boldsymbol{s} \}
\cup \{ \{ j_1, j_2 \} \mid j_1, j_2 \in \mathcal{I}_{2^r-1},
\boldsymbol{h}_{j_1} + \boldsymbol{h}_{j_2} =
\boldsymbol{s} \}$.
Then, the following theorem holds.
\begin{teiri}
\label{teiri:kouho1}
For any $\boldsymbol{s} \in \{0,1\}^r \setminus \{(0,\ldots,0)\}$,
we have the permutation $\sigma$
of $\mathcal{I}_{2^r-1}$
such that
\begin{eqnarray}
V(\boldsymbol{s}) =
\{ \{ \sigma(1) \}, \{ \sigma(2), \sigma(3) \}, \{ \sigma(4), \sigma(5) \} ,
\ldots, \nonumber \\
\{ \sigma(2^r-2), \sigma(2^r-1) \} \}.
\label{siki:sigma1}
\end{eqnarray}
\end{teiri}
\begin{IEEEproof}
Let $\alpha_1$ be an integer such that
$\alpha_1 \in \mathcal{I}_{2^r-1}$
and
\begin{equation}
  \boldsymbol{s} = \boldsymbol{h}_{\alpha_1}.
  \label{one1}
\end{equation}
Then, $\{ \alpha_1 \} \in V(\boldsymbol{s})$.
Further,
let $\alpha_2$ be an integer such that
$\alpha_2 \in \mathcal{I}_{2^r-1} \setminus \{ \alpha_1 \}$,
and
let $\alpha_3$ be an integer such that
$\alpha_3 \in \mathcal{I}_{2^r-1} \setminus \{ \alpha_1, \alpha_2 \}$
and
\begin{equation}
  \boldsymbol{s} = \boldsymbol{h}_{\alpha_2} + \boldsymbol{h}_{\alpha_3}.
  \label{one2}
\end{equation}
Then, $\{ \alpha_2, \alpha_3 \} \in V(\boldsymbol{s})$.
As with $\alpha_2$ and $\alpha_3$,
we have
$\alpha_4, \alpha_5, \ldots, \alpha_{2^r-2}, \alpha_{2^r-1}$.
Then,
$\{ \alpha_4, \alpha_5 \}, \ldots, 
\{ \alpha_{2^r-2}, \alpha_{2^r-1} \} \in V(\boldsymbol{s})$.
Let $\sigma$ be
the permutation of $\mathcal{I}_{2^r-1}$
such that
$\sigma(j) = \alpha_j$ for each $j \in \mathcal{I}_{2^r-1}$.
Then, $\sigma$ satisfies
(\ref{siki:sigma1}).
\end{IEEEproof}
\begin{teiri}
\label{teiri:kouho2}
For any
$\boldsymbol{s}_1, \boldsymbol{s}_2 \in \{0,1\}^r \setminus \{(0,\ldots,0)\},
\boldsymbol{s}_1 \not= \boldsymbol{s}_2$,
we have the permutation $\sigma$
of $\mathcal{I}_{2^r-1}$
such that
\begin{eqnarray}
& & V(\boldsymbol{s}_{1}) \nonumber \\
& = &
\{ \{ \sigma(1) \}, \{ \sigma(2), \sigma(3) \}, \{ \sigma(4), \sigma(5) \} ,
\{ \sigma(6), \sigma(7) \}, \nonumber \\
& & \ldots, \{ \sigma(2^r-4), \sigma(2^r-3) \}, \{ \sigma(2^r-2), \sigma(2^r-1) \}
\}, \nonumber \\
\label{siki:sigma2-1} \\
& & V(\boldsymbol{s}_{2}) \nonumber \\
& = &
\{ \{ \sigma(2) \}, \{ \sigma(1), \sigma(3) \}, \{ \sigma(4), \sigma(6) \} ,
\{ \sigma(5), \sigma(7) \}, \nonumber \\
& & \ldots, \{ \sigma(2^r-4), \sigma(2^r-2) \}, \{ \sigma(2^r-3), \sigma(2^r-1) \}
\}. \nonumber \\
\label{siki:sigma2-2}
\end{eqnarray}
\end{teiri}
\begin{IEEEproof}
Let $\alpha_1$ be an integer such that
$\alpha_1 \in \mathcal{I}_{2^r-1}$
and
\begin{eqnarray}
  \boldsymbol{s}_1 = \boldsymbol{h}_{\alpha_1}. \label{two1}
\end{eqnarray}
Let $\alpha_2$ be an integer such that
$\alpha_2 \in \mathcal{I}_{2^r-1}$
and
\begin{eqnarray}
  \boldsymbol{s}_2 = \boldsymbol{h}_{\alpha_2}. \label{two2}
\end{eqnarray}
Then,
$\{ \alpha_1 \} \in V(\boldsymbol{s}_{1})$,
$\{ \alpha_2 \} \in V(\boldsymbol{s}_{2})$.
Clearly,
$\alpha_1 \not= \alpha_2$.
Let $\alpha_3$ be an integer such that
$\alpha_3 \in \mathcal{I}_{2^r-1} \setminus \{ \alpha_1, \alpha_2 \}$
and
\begin{eqnarray}
 \boldsymbol{s}_1 = \boldsymbol{h}_{\alpha_2} + \boldsymbol{h}_{\alpha_3}.
 \label{two3}
\end{eqnarray}
Then, $\{ \alpha_2, \alpha_3 \} \in V(\boldsymbol{s}_{1})$.
From (\ref{two1}), (\ref{two3}), and (\ref{two2}),
\begin{eqnarray}
\boldsymbol{h}_{\alpha_1} + \boldsymbol{h}_{\alpha_3} 
& = & \boldsymbol{s}_1 + \boldsymbol{h}_{\alpha_3}
= (\boldsymbol{h}_{\alpha_2} + \boldsymbol{h}_{\alpha_3}) + \boldsymbol{h}_{\alpha_3}
\nonumber \\
& = & \boldsymbol{h}_{\alpha_2} = \boldsymbol{s}_2. \label{two1plus32}
\end{eqnarray}
Hence,
$\{ \alpha_1, \alpha_3 \} \in V(\boldsymbol{s}_{2})$.
Let
$\alpha_4$ be an integer such that
$\alpha_4 \in \mathcal{I}_{2^r-1} \setminus \{ \alpha_1 , \ldots , \alpha_3 \}$,
and
let $\alpha_5$ be an integer such that
$\alpha_5 \in
\mathcal{I}_{2^r-1} \setminus \{ \alpha_1, \ldots, \alpha_4 \}$
and
\begin{eqnarray}
  \boldsymbol{s}_1 = \boldsymbol{h}_{\alpha_4} + \boldsymbol{h}_{\alpha_5}.
  \label{two5}
\end{eqnarray}
Then, $\{ \alpha_4, \alpha_5 \} \in V(\boldsymbol{s}_{1})$.
Let $\alpha_6$ be an integer such that
$\alpha_6 \in
\mathcal{I}_{2^r-1} \setminus \{ \alpha_1, \ldots, \alpha_4 \}$
and
\begin{eqnarray}
  \boldsymbol{s}_2 = \boldsymbol{h}_{\alpha_4} + \boldsymbol{h}_{\alpha_6}
  \label{two6}.
\end{eqnarray}
Then,
$\{ \alpha_4, \alpha_6 \} \in V(\boldsymbol{s}_{2})$.
Clearly, $\alpha_5 \not= \alpha_6$.
Let $\alpha_7$ be an integer such that
$\alpha_7 \in \mathcal{I}_{2^r-1} \setminus \{ \alpha_1, \ldots, \alpha_6 \}$
and
\begin{eqnarray}
  \boldsymbol{s}_1 = \boldsymbol{h}_{\alpha_6} + \boldsymbol{h}_{\alpha_7}.
  \label{two7}
\end{eqnarray}
Then, $\{ \alpha_6, \alpha_7 \} \in V(\boldsymbol{s}_{1})$.
From (\ref{two5}) and (\ref{two7}),
\begin{eqnarray}
 \boldsymbol{h}_{\alpha_4} + \boldsymbol{h}_{\alpha_5} =
 \boldsymbol{h}_{\alpha_6} + \boldsymbol{h}_{\alpha_7}.
 \label{two4plus56plus7}
\end{eqnarray}
From (\ref{two4plus56plus7}) and (\ref{two6}),
\begin{eqnarray}
 \boldsymbol{h}_{\alpha_5} + \boldsymbol{h}_{\alpha_7} 
 = \boldsymbol{h}_{\alpha_4} + \boldsymbol{h}_{\alpha_6}
 = \boldsymbol{s}_2. \label{two5plus7}
\end{eqnarray}
Hence,
$\{ \alpha_5, \alpha_7 \} \in V(\boldsymbol{s}_{2})$.
As with $\alpha_4, \alpha_5, \alpha_6, \alpha_7$,
we have
$\alpha_8, \alpha_9, \alpha_{10}, \alpha_{11},
\ldots, \alpha_{2^r-4}, \alpha_{2^r-3} , \alpha_{2^r-2}, \alpha_{2^r-1}$.
Let
$\sigma$ be the permutation of $\mathcal{I}_{2^r-1}$
such that $\sigma(j) = \alpha_{j}$
for each $j \in \mathcal{I}_{2^r-1}$.
Then, $\sigma$ satisfies (\ref{siki:sigma2-1}) and
(\ref{siki:sigma2-2}).
\end{IEEEproof}
\begin{teiri}
\label{teiri:kouho3}
Let $r = 4$.
For any $\boldsymbol{s}_1, \boldsymbol{s}_2, \boldsymbol{s}_3
\in \{0,1\}^4 \setminus \{0000\}
\mbox{ }
(\boldsymbol{s}_i \not= \boldsymbol{s}_{i'} \mbox{ for any }
i, i' \in \mathcal{I}_3, i \not= i')$,
we have the permutation $\sigma$
of $\mathcal{I}_{15}$,
which
satisfies the following conditions:

\noindent
(a)
If $\boldsymbol{s}_1 + \boldsymbol{s}_2 \not= \boldsymbol{s}_3$,
\begin{eqnarray}
V(\boldsymbol{s}_{1}) & = &
\{ \{ \sigma(1) \}, \{ \sigma(2), \sigma(3) \}, \{ \sigma(4), \sigma(5) \} ,
\nonumber \\
& & \{ \sigma(6), \sigma(7) \},
\{ \sigma(8), \sigma(9) \}, \{ \sigma(10), \sigma(11) \},
\nonumber \\
& & \{ \sigma(12), \sigma(13) \} , \{ \sigma(14), \sigma(15) \} \},
\label{siki:sigma3a-1} \\
V(\boldsymbol{s}_{2}) & = &
\{ \{ \sigma(2) \}, \{ \sigma(1), \sigma(3) \}, \{ \sigma(4), \sigma(6) \} ,
\nonumber \\
& & \{ \sigma(5), \sigma(7) \},
\{ \sigma(8), \sigma(10) \}, \{ \sigma(9), \sigma(11) \},
\nonumber \\
& & \{ \sigma(12), \sigma(14) \}, \{ \sigma(13), \sigma(15) \} \},
\label{siki:sigma3a-2} \\
V(\boldsymbol{s}_{3}) & = &
\{ \{ \sigma(4) \}, \{ \sigma(1), \sigma(5) \}, \{ \sigma(2), \sigma(6) \} ,
\nonumber \\
& & \{ \sigma(3), \sigma(7) \},
\{ \sigma(8), \sigma(12) \}, \{ \sigma(9), \sigma(13) \},
\nonumber \\
& & \{ \sigma(10), \sigma(14) \}, \{ \sigma(11), \sigma(15) \} \}.
\label{siki:sigma3a-3}
\end{eqnarray}
(b)
If $\boldsymbol{s}_1 + \boldsymbol{s}_2 = \boldsymbol{s}_3$,
\begin{eqnarray}
V(\boldsymbol{s}_{1}) & = &
\{ \{ \sigma(1) \}, \{ \sigma(2), \sigma(3) \}, \{ \sigma(4), \sigma(5) \} ,
\nonumber \\
& & \{ \sigma(6), \sigma(7) \},
\{ \sigma(8), \sigma(9) \}, \{ \sigma(10), \sigma(11) \},
\nonumber \\
& & \{ \sigma(12), \sigma(13) \} , \{ \sigma(14), \sigma(15) \} \},
\label{siki:sigma3b-1} \\
V(\boldsymbol{s}_{2}) & = &
\{ \{ \sigma(2) \}, \{ \sigma(1), \sigma(3) \}, \{ \sigma(4), \sigma(6) \} ,
\nonumber \\
& & \{ \sigma(5), \sigma(7) \},
\{ \sigma(8), \sigma(10) \}, \{ \sigma(9), \sigma(11) \},
\nonumber \\
& & \{ \sigma(12), \sigma(14) \}, \{ \sigma(13), \sigma(15) \} \},
\label{siki:sigma3b-2} \\
V(\boldsymbol{s}_{3}) & = &
\{ \{ \sigma(3) \}, \{ \sigma(1), \sigma(2) \}, \{ \sigma(4), \sigma(7) \} ,
\nonumber \\
& & \{ \sigma(5), \sigma(6) \},
\{ \sigma(8), \sigma(11) \}, \{ \sigma(9), \sigma(10) \},
\nonumber \\
& & \{ \sigma(12), \sigma(15) \}, \{ \sigma(13), \sigma(14) \} \}.
\label{siki:sigma3b-3}
\end{eqnarray}
\end{teiri}
\begin{IEEEproof}
Let $\alpha_1$ be an integer such that
$\alpha_1 \in \mathcal{I}_{15}$ and
\begin{eqnarray}
  \boldsymbol{s}_1 = \boldsymbol{h}_{\alpha_1}. \label{three1}
\end{eqnarray}
Let $\alpha_2$ be an integer such that
$\alpha_2 \in \mathcal{I}_{15}$ and
\begin{eqnarray}
  \boldsymbol{s}_2 = \boldsymbol{h}_{\alpha_2}. \label{three2}
\end{eqnarray}
Then, $\{ \alpha_1 \} \in V(\boldsymbol{s}_{1})$,
$\{ \alpha_2 \} \in V(\boldsymbol{s}_{2})$.
Clearly,
$\alpha_1 \not= \alpha_2$.
Let $\alpha_3$ be an integer such that
$\alpha_3 \in \mathcal{I}_{15} \setminus \{ \alpha_1, \alpha_2 \}$
and
\begin{eqnarray}
\boldsymbol{s}_1 = \boldsymbol{h}_{\alpha_2} + \boldsymbol{h}_{\alpha_3}.
\label{three3}
\end{eqnarray}
Then, $\{ \alpha_2, \alpha_3 \} \in V(\boldsymbol{s}_{1})$.
From (\ref{three1}), (\ref{three3}), and (\ref{three2}),
\begin{eqnarray}
  \boldsymbol{h}_{\alpha_1} + \boldsymbol{h}_{\alpha_3}
  & = & \boldsymbol{s}_1 + \boldsymbol{h}_{\alpha_3}
  = (\boldsymbol{h}_{\alpha_2} + \boldsymbol{h}_{\alpha_3})
  + \boldsymbol{h}_{\alpha_3}
  \nonumber \\
  & = & \boldsymbol{h}_{\alpha_2}
  = \boldsymbol{s}_2.
  \label{three1plus32}
\end{eqnarray}
Hence, $\{ \alpha_1, \alpha_3 \} \in V(\boldsymbol{s}_{2})$.

(a)
Let $\alpha_4$ be an integer such that
$\alpha_4 \in \mathcal{I}_{15}$ and
\begin{eqnarray}
  \boldsymbol{s}_3 = \boldsymbol{h}_{\alpha_4}.
  \label{threea4}
\end{eqnarray}
Then, $\{ \alpha_4 \} \in V(\boldsymbol{s}_{3})$.
Clearly,
$\alpha_4 \not\in \{ \alpha_1, \alpha_2 \}$.
From
(\ref{threea4}), (\ref{three3}), and (\ref{three2}),
\begin{eqnarray}
\boldsymbol{h}_{\alpha_4} = \boldsymbol{s}_3
\not= \boldsymbol{s}_1 + \boldsymbol{s}_2
= (\boldsymbol{h}_{\alpha_2} + \boldsymbol{h}_{\alpha_3}) +
\boldsymbol{h}_{\alpha_2} = \boldsymbol{h}_{\alpha_3}.
\label{threea4not3}
\end{eqnarray}
Therefore,
$\alpha_4 \not= \alpha_3$.
Let $\alpha_5$ be an integer such that
$\alpha_5 \in \mathcal{I}_{15} \setminus \{ \alpha_1, \ldots, \alpha_4 \}$
and
\begin{eqnarray}
  \boldsymbol{s}_1 = \boldsymbol{h}_{\alpha_4} + \boldsymbol{h}_{\alpha_5}.
  \label{threea5}
\end{eqnarray}
Then, $\{ \alpha_4, \alpha_5 \} \in V(\boldsymbol{s}_{1})$.
From (\ref{three1}) and (\ref{threea5}),
\begin{eqnarray}
\boldsymbol{h}_{\alpha_1} = \boldsymbol{h}_{\alpha_4} + \boldsymbol{h}_{\alpha_5}.
\label{threea14plus5}
\end{eqnarray}
From (\ref{threea14plus5}) and (\ref{threea4}),
\begin{eqnarray}
\boldsymbol{h}_{\alpha_1} + \boldsymbol{h}_{\alpha_5} = \boldsymbol{h}_{\alpha_4}
= \boldsymbol{s}_3. \label{threea1plus53}
\end{eqnarray}
Hence,
$\{ \alpha_1, \alpha_5 \} \in V(\boldsymbol{s}_{3})$.
Let $\alpha_6$ be an integer such that
$\alpha_6 \in \mathcal{I}_{15} \setminus \{ \alpha_1, \ldots, \alpha_5 \}$
and
\begin{eqnarray}
\boldsymbol{s}_2 = \boldsymbol{h}_{\alpha_4} + \boldsymbol{h}_{\alpha_6}.
\label{threea6}
\end{eqnarray}
Then $\{ \alpha_4, \alpha_6 \} \in V(\boldsymbol{s}_{2})$.
From (\ref{three2}) and (\ref{threea6}),
\begin{eqnarray}
  \boldsymbol{h}_{\alpha_2} = \boldsymbol{h}_{\alpha_4} + \boldsymbol{h}_{\alpha_6}.
  \label{threea24plus6}
\end{eqnarray}
From (\ref{threea24plus6}) and (\ref{threea4}),
\begin{eqnarray}
  \boldsymbol{h}_{\alpha_2} + \boldsymbol{h}_{\alpha_6}
  = \boldsymbol{h}_{\alpha_4}
  = \boldsymbol{s}_3.
  \label{threea2plus63}
\end{eqnarray}
Therefore,
$\{ \alpha_2, \alpha_6 \} \in V(\boldsymbol{s}_{3})$.
Let $\alpha_7$
be an integer such that
$\alpha_7 \in \mathcal{I}_{15} \setminus \{ \alpha_1, \ldots, \alpha_6 \}$
and
\begin{eqnarray}
\boldsymbol{s}_1 = \boldsymbol{h}_{\alpha_6} + \boldsymbol{h}_{\alpha_7}.
\label{threea7}
\end{eqnarray}
Then, $\{ \alpha_6, \alpha_7 \} \in V(\boldsymbol{s}_{1})$.
From (\ref{threea5}) and (\ref{threea7}),
\begin{eqnarray}
\boldsymbol{h}_{\alpha_4} + \boldsymbol{h}_{\alpha_5}
= \boldsymbol{h}_{\alpha_6} + \boldsymbol{h}_{\alpha_7}.
\label{threea4plus56plus7}
\end{eqnarray}
From (\ref{threea4plus56plus7}) and (\ref{threea6}),
\begin{eqnarray}
\boldsymbol{h}_{\alpha_5} + \boldsymbol{h}_{\alpha_7} =
\boldsymbol{h}_{\alpha_4} + \boldsymbol{h}_{\alpha_6}
= \boldsymbol{s}_2. \label{threea5plus72}
\end{eqnarray}
Hence,
$\{ \alpha_5, \alpha_7 \} \in V(\boldsymbol{s}_{2})$.
From (\ref{three3}) and (\ref{threea7}),
\begin{eqnarray}
\boldsymbol{h}_{\alpha_2} + \boldsymbol{h}_{\alpha_3} =
\boldsymbol{h}_{\alpha_6} + \boldsymbol{h}_{\alpha_7}.
\label{threea2plus36plus7}
\end{eqnarray}
From (\ref{threea2plus36plus7}) and (\ref{threea2plus63}),
\begin{eqnarray}
\boldsymbol{h}_{\alpha_3} + \boldsymbol{h}_{\alpha_7} =
\boldsymbol{h}_{\alpha_2} + \boldsymbol{h}_{\alpha_6}
= \boldsymbol{s}_3.
\label{threea3plus73}
\end{eqnarray}
Therefore,
$\{ \alpha_3, \alpha_7 \} \in V(\boldsymbol{s}_{3})$.
Let $\alpha_8$ be an integer such that
$\alpha_8 \in \mathcal{I}_{15} \setminus \{ \alpha_1, \ldots, \alpha_7 \}$,
and
let $\alpha_9$
be an integer such that
$\alpha_9 \in \mathcal{I}_{15} \setminus \{ \alpha_1, \ldots, \alpha_8 \}$
and
\begin{eqnarray}
\boldsymbol{s}_1 = \boldsymbol{h}_{\alpha_8} + \boldsymbol{h}_{\alpha_9}.
\label{threea9}
\end{eqnarray}
Then, $\{ \alpha_8, \alpha_9 \} \in V(\boldsymbol{s}_{1})$.
Let $\alpha_{10}$
be an integer such that
$\alpha_{10} \in \mathcal{I}_{15} \setminus \{ \alpha_1, \ldots, \alpha_9 \}$
and
\begin{eqnarray}
  \boldsymbol{s}_2 = \boldsymbol{h}_{\alpha_8} + \boldsymbol{h}_{\alpha_{10}}.
  \label{threea10}
\end{eqnarray}
Then, $\{ \alpha_8, \alpha_{10} \} \in V(\boldsymbol{s}_{2})$.
Let $\alpha_{11}$
be an integer such that
$\alpha_{11} \in \mathcal{I}_{15} \setminus \{ \alpha_1, \ldots, \alpha_{10} \}$
and
\begin{eqnarray}
\boldsymbol{s}_1 = \boldsymbol{h}_{\alpha_{10}} + \boldsymbol{h}_{\alpha_{11}}.
\label{threea11}
\end{eqnarray}
Then, $\{ \alpha_{10}, \alpha_{11} \} \in V(\boldsymbol{s}_{1})$.
From (\ref{threea9}) and (\ref{threea11}),
\begin{eqnarray}
\boldsymbol{h}_{\alpha_8} + \boldsymbol{h}_{\alpha_{9}} =
\boldsymbol{h}_{\alpha_{10}} + \boldsymbol{h}_{\alpha_{11}}.
\label{threea8plus910plus11}
\end{eqnarray}
From (\ref{threea8plus910plus11}) and (\ref{threea10}),
\begin{eqnarray}
\boldsymbol{h}_{\alpha_9} + \boldsymbol{h}_{\alpha_{11}} =
\boldsymbol{h}_{\alpha_8} + \boldsymbol{h}_{\alpha_{10}}
= \boldsymbol{s}_2.
\label{threea9plus112}
\end{eqnarray}
Hence,
$\{ \alpha_9, \alpha_{11} \} \in V(\boldsymbol{s}_{2})$.
Let $\alpha_{12}$
be an integer such that
$\alpha_{12} \in 
\mathcal{I}_{15} \setminus \{ \alpha_1, \ldots, \alpha_{10} \}$
and
\begin{eqnarray}
\boldsymbol{s}_3 = \boldsymbol{h}_{\alpha_{8}} + \boldsymbol{h}_{\alpha_{12}}.
\label{threea12}
\end{eqnarray}
Then, $\{ \alpha_{8}, \alpha_{12} \} \in V(\boldsymbol{s}_3)$.
From (\ref{threea9}) and (\ref{threea9plus112}),
\begin{eqnarray}
\boldsymbol{s}_1 + \boldsymbol{s}_2
= (\boldsymbol{h}_{\alpha_8} + \boldsymbol{h}_{\alpha_9}) +
(\boldsymbol{h}_{\alpha_9} + \boldsymbol{h}_{\alpha_{11}})
= \boldsymbol{h}_{\alpha_8} + \boldsymbol{h}_{\alpha_{11}}.
\label{threea1plus28plus11}
\end{eqnarray}
From (\ref{threea12}) and (\ref{threea1plus28plus11}),
\begin{eqnarray}
  \boldsymbol{h}_{\alpha_{8}} + \boldsymbol{h}_{\alpha_{12}}
  = \boldsymbol{s}_3 \not= \boldsymbol{s}_1 + \boldsymbol{s}_2
  = \boldsymbol{h}_{\alpha_8} + \boldsymbol{h}_{\alpha_{11}}.
  \label{threea12not11}
\end{eqnarray}
Therefore,
$\alpha_{12} \not= \alpha_{11}$.
Let $\alpha_{13}$
be an integer such that
$\alpha_{13} \in 
\mathcal{I}_{15} \setminus \{ \alpha_1, \ldots, \alpha_{12} \}$
and
\begin{eqnarray}
\boldsymbol{s}_1 = \boldsymbol{h}_{\alpha_{12}} + \boldsymbol{h}_{\alpha_{13}}.
\label{threea13}
\end{eqnarray}
Then, $\{ \alpha_{12}, \alpha_{13} \} \in V(\boldsymbol{s}_1)$.
From (\ref{threea9}) and (\ref{threea13}),
\begin{eqnarray}
\boldsymbol{h}_{\alpha_{8}} + \boldsymbol{h}_{\alpha_{9}}
= \boldsymbol{h}_{\alpha_{12}} + \boldsymbol{h}_{\alpha_{13}}.
\label{threea8plus912plus13}
\end{eqnarray}
From (\ref{threea8plus912plus13}) and (\ref{threea12}),
\begin{eqnarray}
\boldsymbol{h}_{\alpha_{9}} + \boldsymbol{h}_{\alpha_{13}}
= \boldsymbol{h}_{\alpha_{8}} + \boldsymbol{h}_{\alpha_{12}}
= \boldsymbol{s}_3. \label{threea9plus133}
\end{eqnarray}
Hence,
$\{ \alpha_9, \alpha_{13} \} \in V(\boldsymbol{s}_3)$.
Let $\alpha_{14}$ be an integer such that
$\alpha_{14} \in 
\mathcal{I}_{15} \setminus \{ \alpha_1, \ldots, \alpha_{13} \}$
and
\begin{eqnarray}
  \boldsymbol{s}_2 = \boldsymbol{h}_{\alpha_{12}} + \boldsymbol{h}_{\alpha_{14}}.
  \label{threea14}
\end{eqnarray}
Then $\{ \alpha_{12}, \alpha_{14} \} \in V(\boldsymbol{s}_2)$.
From (\ref{threea10}) and (\ref{threea14}),
\begin{eqnarray}
\boldsymbol{h}_{\alpha_{8}} + \boldsymbol{h}_{\alpha_{10}}
= \boldsymbol{h}_{\alpha_{12}} + \boldsymbol{h}_{\alpha_{14}}.
\label{threea8plus1012plus14}
\end{eqnarray}
From (\ref{threea8plus1012plus14}) and (\ref{threea12}),
\begin{eqnarray}
\boldsymbol{h}_{\alpha_{10}} + \boldsymbol{h}_{\alpha_{14}}
= \boldsymbol{h}_{\alpha_{8}} + \boldsymbol{h}_{\alpha_{12}}
= \boldsymbol{s}_3.
\label{threea10plus143}
\end{eqnarray}
Therefore,
$\{ \alpha_{10}, \alpha_{14} \} \in V(\boldsymbol{s}_3)$.
Let $\alpha_{15}$ be an integer such that
$\alpha_{15} \in 
\mathcal{I}_{15} \setminus \{ \alpha_1, \ldots, \alpha_{14} \}$
and
\begin{eqnarray}
\boldsymbol{s}_1 = \boldsymbol{h}_{\alpha_{14}} + \boldsymbol{h}_{\alpha_{15}}.
\label{threea15}
\end{eqnarray}
Then, $\{ \alpha_{14}, \alpha_{15} \} \in V(\boldsymbol{s}_1)$.
From (\ref{threea13}) and (\ref{threea15}),
\begin{eqnarray}
\boldsymbol{h}_{\alpha_{12}} + \boldsymbol{h}_{\alpha_{13}}
= \boldsymbol{h}_{\alpha_{14}} + \boldsymbol{h}_{\alpha_{15}}.
\label{threea12plus1314plus15}
\end{eqnarray}
From (\ref{threea12plus1314plus15}) and
(\ref{threea14}),
\begin{eqnarray}
\boldsymbol{h}_{\alpha_{13}} + \boldsymbol{h}_{\alpha_{15}}
= \boldsymbol{h}_{\alpha_{12}} + \boldsymbol{h}_{\alpha_{14}}
= \boldsymbol{s}_2.
\label{threea13plus152}
\end{eqnarray}
Hence, $\{ \alpha_{13}, \alpha_{15} \} \in
V(\boldsymbol{s}_2)$.
From (\ref{threea11}) and (\ref{threea15}),
\begin{eqnarray}
\boldsymbol{h}_{\alpha_{10}} + \boldsymbol{h}_{\alpha_{11}}
= \boldsymbol{h}_{\alpha_{14}} + \boldsymbol{h}_{\alpha_{15}}.
\label{threea10plus1114plus15}
\end{eqnarray}
From (\ref{threea10plus1114plus15})
and (\ref{threea10plus143}),
\begin{eqnarray}
\boldsymbol{h}_{\alpha_{11}} + \boldsymbol{h}_{\alpha_{15}}
= \boldsymbol{h}_{\alpha_{10}} + \boldsymbol{h}_{\alpha_{14}}
= \boldsymbol{s}_3. \label{threea11plus153}
\end{eqnarray}
Therefore,
$\{ \alpha_{11}, \alpha_{15} \} \in V(\boldsymbol{s}_3)$.
Let $\sigma$ be the permutation of $\mathcal{I}_{15}$
such that $\sigma(j) = \alpha_j$
for each $j \in \mathcal{I}_{15}$.
Then, $\sigma$ satisfies (\ref{siki:sigma3a-1}),
(\ref{siki:sigma3a-2}), and (\ref{siki:sigma3a-3}).

(b)
From (\ref{three3}) and (\ref{three2}),
\begin{eqnarray}
\boldsymbol{s}_3 = \boldsymbol{s}_1 + \boldsymbol{s}_2
= (\boldsymbol{h}_{\alpha_2} + \boldsymbol{h}_{\alpha_3})
+ \boldsymbol{h}_{\alpha_2}
= \boldsymbol{h}_{\alpha_3}. \label{threeb33}
\end{eqnarray}
Hence, $\{ \alpha_3 \} \in V(\boldsymbol{s}_3)$.
From (\ref{three1}) and (\ref{three2}),
\begin{eqnarray}
\boldsymbol{h}_{\alpha_1} + \boldsymbol{h}_{\alpha_2}
= \boldsymbol{s}_1 + \boldsymbol{s}_2
= \boldsymbol{s}_3.
\label{threeb1plus23}
\end{eqnarray}
Therefore,
$\{ \alpha_1, \alpha_2 \} \in V(\boldsymbol{s}_3)$.
Let $\alpha_4$ be an integer such that
$\alpha_4 \in \mathcal{I}_{15} \setminus \{ \alpha_1, \ldots, \alpha_3 \}$,
and let $\alpha_5$ be an integer such that
$\alpha_{5} \in 
\mathcal{I}_{15} \setminus \{ \alpha_1, \ldots, \alpha_{4} \}$
and
\begin{eqnarray}
\boldsymbol{s}_1 = \boldsymbol{h}_{\alpha_{4}} + \boldsymbol{h}_{\alpha_{5}}.
\label{threeb5}
\end{eqnarray}
Then, $\{ \alpha_4, \alpha_5 \} \in V(\boldsymbol{s}_1)$.
Let $\alpha_6$ be an integer such that
$\alpha_6 \in \mathcal{I}_{15} \setminus \{ \alpha_1, \ldots, \alpha_5 \}$
and
\begin{eqnarray}
\boldsymbol{s}_2 = \boldsymbol{h}_{\alpha_{4}} + \boldsymbol{h}_{\alpha_{6}}.
\label{threeb6}
\end{eqnarray}
Then,
$\{ \alpha_4, \alpha_6 \} \in V(\boldsymbol{s}_2)$.
From (\ref{threeb5}) and (\ref{threeb6}),
\begin{eqnarray}
  \boldsymbol{s}_3 & = & \boldsymbol{s}_1 + \boldsymbol{s}_2
  = (\boldsymbol{h}_{\alpha_{4}} + \boldsymbol{h}_{\alpha_{5}})
  + (\boldsymbol{h}_{\alpha_{4}} + \boldsymbol{h}_{\alpha_{6}})
  \nonumber \\
  & = & \boldsymbol{h}_{\alpha_{5}} + \boldsymbol{h}_{\alpha_{6}}.
  \label{threeb35plus6}
\end{eqnarray}
Hence,
$\{ \alpha_5, \alpha_6 \} \in V(\boldsymbol{s}_3)$.
Let $\alpha_7$ be an integer such that
$\alpha_7 \in \mathcal{I}_{15} \setminus \{ \alpha_1, \ldots, \alpha_6 \}$
and
\begin{eqnarray}
\boldsymbol{s}_1 = \boldsymbol{h}_{\alpha_{6}} + \boldsymbol{h}_{\alpha_{7}}.
\label{threeb7}
\end{eqnarray}
Then, $\{ \alpha_6, \alpha_7 \} \in V(\boldsymbol{s}_1)$.
From (\ref{threeb5}) and (\ref{threeb7}),
\begin{eqnarray}
  \boldsymbol{h}_{\alpha_{4}} + \boldsymbol{h}_{\alpha_{5}}
  = \boldsymbol{h}_{\alpha_{6}} + \boldsymbol{h}_{\alpha_{7}}.
  \label{threeb4plus56plus7}
\end{eqnarray}
From (\ref{threeb4plus56plus7}) and (\ref{threeb6}),
\begin{eqnarray}
  \boldsymbol{h}_{\alpha_{5}} + \boldsymbol{h}_{\alpha_{7}}
  = \boldsymbol{h}_{\alpha_{4}} + \boldsymbol{h}_{\alpha_{6}}
  = \boldsymbol{s}_2.
  \label{threeb5plus72}
\end{eqnarray}
Therefore,
$\{ \alpha_5, \alpha_7 \} \in V(\boldsymbol{s}_2)$.
From (\ref{threeb4plus56plus7}) and (\ref{threeb35plus6}),
\begin{eqnarray}
  \boldsymbol{h}_{\alpha_{4}} + \boldsymbol{h}_{\alpha_{7}}
  = \boldsymbol{h}_{\alpha_{5}} + \boldsymbol{h}_{\alpha_{6}}
  = \boldsymbol{s}_3.
  \label{threeb4plus73}
\end{eqnarray}
Hence,
$\{ \alpha_4, \alpha_7 \} \in V(\boldsymbol{s}_3)$.
Let $\alpha_8$ be an integer such that
$\alpha_8 \in \mathcal{I}_{15} \setminus \{ \alpha_1, \ldots, \alpha_7 \}$,
and let
$\alpha_9$ be an integer such that
$\alpha_9 \in \mathcal{I}_{15} \setminus \{ \alpha_1, \ldots, \alpha_8 \}$
and
\begin{eqnarray}
\boldsymbol{s}_1 = \boldsymbol{h}_{\alpha_{8}} + \boldsymbol{h}_{\alpha_{9}}.
\label{threeb9}
\end{eqnarray}
Then, $\{ \alpha_8, \alpha_9 \} \in V(\boldsymbol{s}_1)$.
Let $\alpha_{10}$ be an integer such that
$\alpha_{10} \in \mathcal{I}_{15} \setminus \{ \alpha_1, \ldots, \alpha_9 \}$
and
\begin{eqnarray}
\boldsymbol{s}_2 = \boldsymbol{h}_{\alpha_{8}} + \boldsymbol{h}_{\alpha_{10}}.
\label{threeb10}
\end{eqnarray}
Then, $\{ \alpha_8, \alpha_{10} \} \in V(\boldsymbol{s}_2)$.
From (\ref{threeb9}) and (\ref{threeb10}),
\begin{eqnarray}
  \boldsymbol{s}_3 & = & \boldsymbol{s}_1 + \boldsymbol{s}_2
  = (\boldsymbol{h}_{\alpha_{8}} + \boldsymbol{h}_{\alpha_{9}})
  + (\boldsymbol{h}_{\alpha_{8}} + \boldsymbol{h}_{\alpha_{10}})
  \nonumber \\
  & = & \boldsymbol{h}_{\alpha_{9}} + \boldsymbol{h}_{\alpha_{10}}.
  \label{threeb39plus10}
\end{eqnarray}
Therefore,
$\{ \alpha_9, \alpha_{10} \} \in V(\boldsymbol{s}_3)$.
Let $\alpha_{11}$ be an integer such that
$\alpha_{11} \in \mathcal{I}_{15} \setminus \{ \alpha_1, \ldots, \alpha_{10} \}$
and
\begin{eqnarray}
\boldsymbol{s}_1 = \boldsymbol{h}_{\alpha_{10}} + \boldsymbol{h}_{\alpha_{11}}.
\label{threeb11}
\end{eqnarray}
Then, $\{ \alpha_{10}, \alpha_{11} \} \in V(\boldsymbol{s}_1)$.
From (\ref{threeb9}) and (\ref{threeb11}),
\begin{eqnarray}
  \boldsymbol{h}_{\alpha_{8}} + \boldsymbol{h}_{\alpha_{9}}
  = \boldsymbol{h}_{\alpha_{10}} + \boldsymbol{h}_{\alpha_{11}}.
  \label{threeb8plus910plus11}
\end{eqnarray}
From (\ref{threeb8plus910plus11}) and (\ref{threeb10}),
\begin{eqnarray}
  \boldsymbol{h}_{\alpha_{9}} + \boldsymbol{h}_{\alpha_{11}}
  = \boldsymbol{h}_{\alpha_{8}} + \boldsymbol{h}_{\alpha_{10}}
  = \boldsymbol{s}_2.
  \label{threeb9plus112}
\end{eqnarray}
Hence,
$\{ \alpha_{9}, \alpha_{11} \} \in V(\boldsymbol{s}_2)$.
From (\ref{threeb8plus910plus11}) and (\ref{threeb39plus10}),
\begin{eqnarray}
  \boldsymbol{h}_{\alpha_{8}} + \boldsymbol{h}_{\alpha_{11}}
  = \boldsymbol{h}_{\alpha_{9}} + \boldsymbol{h}_{\alpha_{10}}
  = \boldsymbol{s}_3.
  \label{threeb8plus113}
\end{eqnarray}
Therefore,
$\{ \alpha_{8}, \alpha_{11} \} \in V(\boldsymbol{s}_3)$.
Let $\alpha_{12}$ be an integer such that
$\alpha_{12} \in \mathcal{I}_{15} \setminus \{ \alpha_1, \ldots, \alpha_{11} \}$,
and let
$\alpha_{13}$ be an integer such that
$\alpha_{13} \in \mathcal{I}_{15} \setminus \{ \alpha_1, \ldots, \alpha_{12} \}$
and
\begin{eqnarray}
\boldsymbol{s}_1 = \boldsymbol{h}_{\alpha_{12}} + \boldsymbol{h}_{\alpha_{13}}.
\label{threeb13}
\end{eqnarray}
Then, $\{ \alpha_{12}, \alpha_{13} \} \in V(\boldsymbol{s}_1)$.
Let $\alpha_{14}$ be an integer such that
$\alpha_{14} \in \mathcal{I}_{15} \setminus \{ \alpha_1, \ldots, \alpha_{13} \}$
and
\begin{eqnarray}
\boldsymbol{s}_2 = \boldsymbol{h}_{\alpha_{12}} + \boldsymbol{h}_{\alpha_{14}}.
\label{threeb14}
\end{eqnarray}
Then, $\{ \alpha_{12}, \alpha_{14} \} \in V(\boldsymbol{s}_2)$.
From (\ref{threeb13}) and (\ref{threeb14}),
\begin{eqnarray}
  \boldsymbol{s}_3 & = & \boldsymbol{s}_1 + \boldsymbol{s}_2
  = (\boldsymbol{h}_{\alpha_{12}} + \boldsymbol{h}_{\alpha_{13}})
  + (\boldsymbol{h}_{\alpha_{12}} + \boldsymbol{h}_{\alpha_{14}})
  \nonumber \\
  & = & \boldsymbol{h}_{\alpha_{13}} + \boldsymbol{h}_{\alpha_{14}}.
  \label{threeb313plus14}
\end{eqnarray}
Hence,
$\{ \alpha_{13}, \alpha_{14} \} \in V(\boldsymbol{s}_3)$.
Let $\alpha_{15}$ be an integer such that
$\alpha_{15} \in \mathcal{I}_{15} \setminus \{ \alpha_1, \ldots, \alpha_{14} \}$
and
\begin{eqnarray}
\boldsymbol{s}_1 = \boldsymbol{h}_{\alpha_{14}} + \boldsymbol{h}_{\alpha_{15}}.
\label{threeb15}
\end{eqnarray}
Then,
$\{ \alpha_{14}, \alpha_{15} \} \in V(\boldsymbol{s}_1)$.
From (\ref{threeb13}) and (\ref{threeb15}),
\begin{eqnarray}
  \boldsymbol{h}_{\alpha_{12}} + \boldsymbol{h}_{\alpha_{13}}
  = \boldsymbol{h}_{\alpha_{14}} + \boldsymbol{h}_{\alpha_{15}}.
  \label{threeb12plus1314plus15}
\end{eqnarray}
From (\ref{threeb12plus1314plus15}) and (\ref{threeb14}),
\begin{eqnarray}
  \boldsymbol{h}_{\alpha_{13}} + \boldsymbol{h}_{\alpha_{15}}
  = \boldsymbol{h}_{\alpha_{12}} + \boldsymbol{h}_{\alpha_{14}}
  = \boldsymbol{s}_2.
  \label{threeb13plus152}
\end{eqnarray}
Therefore,
$\{ \alpha_{13}, \alpha_{15} \} \in V(\boldsymbol{s}_2)$.
From (\ref{threeb12plus1314plus15}) and (\ref{threeb313plus14}),
\begin{eqnarray}
  \boldsymbol{h}_{\alpha_{12}} + \boldsymbol{h}_{\alpha_{15}}
  = \boldsymbol{h}_{\alpha_{13}} + \boldsymbol{h}_{\alpha_{14}}
  = \boldsymbol{s}_3.
  \label{threeb12plus153}
\end{eqnarray}
Hence,
$\{ \alpha_{12}, \alpha_{15} \} \in V(\boldsymbol{s}_3)$.
Let $\sigma$ be the permutation of $\mathcal{I}_{15}$
such that $\sigma(j) = \alpha_{j}$ for each
$j \in \mathcal{I}_{15}$.
Then, $\sigma$ satisfies (\ref{siki:sigma3b-1}),
(\ref{siki:sigma3b-2}), and (\ref{siki:sigma3b-3}).
\end{IEEEproof}

\begin{teiri}
\label{teiri:kouho4}
Let $r = 4$.
For any $\boldsymbol{s}_1, \boldsymbol{s}_2, \boldsymbol{s}_3,
\boldsymbol{s}_4
\in \{0,1\}^4 \setminus \{0000\}
\mbox{ }
(\boldsymbol{s}_i \not= \boldsymbol{s}_{i'} \mbox{ for any }
i, i' \in \mathcal{I}_4, i \not= i')$,
we have the permutation $\sigma$
of $\mathcal{I}_{15}$,
which
satisfies the following conditions:

\noindent
(a)
If $\boldsymbol{s}_1, \boldsymbol{s}_2, \boldsymbol{s}_3,$
and $\boldsymbol{s}_4$ are linearly independent,
\begin{eqnarray}
V(\boldsymbol{s}_{1}) & = &
\{ \{ \sigma(1) \}, \{ \sigma(2), \sigma(3) \},
\{ \sigma(4), \sigma(5) \}, \nonumber \\
& & \{ \sigma(6), \sigma(7) \},
\{ \sigma(8), \sigma(9) \}, \{ \sigma(10), \sigma(11) \},
\nonumber \\
& & \{ \sigma(12), \sigma(13) \},
\{ \sigma(14), \sigma(15) \} \},
\label{siki:sigma4a-1} \\
V(\boldsymbol{s}_{2}) & = &
\{ \{ \sigma(2) \}, \{ \sigma(1), \sigma(3) \},
\{ \sigma(4), \sigma(6) \}, \nonumber \\
& & \{ \sigma(5), \sigma(7) \},
\{ \sigma(8), \sigma(10) \}, \{ \sigma(9), \sigma(11) \},
\nonumber \\
& & \{ \sigma(12), \sigma(14) \}, \{ \sigma(13), \sigma(15) \} \},
\label{siki:sigma4a-2} \\
V(\boldsymbol{s}_{3}) & = &
\{ \{ \sigma(4) \}, \{ \sigma(1), \sigma(5) \},
\{ \sigma(2), \sigma(6) \}, \nonumber \\
& & \{ \sigma(3), \sigma(7) \},
\{ \sigma(8), \sigma(12) \}, \{ \sigma(9), \sigma(13) \},
\nonumber \\
& & \{ \sigma(10), \sigma(14) \}, \{ \sigma(11), \sigma(15) \} \},
\label{siki:sigma4a-3} \\
V(\boldsymbol{s}_{4}) & = &
\{ \{ \sigma(8) \}, \{ \sigma(1), \sigma(9) \},
\{ \sigma(2), \sigma(10) \}, \nonumber \\
& & \{ \sigma(3), \sigma(11) \},
\{ \sigma(4), \sigma(12) \}, \{ \sigma(5), \sigma(13) \},
\nonumber \\
& & \{ \sigma(6), \sigma(14) \}, \{ \sigma(7), \sigma(15) \} \}.
\label{siki:sigma4a-4}
\end{eqnarray}
(b)
If there exists the permutation $\tau$ of $\mathcal{I}_4$
such that
$\boldsymbol{s}_{\tau(1)} + \boldsymbol{s}_{\tau(2)}
= \boldsymbol{s}_{\tau(3)}$,
\begin{eqnarray}
V(\boldsymbol{s}_{\tau(1)}) & = &
\{ \{ \sigma(1) \}, \{ \sigma(2), \sigma(3) \},
\{ \sigma(4), \sigma(5) \}, \nonumber \\
& & \{ \sigma(6), \sigma(7) \},
\{ \sigma(8), \sigma(9) \}, \{ \sigma(10), \sigma(11) \},
\nonumber \\
& & \{ \sigma(12), \sigma(13) \} , \{ \sigma(14), \sigma(15) \} \},
\label{siki:sigma4b-1} \\
V(\boldsymbol{s}_{\tau(2)}) & = &
\{ \{ \sigma(2) \}, \{ \sigma(1), \sigma(3) \},
\{ \sigma(4), \sigma(6) \}, \nonumber \\
& & \{ \sigma(5), \sigma(7) \},
\{ \sigma(8), \sigma(10) \}, \{ \sigma(9), \sigma(11) \},
\nonumber \\
& & \{ \sigma(12), \sigma(14) \}, \{ \sigma(13), \sigma(15) \} \},
\label{siki:sigma4b-2} \\
V(\boldsymbol{s}_{\tau(3)}) & = &
\{ \{ \sigma(3) \}, \{ \sigma(1), \sigma(2) \},
\{ \sigma(4), \sigma(7) \}, \nonumber \\
& & \{ \sigma(5), \sigma(6) \},
\{ \sigma(8), \sigma(11) \}, \{ \sigma(9), \sigma(10) \},
\nonumber \\
& & \{ \sigma(12), \sigma(15) \}, \{ \sigma(13), \sigma(14) \} \},
\label{siki:sigma4b-3} \\
V(\boldsymbol{s}_{\tau(4)}) & = &
\{ \{ \sigma(4) \}, \{ \sigma(1), \sigma(5) \},
\{ \sigma(2), \sigma(6) \}, \nonumber \\
& & \{ \sigma(3), \sigma(7) \},
\{ \sigma(8), \sigma(12) \}, \{ \sigma(9), \sigma(13) \},
\nonumber \\
& & \{ \sigma(10), \sigma(14) \}, \{ \sigma(11), \sigma(15) \} \}.
\label{siki:sigma4b-4}
\end{eqnarray}
(c)
If
$\boldsymbol{s}_1 + \boldsymbol{s}_2 =
\boldsymbol{s}_3 + \boldsymbol{s}_4$,
\begin{eqnarray}
V(\boldsymbol{s}_{1}) & = &
\{ \{ \sigma(1) \}, \{ \sigma(2), \sigma(3) \},
\{ \sigma(4), \sigma(5) \}, \nonumber \\
& & \{ \sigma(6), \sigma(7) \},
\{ \sigma(8), \sigma(9) \}, \{ \sigma(10), \sigma(11) \},
\nonumber \\
& & \{ \sigma(12), \sigma(13) \} , \{ \sigma(14), \sigma(15) \} \},
\label{siki:sigma4c-1} \\
V(\boldsymbol{s}_{2}) & = &
\{ \{ \sigma(2) \}, \{ \sigma(1), \sigma(3) \},
\{ \sigma(4), \sigma(6) \}, \nonumber \\
& & \{ \sigma(5), \sigma(7) \},
\{ \sigma(8), \sigma(10) \}, \{ \sigma(9), \sigma(11) \},
\nonumber \\
& & \{ \sigma(12), \sigma(14) \}, \{ \sigma(13), \sigma(15) \} \},
\label{siki:sigma4c-2} \\
V(\boldsymbol{s}_{3}) & = &
\{ \{ \sigma(4) \}, \{ \sigma(1), \sigma(5) \},
\{ \sigma(2), \sigma(6) \}, \nonumber \\
& & \{ \sigma(3), \sigma(7) \},
\{ \sigma(8), \sigma(12) \}, \{ \sigma(9), \sigma(13) \},
\nonumber \\
& & \{ \sigma(10), \sigma(14) \}, \{ \sigma(11), \sigma(15) \} \},
\label{siki:sigma4c-3} \\
V(\boldsymbol{s}_{4}) & = &
\{ \{ \sigma(7) \}, \{ \sigma(1), \sigma(6) \},
\{ \sigma(2), \sigma(5) \}, \nonumber \\
& & \{ \sigma(3), \sigma(4) \},
\{ \sigma(8), \sigma(15) \}, \{ \sigma(9), \sigma(14) \},
\nonumber \\
& & \{ \sigma(10), \sigma(13) \}, \{ \sigma(11), \sigma(12) \} \}.
\label{siki:sigma4c-4}
\end{eqnarray}
\end{teiri}
\begin{IEEEproof}
Let $\alpha_1$ be an integer such that
$\alpha_1 \in \mathcal{I}_{15}$ and
\begin{eqnarray}
  \boldsymbol{s}_1 = \boldsymbol{h}_{\alpha_1}. \label{four1}
\end{eqnarray}
Let $\alpha_2$ be an integer such that
$\alpha_2 \in \mathcal{I}_{15}$ and
\begin{eqnarray}
  \boldsymbol{s}_2 = \boldsymbol{h}_{\alpha_2}. \label{four2}
\end{eqnarray}
Then, $\{ \alpha_1 \} \in V(\boldsymbol{s}_{1})$,
$\{ \alpha_2 \} \in V(\boldsymbol{s}_{2})$.
Clearly,
$\alpha_1 \not= \alpha_2$.
Let $\alpha_3$ be an integer such that
$\alpha_3 \in \mathcal{I}_{15} \setminus \{ \alpha_1, \alpha_2 \}$
and
\begin{eqnarray}
\boldsymbol{s}_1 = \boldsymbol{h}_{\alpha_2} + \boldsymbol{h}_{\alpha_3}.
\label{four3}
\end{eqnarray}
Then, $\{ \alpha_2, \alpha_3 \} \in V(\boldsymbol{s}_{1})$.
From (\ref{four1}), (\ref{four3}), and (\ref{four2}),
\begin{eqnarray}
  \boldsymbol{h}_{\alpha_1} + \boldsymbol{h}_{\alpha_3}
  & = & \boldsymbol{s}_1 + \boldsymbol{h}_{\alpha_3}
  = (\boldsymbol{h}_{\alpha_2} + \boldsymbol{h}_{\alpha_3})
  + \boldsymbol{h}_{\alpha_3}
  \nonumber \\
  & = & \boldsymbol{h}_{\alpha_2}
  = \boldsymbol{s}_2.
  \label{four1plus32}
\end{eqnarray}
Hence, $\{ \alpha_1, \alpha_3 \} \in V(\boldsymbol{s}_{2})$.

(a)
Let $\alpha_4$ be an integer such that
$\alpha_4 \in \mathcal{I}_{15}$ and
\begin{eqnarray}
  \boldsymbol{s}_3 = \boldsymbol{h}_{\alpha_4}.
  \label{foura4}
\end{eqnarray}
Then, $\{ \alpha_4 \} \in V(\boldsymbol{s}_3)$.
Clearly,
$\alpha_4 \not\in \{ \alpha_1, \alpha_2 \}$.
From (\ref{foura4}), (\ref{four3}), and (\ref{four2}),
\begin{eqnarray}
  \boldsymbol{h}_{\alpha_4} = \boldsymbol{s}_3
  \not= \boldsymbol{s}_1 + \boldsymbol{s}_2
  = (\boldsymbol{h}_{\alpha_2} + \boldsymbol{h}_{\alpha_3})
  + \boldsymbol{h}_{\alpha_2}
  = \boldsymbol{h}_{\alpha_3}.
  \label{foura4not3}
\end{eqnarray}
Therefore,
$\alpha_4 \not= \alpha_3$.
Let $\alpha_5$ be an integer such that
$\alpha_5 \in \mathcal{I}_{15} \setminus
\{ \alpha_1, \ldots, \alpha_4 \}$
and
\begin{eqnarray}
  \boldsymbol{s}_1 =
  \boldsymbol{h}_{\alpha_4} + \boldsymbol{h}_{\alpha_5}.
  \label{foura5}
\end{eqnarray}
Then,
$\{ \alpha_4, \alpha_5 \} \in V(\boldsymbol{s}_1)$.
From (\ref{four1}) and (\ref{foura5}),
\begin{eqnarray}
  \boldsymbol{h}_{\alpha_1} =
  \boldsymbol{h}_{\alpha_4} + \boldsymbol{h}_{\alpha_5}.
  \label{foura14plus5}
\end{eqnarray}
From (\ref{foura14plus5}) and (\ref{foura4}),
\begin{eqnarray}
  \boldsymbol{h}_{\alpha_1} + \boldsymbol{h}_{\alpha_5}
  = \boldsymbol{h}_{\alpha_4} = \boldsymbol{s}_3.
  \label{foura1plus53}
\end{eqnarray}
Hence,
$\{ \alpha_1, \alpha_5 \} \in V(\boldsymbol{s}_3)$.
Let $\alpha_6$ be an integer such that
$\alpha_6 \in \mathcal{I}_{15} \setminus
\{ \alpha_1, \ldots, \alpha_5 \}$
and
\begin{eqnarray}
  \boldsymbol{s}_2 =
  \boldsymbol{h}_{\alpha_4} + \boldsymbol{h}_{\alpha_6}.
  \label{foura6}
\end{eqnarray}
Then,
$\{ \alpha_4, \alpha_6 \} \in V(\boldsymbol{s}_2)$.
From (\ref{four2}) and (\ref{foura6}),
\begin{eqnarray}
  \boldsymbol{h}_{\alpha_2} =
  \boldsymbol{h}_{\alpha_4} + \boldsymbol{h}_{\alpha_6}.
  \label{foura24plus6}
\end{eqnarray}
From (\ref{foura24plus6}) and
(\ref{foura4}),
\begin{eqnarray}
  \boldsymbol{h}_{\alpha_2} + \boldsymbol{h}_{\alpha_6}
  = \boldsymbol{h}_{\alpha_4}
  = \boldsymbol{s}_3.
  \label{foura2plus63}
\end{eqnarray}
Therefore,
$\{ \alpha_2, \alpha_6 \} \in V(\boldsymbol{s}_3)$.
Let $\alpha_7$ be an integer such that
$\alpha_7 \in \mathcal{I}_{15} \setminus
\{ \alpha_1, \ldots, \alpha_6 \}$ and
\begin{eqnarray}
  \boldsymbol{s}_1 =
  \boldsymbol{h}_{\alpha_6} + \boldsymbol{h}_{\alpha_7}.
  \label{foura7}
\end{eqnarray}
Then,
$\{ \alpha_6, \alpha_7 \} \in V(\boldsymbol{s}_1)$.
From (\ref{foura5}) and (\ref{foura7}),
\begin{eqnarray}
  \boldsymbol{h}_{\alpha_4} + \boldsymbol{h}_{\alpha_5}
  = \boldsymbol{h}_{\alpha_6} + \boldsymbol{h}_{\alpha_7}.
  \label{foura4plus56plus7}
\end{eqnarray}
From (\ref{foura4plus56plus7})
and (\ref{foura6}),
\begin{eqnarray}
  \boldsymbol{h}_{\alpha_5} + \boldsymbol{h}_{\alpha_7}
  = \boldsymbol{h}_{\alpha_4} + \boldsymbol{h}_{\alpha_6}
  = \boldsymbol{s}_2.
  \label{foura5plus72}
\end{eqnarray}
Hence,
$\{ \alpha_5, \alpha_7 \} \in V(\boldsymbol{s}_2)$.
From (\ref{four3}) and (\ref{foura7}),
\begin{eqnarray}
  \boldsymbol{h}_{\alpha_2} + \boldsymbol{h}_{\alpha_3}
  = \boldsymbol{h}_{\alpha_6} + \boldsymbol{h}_{\alpha_7}.
  \label{foura2plus36plus7}
\end{eqnarray}
From (\ref{foura2plus36plus7})
and (\ref{foura2plus63}),
\begin{eqnarray}
  \boldsymbol{h}_{\alpha_3} + \boldsymbol{h}_{\alpha_7}
  = \boldsymbol{h}_{\alpha_2} + \boldsymbol{h}_{\alpha_6}
  = \boldsymbol{s}_3.
  \label{foura3plus73}
\end{eqnarray}
Therefore,
$\{ \alpha_3, \alpha_7 \} \in V(\boldsymbol{s}_3)$.
Let $\alpha_8$ be an integer such that
$\alpha_8 \in \mathcal{I}_{15}$ and
\begin{eqnarray}
  \boldsymbol{s}_4 = \boldsymbol{h}_{\alpha_8}.
  \label{foura8}
\end{eqnarray}
Then, $\{ \alpha_8 \} \in V(\boldsymbol{s}_4)$.
Clearly,
$\alpha_8 \not\in \{ \alpha_1, \ldots, \alpha_4 \}$.
From (\ref{foura8}), (\ref{foura5}), and
(\ref{foura4}),
\begin{eqnarray}
  \boldsymbol{h}_{\alpha_8} = \boldsymbol{s}_4
  \not= \boldsymbol{s}_1 + \boldsymbol{s}_3
  = (\boldsymbol{h}_{\alpha_4} + \boldsymbol{h}_{\alpha_5})
  + \boldsymbol{h}_{\alpha_4}
  = \boldsymbol{h}_{\alpha_5}.
  \label{foura8not5}
\end{eqnarray}
From (\ref{foura8}), (\ref{foura6}), and
(\ref{foura4}),
\begin{eqnarray}
  \boldsymbol{h}_{\alpha_8} = \boldsymbol{s}_4
  \not= \boldsymbol{s}_2 + \boldsymbol{s}_3
  = (\boldsymbol{h}_{\alpha_4} + \boldsymbol{h}_{\alpha_6})
  + \boldsymbol{h}_{\alpha_4}
  = \boldsymbol{h}_{\alpha_6}.
  \label{foura8not6}
\end{eqnarray}
From (\ref{foura8}), (\ref{foura7}), (\ref{foura6}),
and (\ref{foura4}),
\begin{eqnarray}
  & & \boldsymbol{h}_{\alpha_8} = \boldsymbol{s}_4
  \nonumber \\
  & \not= &
  \boldsymbol{s}_1 + \boldsymbol{s}_2 + \boldsymbol{s}_3
  = (\boldsymbol{h}_{\alpha_6} + \boldsymbol{h}_{\alpha_7})
  + (\boldsymbol{h}_{\alpha_4} + \boldsymbol{h}_{\alpha_6})
  + \boldsymbol{h}_{\alpha_4}
  \nonumber \\
  & & = \boldsymbol{h}_{\alpha_7}.
  \label{foura8not7}
\end{eqnarray}
From (\ref{foura8not5}), (\ref{foura8not6}),
and (\ref{foura8not7}),
$\alpha_8 \not\in \{ \alpha_1, \ldots, \alpha_7 \}$.
Let $\alpha_9$ be an integer such that
$\alpha_9 \in \mathcal{I}_{15}
\setminus \{ \alpha_1, \ldots, \alpha_8 \}$ and
\begin{eqnarray}
  \boldsymbol{s}_1 =
  \boldsymbol{h}_{\alpha_8} + \boldsymbol{h}_{\alpha_9}.
  \label{foura9}
\end{eqnarray}
Then, $\{ \alpha_8, \alpha_9 \} \in
V(\boldsymbol{s}_1)$.
From (\ref{four1}) and (\ref{foura9}),
\begin{eqnarray}
  \boldsymbol{h}_{\alpha_1}
  = \boldsymbol{h}_{\alpha_8} + \boldsymbol{h}_{\alpha_9}.
  \label{foura18plus9}
\end{eqnarray}
From (\ref{foura18plus9}) and (\ref{foura8}),
\begin{eqnarray}
  \boldsymbol{h}_{\alpha_1} + \boldsymbol{h}_{\alpha_9}
  = \boldsymbol{h}_{\alpha_8}
  = \boldsymbol{s}_4.
  \label{foura1plus94}
\end{eqnarray}
Hence,
$\{ \alpha_1, \alpha_9 \} \in V(\boldsymbol{s}_4)$.
Let $\alpha_{10}$ be an integer such that
$\alpha_{10} \in \mathcal{I}_{15}
\setminus \{ \alpha_1, \ldots, \alpha_9 \}$ and
\begin{eqnarray}
  \boldsymbol{s}_2 =
  \boldsymbol{h}_{\alpha_8} + \boldsymbol{h}_{\alpha_{10}}.
  \label{foura10}
\end{eqnarray}
Then,
$\{ \alpha_8, \alpha_{10} \} \in V(\boldsymbol{s}_2)$.
From (\ref{four2}) and (\ref{foura10}),
\begin{eqnarray}
  \boldsymbol{h}_{\alpha_2} =
  \boldsymbol{h}_{\alpha_8} + \boldsymbol{h}_{\alpha_{10}}.
  \label{foura28plus10}
\end{eqnarray}
From (\ref{foura28plus10})
and (\ref{foura8}),
\begin{eqnarray}
  \boldsymbol{h}_{\alpha_2} + \boldsymbol{h}_{\alpha_{10}}
  = \boldsymbol{h}_{\alpha_8}
  = \boldsymbol{s}_4.
  \label{foura2plus104}
\end{eqnarray}
Therefore,
$\{ \alpha_2, \alpha_{10} \} \in V(\boldsymbol{s}_4)$.
Let $\alpha_{11}$ be an integer such that
$\alpha_{11} \in \mathcal{I}_{15}
\setminus \{ \alpha_1, \ldots, \alpha_{10} \}$ and
\begin{eqnarray}
  \boldsymbol{s}_1 =
  \boldsymbol{h}_{\alpha_{10}} + \boldsymbol{h}_{\alpha_{11}}.
  \label{foura11}
\end{eqnarray}
Then,
$\{ \alpha_{10}, \alpha_{11} \} \in V(\boldsymbol{s}_1)$.
From (\ref{foura9}) and (\ref{foura11}),
\begin{eqnarray}
  \boldsymbol{h}_{\alpha_8} + \boldsymbol{h}_{\alpha_9}
  = \boldsymbol{h}_{\alpha_{10}} + \boldsymbol{h}_{\alpha_{11}}.
  \label{foura8plus910plus11}
\end{eqnarray}
From (\ref{foura8plus910plus11}) and
(\ref{foura10}),
\begin{eqnarray}
  \boldsymbol{h}_{\alpha_9} + \boldsymbol{h}_{\alpha_{11}}
  = \boldsymbol{h}_{\alpha_8} + \boldsymbol{h}_{\alpha_{10}}
  = \boldsymbol{s}_2.
  \label{foura9plus112}
\end{eqnarray}
Hence,
$\{ \alpha_9, \alpha_{11} \} \in V(\boldsymbol{s}_2)$.
From (\ref{four1plus32}) and (\ref{foura9plus112}),
\begin{eqnarray}
  \boldsymbol{h}_{\alpha_1} + \boldsymbol{h}_{\alpha_3}
  = \boldsymbol{h}_{\alpha_9} + \boldsymbol{h}_{\alpha_{11}}.
  \label{foura1plus39plus11}
\end{eqnarray}
From (\ref{foura1plus39plus11}) and (\ref{foura1plus94}),
\begin{eqnarray}
  \boldsymbol{h}_{\alpha_3} + \boldsymbol{h}_{\alpha_{11}}
  = \boldsymbol{h}_{\alpha_1} + \boldsymbol{h}_{\alpha_9}
  = \boldsymbol{s}_4.
  \label{foura3plus114}
\end{eqnarray}
Therefore,
$\{ \alpha_3, \alpha_{11} \} \in V(\boldsymbol{s}_4)$.
Let $\alpha_{12}$ be an integer such that
$\alpha_{12} \in
\mathcal{I}_{15} \setminus \{ \alpha_1, \ldots, \alpha_{10} \}$
and
\begin{eqnarray}
  \boldsymbol{s}_3 =
  \boldsymbol{h}_{\alpha_{8}} + \boldsymbol{h}_{\alpha_{12}}.
  \label{foura12}
\end{eqnarray}
Then, $\{ \alpha_8, \alpha_{12} \} \in V(\boldsymbol{s}_3)$.
From (\ref{foura12}), (\ref{foura9}), and (\ref{foura9plus112}),
\begin{eqnarray}
  & & \boldsymbol{h}_{\alpha_{8}} + \boldsymbol{h}_{\alpha_{12}}
  = \boldsymbol{s}_3
  \nonumber \\
  & \not= & \boldsymbol{s}_1 + \boldsymbol{s}_2
  = (\boldsymbol{h}_{\alpha_8} + \boldsymbol{h}_{\alpha_9}) +
  (\boldsymbol{h}_{\alpha_9} + \boldsymbol{h}_{\alpha_{11}})
  \nonumber \\
  & & = \boldsymbol{h}_{\alpha_8} + \boldsymbol{h}_{\alpha_{11}}.
  \label{foura8plus12not8plus11}
\end{eqnarray}
Therefore, $\alpha_{12} \not= \alpha_{11}$.
From (\ref{foura4}) and (\ref{foura12}),
\begin{eqnarray}
  \boldsymbol{h}_{\alpha_4}
  = \boldsymbol{h}_{\alpha_{8}} + \boldsymbol{h}_{\alpha_{12}}.
  \label{foura48plus12}
\end{eqnarray}
From (\ref{foura48plus12}) and (\ref{foura8}),
\begin{eqnarray}
  \boldsymbol{h}_{\alpha_4} + \boldsymbol{h}_{\alpha_{12}}
  = \boldsymbol{h}_{\alpha_{8}}
  = \boldsymbol{s}_4.
  \label{foura4plus124}
\end{eqnarray}
Hence,
$\{ \alpha_4, \alpha_{12} \} \in V(\boldsymbol{s}_4)$.
Let $\alpha_{13}$ be an integer such that
$\alpha_{13} \in
\mathcal{I}_{15} \setminus \{ \alpha_1, \ldots, \alpha_{12} \}$
and
\begin{eqnarray}
  \boldsymbol{s}_1 =
  \boldsymbol{h}_{\alpha_{12}} + \boldsymbol{h}_{\alpha_{13}}.
  \label{foura13}
\end{eqnarray}
Then,
$\{ \alpha_{12}, \alpha_{13} \} \in V(\boldsymbol{s}_1)$.
From (\ref{foura9}) and (\ref{foura13}),
\begin{eqnarray}
  \boldsymbol{h}_{\alpha_8} + \boldsymbol{h}_{\alpha_9}
  = \boldsymbol{h}_{\alpha_{12}} + \boldsymbol{h}_{\alpha_{13}}.
  \label{foura8plus912plus13}
\end{eqnarray}
From (\ref{foura8plus912plus13})
and (\ref{foura12}),
\begin{eqnarray}
  \boldsymbol{h}_{\alpha_9} + \boldsymbol{h}_{\alpha_{13}}
  = \boldsymbol{h}_{\alpha_8} + \boldsymbol{h}_{\alpha_{12}}
  = \boldsymbol{s}_3.
  \label{foura9plus133}
\end{eqnarray}
Therefore,
$\{ \alpha_{9}, \alpha_{13} \} \in V(\boldsymbol{s}_3)$.
From (\ref{foura5}) and (\ref{foura13}),
\begin{eqnarray}
  \boldsymbol{h}_{\alpha_4} + \boldsymbol{h}_{\alpha_5}
  = \boldsymbol{h}_{\alpha_{12}} + \boldsymbol{h}_{\alpha_{13}}.
  \label{foura4plus512plus13}
\end{eqnarray}
From (\ref{foura4plus512plus13}) and (\ref{foura4plus124}),
\begin{eqnarray}
  \boldsymbol{h}_{\alpha_5} + \boldsymbol{h}_{\alpha_{13}}
  = \boldsymbol{h}_{\alpha_4} + \boldsymbol{h}_{\alpha_{12}}
  = \boldsymbol{s}_4.
  \label{foura5plus134}
\end{eqnarray}
Hence,
$\{ \alpha_{5}, \alpha_{13} \} \in V(\boldsymbol{s}_4)$.
Let $\alpha_{14}$ be an integer such that
$\alpha_{14} \in
\mathcal{I}_{15} \setminus \{ \alpha_1, \ldots, \alpha_{13} \}$
and
\begin{eqnarray}
  \boldsymbol{s}_2 =
  \boldsymbol{h}_{\alpha_{12}} + \boldsymbol{h}_{\alpha_{14}}.
  \label{foura14}
\end{eqnarray}
Then,
$\{ \alpha_{12}, \alpha_{14} \} \in V(\boldsymbol{s}_2)$.
From (\ref{foura10}) and (\ref{foura14}),
\begin{eqnarray}
  \boldsymbol{h}_{\alpha_8} + \boldsymbol{h}_{\alpha_{10}}
  = \boldsymbol{h}_{\alpha_{12}} + \boldsymbol{h}_{\alpha_{14}}.
  \label{foura8plus1012plus14}
\end{eqnarray}
From (\ref{foura8plus1012plus14})
and (\ref{foura12}),
\begin{eqnarray}
  \boldsymbol{h}_{\alpha_{10}} + \boldsymbol{h}_{\alpha_{14}}
  = \boldsymbol{h}_{\alpha_8} + \boldsymbol{h}_{\alpha_{12}}
  = \boldsymbol{s}_3.
  \label{foura10plus143}
\end{eqnarray}
Therefore,
$\{ \alpha_{10}, \alpha_{14} \} \in V(\boldsymbol{s}_3)$.
From (\ref{foura6}) and (\ref{foura14}),
\begin{eqnarray}
  \boldsymbol{h}_{\alpha_4} + \boldsymbol{h}_{\alpha_6}
  = \boldsymbol{h}_{\alpha_{12}} + \boldsymbol{h}_{\alpha_{14}}.
  \label{foura4plus612plus14}
\end{eqnarray}
From (\ref{foura4plus612plus14}) and (\ref{foura4plus124}),
\begin{eqnarray}
  \boldsymbol{h}_{\alpha_6} + \boldsymbol{h}_{\alpha_{14}}
  = \boldsymbol{h}_{\alpha_4} + \boldsymbol{h}_{\alpha_{12}}
  = \boldsymbol{s}_4.
  \label{foura6plus144}
\end{eqnarray}
Hence,
$\{ \alpha_{6}, \alpha_{14} \} \in V(\boldsymbol{s}_4)$.
Let $\alpha_{15}$ be an integer such that
$\alpha_{15} \in
\mathcal{I}_{15} \setminus \{ \alpha_1, \ldots, \alpha_{14} \}$
and
\begin{eqnarray}
  \boldsymbol{s}_1 =
  \boldsymbol{h}_{\alpha_{14}} + \boldsymbol{h}_{\alpha_{15}}.
  \label{foura15}
\end{eqnarray}
Then,
$\{ \alpha_{14}, \alpha_{15} \} \in V(\boldsymbol{s}_1)$.
From (\ref{foura13}) and (\ref{foura15}),
\begin{eqnarray}
  \boldsymbol{h}_{\alpha_{12}} + \boldsymbol{h}_{\alpha_{13}}
  = \boldsymbol{h}_{\alpha_{14}} + \boldsymbol{h}_{\alpha_{15}}.
  \label{foura12plus1314plus15}
\end{eqnarray}
From (\ref{foura12plus1314plus15})
and (\ref{foura14}),
\begin{eqnarray}
  \boldsymbol{h}_{\alpha_{13}} + \boldsymbol{h}_{\alpha_{15}}
  = \boldsymbol{h}_{\alpha_{12}} + \boldsymbol{h}_{\alpha_{14}}
  = \boldsymbol{s}_2.
  \label{foura13plus152}
\end{eqnarray}
Therefore,
$\{ \alpha_{13}, \alpha_{15} \} \in V(\boldsymbol{s}_2)$.
From (\ref{foura11}) and (\ref{foura15}),
\begin{eqnarray}
  \boldsymbol{h}_{\alpha_{10}} + \boldsymbol{h}_{\alpha_{11}}
  = \boldsymbol{h}_{\alpha_{14}} + \boldsymbol{h}_{\alpha_{15}}.
  \label{foura10plus1114plus15}
\end{eqnarray}
From (\ref{foura10plus1114plus15})
and (\ref{foura10plus143}),
\begin{eqnarray}
  \boldsymbol{h}_{\alpha_{11}} + \boldsymbol{h}_{\alpha_{15}}
  = \boldsymbol{h}_{\alpha_{10}} + \boldsymbol{h}_{\alpha_{14}}
  = \boldsymbol{s}_3.
  \label{foura11plus153}
\end{eqnarray}
Hence,
$\{ \alpha_{11}, \alpha_{15} \} \in V(\boldsymbol{s}_3)$.
From (\ref{foura7})
and (\ref{foura15}),
\begin{eqnarray}
  \boldsymbol{h}_{\alpha_6} + \boldsymbol{h}_{\alpha_7}
  = \boldsymbol{h}_{\alpha_{14}} + \boldsymbol{h}_{\alpha_{15}}.
  \label{foura6plus714plus15}
\end{eqnarray}
From (\ref{foura6plus714plus15})
and (\ref{foura6plus144}),
\begin{eqnarray}
  \boldsymbol{h}_{\alpha_7} + \boldsymbol{h}_{\alpha_{15}}
  = \boldsymbol{h}_{\alpha_6} + \boldsymbol{h}_{\alpha_{14}}
  = \boldsymbol{s}_4.
  \label{foura7plus154}
\end{eqnarray}
Therefore,
$\{ \alpha_{7}, \alpha_{15} \} \in V(\boldsymbol{s}_4)$.
Let $\sigma$ be the permutation of
$\mathcal{I}_{15}$ such that
$\sigma(j) = \alpha_j$ for each
$j \in \mathcal{I}_{15}$.
Then, $\sigma$ satisfies
(\ref{siki:sigma4a-1}), (\ref{siki:sigma4a-2}),
(\ref{siki:sigma4a-3}), and
(\ref{siki:sigma4a-4}).

(b)
For simplicity,
suppose
$\boldsymbol{s}_1 + \boldsymbol{s}_2 = \boldsymbol{s}_3$.
From (\ref{four3}) and (\ref{four2}),
\begin{eqnarray}
  \boldsymbol{s}_3 = \boldsymbol{s}_1 + \boldsymbol{s}_2
  = (\boldsymbol{h}_{\alpha_2} + \boldsymbol{h}_{\alpha_3})
  + \boldsymbol{h}_{\alpha_2}
  = \boldsymbol{h}_{\alpha_3}.
  \label{fourb33}
\end{eqnarray}
Hence,
$\{ \alpha_{3} \} \in V(\boldsymbol{s}_3)$.
From (\ref{four1}) and (\ref{four2}),
\begin{eqnarray}
  \boldsymbol{h}_{\alpha_1} + \boldsymbol{h}_{\alpha_2}
  = \boldsymbol{s}_1 + \boldsymbol{s}_2
  = \boldsymbol{s}_3.
  \label{fourb1plus23}
\end{eqnarray}
Therefore,
$\{ \alpha_{1}, \alpha_{2} \} \in V(\boldsymbol{s}_3)$.
Let $\alpha_4$ be an integer such that
$\alpha_4 \in \mathcal{I}_{15}$ and
\begin{eqnarray}
  \boldsymbol{s}_4 =
  \boldsymbol{h}_{\alpha_4}.
  \label{fourb4}
\end{eqnarray}
Then, $\{ \alpha_4 \} \in V(\boldsymbol{s}_4)$.
Clearly,
$\alpha_4 \not\in \{ \alpha_1, \alpha_2, \alpha_3 \}$.
Let $\alpha_5$ be an integer such that
$\alpha_5 \in \mathcal{I}_{15}
\setminus \{ \alpha_1, \ldots, \alpha_4 \}$
and
\begin{eqnarray}
  \boldsymbol{s}_1 = \boldsymbol{h}_{\alpha_4}
  + \boldsymbol{h}_{\alpha_5}.
  \label{fourb5}
\end{eqnarray}
Then, $\{ \alpha_4, \alpha_5 \} \in V(\boldsymbol{s}_1)$.
From (\ref{four1}) and (\ref{fourb5}),
\begin{eqnarray}
  \boldsymbol{h}_{\alpha_1}
  = \boldsymbol{h}_{\alpha_4} + \boldsymbol{h}_{\alpha_5}.
  \label{fourb14plus5}
\end{eqnarray}
From (\ref{fourb14plus5}) and (\ref{fourb4}),
\begin{eqnarray}
  \boldsymbol{h}_{\alpha_1} + \boldsymbol{h}_{\alpha_5}
  = \boldsymbol{h}_{\alpha_4}
  = \boldsymbol{s}_4.
  \label{fourb1plus54}
\end{eqnarray}
Hence,
$\{ \alpha_1, \alpha_5 \} \in V(\boldsymbol{s}_4)$.
Let $\alpha_6$ be an integer such that
$\alpha_6 \in \mathcal{I}_{15}
\setminus \{ \alpha_1, \ldots, \alpha_5 \}$
and
\begin{eqnarray}
  \boldsymbol{s}_2 =
  \boldsymbol{h}_{\alpha_4}
  + \boldsymbol{h}_{\alpha_6}.
  \label{fourb6}
\end{eqnarray}
Then,
$\{ \alpha_4, \alpha_6 \} \in V(\boldsymbol{s}_2)$.
From (\ref{fourb5}) and (\ref{fourb6}),
\begin{eqnarray}
  \boldsymbol{s}_3 & = &
  \boldsymbol{s}_1 + \boldsymbol{s}_2
  = (\boldsymbol{h}_{\alpha_4} + \boldsymbol{h}_{\alpha_5})
  + (\boldsymbol{h}_{\alpha_4} + \boldsymbol{h}_{\alpha_6})
  \nonumber \\
  & = & \boldsymbol{h}_{\alpha_5} + \boldsymbol{h}_{\alpha_6}.
  \label{fourb35plus6}
\end{eqnarray}
Therefore,
$\{ \alpha_5, \alpha_6 \} \in V(\boldsymbol{s}_3)$.
From (\ref{four2}) and (\ref{fourb6}),
\begin{eqnarray}
  \boldsymbol{h}_{\alpha_2}
  = \boldsymbol{h}_{\alpha_4} + \boldsymbol{h}_{\alpha_6}.
  \label{fourb24plus6}
\end{eqnarray}
From (\ref{fourb24plus6}) and (\ref{fourb4}),
\begin{eqnarray}
  \boldsymbol{h}_{\alpha_2} + \boldsymbol{h}_{\alpha_6}
  = \boldsymbol{h}_{\alpha_4}
  = \boldsymbol{s}_4.
  \label{fourb2plus64}
\end{eqnarray}
Hence,
$\{ \alpha_2, \alpha_6 \} \in V(\boldsymbol{s}_4)$.
Let $\alpha_7$ be an integer such that
$\alpha_7 \in \mathcal{I}_{15}
\setminus \{ \alpha_1, \ldots, \alpha_6 \}$
and
\begin{eqnarray}
  \boldsymbol{s}_1 =
  \boldsymbol{h}_{\alpha_6} + \boldsymbol{h}_{\alpha_7}.
  \label{fourb7}
\end{eqnarray}
Then,
$\{ \alpha_6, \alpha_7 \} \in V(\boldsymbol{s}_1)$.
From (\ref{fourb5}) and (\ref{fourb7}),
\begin{eqnarray}
  \boldsymbol{h}_{\alpha_4} + \boldsymbol{h}_{\alpha_5}
  = \boldsymbol{h}_{\alpha_6} + \boldsymbol{h}_{\alpha_7}.
  \label{fourb4plus56plus7}
\end{eqnarray}
From (\ref{fourb4plus56plus7})
and (\ref{fourb6}),
\begin{eqnarray}
  \boldsymbol{h}_{\alpha_5} + \boldsymbol{h}_{\alpha_7}
  = \boldsymbol{h}_{\alpha_4} + \boldsymbol{h}_{\alpha_6}
  = \boldsymbol{s}_2.
  \label{fourb5plus72}
\end{eqnarray}
Therefore,
$\{ \alpha_5, \alpha_7 \} \in V(\boldsymbol{s}_2)$.
From (\ref{fourb4plus56plus7}) and (\ref{fourb35plus6}),
\begin{eqnarray}
  \boldsymbol{h}_{\alpha_4} + \boldsymbol{h}_{\alpha_7}
  = \boldsymbol{h}_{\alpha_5} + \boldsymbol{h}_{\alpha_6}
  = \boldsymbol{s}_3.
  \label{fourb4plus73}
\end{eqnarray}
Hence,
$\{ \alpha_4, \alpha_7 \} \in V(\boldsymbol{s}_3)$.
From (\ref{four3}) and (\ref{fourb7}),
\begin{eqnarray}
  \boldsymbol{h}_{\alpha_2} + \boldsymbol{h}_{\alpha_3}
  = \boldsymbol{h}_{\alpha_6} + \boldsymbol{h}_{\alpha_7}.
  \label{fourb2plus36plus7}
\end{eqnarray}
From (\ref{fourb2plus36plus7}) and (\ref{fourb2plus64}),
\begin{eqnarray}
  \boldsymbol{h}_{\alpha_3} + \boldsymbol{h}_{\alpha_7}
  = \boldsymbol{h}_{\alpha_2} + \boldsymbol{h}_{\alpha_6}
  = \boldsymbol{s}_4.
  \label{fourb3plus74}
\end{eqnarray}
Therefore,
$\{ \alpha_3, \alpha_7 \} \in V(\boldsymbol{s}_4)$.
Let $\alpha_8$ be an integer such that
$\alpha_8 \in \mathcal{I}_{15}
\setminus \{ \alpha_1, \ldots, \alpha_7 \}$,
and
let $\alpha_9$ be an integer such that
$\alpha_9 \in \mathcal{I}_{15}
\setminus \{ \alpha_1, \ldots, \alpha_8 \}$
and
\begin{eqnarray}
  \boldsymbol{s}_1 =
  \boldsymbol{h}_{\alpha_8} + \boldsymbol{h}_{\alpha_9}.
  \label{fourb9}
\end{eqnarray}
Then,
$\{ \alpha_8, \alpha_9 \} \in V(\boldsymbol{s}_1)$.
Let $\alpha_{10}$ be an integer such that
$\alpha_{10} \in \mathcal{I}_{15}
\setminus \{ \alpha_1, \ldots, \alpha_9 \}$
and
\begin{eqnarray}
  \boldsymbol{s}_2 =
  \boldsymbol{h}_{\alpha_8} + \boldsymbol{h}_{\alpha_{10}}.
  \label{fourb10}
\end{eqnarray}
Then,
$\{ \alpha_8, \alpha_{10} \} \in V(\boldsymbol{s}_2)$.
From (\ref{fourb9}) and (\ref{fourb10}),
\begin{eqnarray}
  \boldsymbol{s}_3 & = &
  \boldsymbol{s}_1 + \boldsymbol{s}_2
  = (\boldsymbol{h}_{\alpha_8} + \boldsymbol{h}_{\alpha_9})
  + (\boldsymbol{h}_{\alpha_8} + \boldsymbol{h}_{\alpha_{10}})
  \nonumber \\
  & = & \boldsymbol{h}_{\alpha_9} + \boldsymbol{h}_{\alpha_{10}}.
  \label{fourb39plus10}
\end{eqnarray}
Hence,
$\{ \alpha_9, \alpha_{10} \} \in V(\boldsymbol{s}_3)$.
Let $\alpha_{11}$ be an integer such that
$\alpha_{11} \in \mathcal{I}_{15}
\setminus \{ \alpha_1, \ldots, \alpha_{10} \}$
and
\begin{eqnarray}
  \boldsymbol{s}_1 =
  \boldsymbol{h}_{\alpha_{10}} + \boldsymbol{h}_{\alpha_{11}}.
  \label{fourb11}
\end{eqnarray}
Then,
$\{ \alpha_{10}, \alpha_{11} \} \in V(\boldsymbol{s}_1)$.
From (\ref{fourb9}) and (\ref{fourb11}),
\begin{eqnarray}
  \boldsymbol{h}_{\alpha_8} + \boldsymbol{h}_{\alpha_9}
  = \boldsymbol{h}_{\alpha_{10}} + \boldsymbol{h}_{\alpha_{11}}.
  \label{fourb8plus910plus11}
\end{eqnarray}
From (\ref{fourb8plus910plus11}) and
(\ref{fourb10}),
\begin{eqnarray}
  \boldsymbol{h}_{\alpha_9} + \boldsymbol{h}_{\alpha_{11}}
  = \boldsymbol{h}_{\alpha_8} + \boldsymbol{h}_{\alpha_{10}}
  = \boldsymbol{s}_2.
  \label{fourb9plus112}
\end{eqnarray}
Therefore,
$\{ \alpha_{9}, \alpha_{11} \} \in V(\boldsymbol{s}_2)$.
From (\ref{fourb8plus910plus11}) and
(\ref{fourb39plus10}),
\begin{eqnarray}
  \boldsymbol{h}_{\alpha_8} + \boldsymbol{h}_{\alpha_{11}}
  = \boldsymbol{h}_{\alpha_9} + \boldsymbol{h}_{\alpha_{10}}
  = \boldsymbol{s}_3.
  \label{fourb8plus113}
\end{eqnarray}
Hence,
$\{ \alpha_{8}, \alpha_{11} \} \in V(\boldsymbol{s}_3)$.
Let $\alpha_{12}$ be an integer such that
$\alpha_{12} \in \mathcal{I}_{15}
\setminus \{ \alpha_1, \ldots, \alpha_{11} \}$
and
\begin{eqnarray}
  \boldsymbol{s}_4 =
  \boldsymbol{h}_{\alpha_{8}} + \boldsymbol{h}_{\alpha_{12}}.
  \label{fourb12}
\end{eqnarray}
Then,
$\{ \alpha_{8}, \alpha_{12} \} \in V(\boldsymbol{s}_4)$.
Let $\alpha_{13}$ be an integer such that
$\alpha_{13} \in \mathcal{I}_{15}
\setminus \{ \alpha_1, \ldots, \alpha_{12} \}$
and
\begin{eqnarray}
  \boldsymbol{s}_1 =
  \boldsymbol{h}_{\alpha_{12}} + \boldsymbol{h}_{\alpha_{13}}.
  \label{fourb13}
\end{eqnarray}
Then,
$\{ \alpha_{12}, \alpha_{13} \} \in V(\boldsymbol{s}_1)$.
From (\ref{fourb9}) and (\ref{fourb13}),
\begin{eqnarray}
  \boldsymbol{h}_{\alpha_8} + \boldsymbol{h}_{\alpha_9}
  = \boldsymbol{h}_{\alpha_{12}} + \boldsymbol{h}_{\alpha_{13}}.
  \label{fourb8plus912plus13}
\end{eqnarray}
From (\ref{fourb8plus912plus13})
and (\ref{fourb12}),
\begin{eqnarray}
  \boldsymbol{h}_{\alpha_9} + \boldsymbol{h}_{\alpha_{13}}
  = \boldsymbol{h}_{\alpha_8} + \boldsymbol{h}_{\alpha_{12}}
  = \boldsymbol{s}_4.
  \label{fourb9plus134}
\end{eqnarray}
Therefore,
$\{ \alpha_{9}, \alpha_{13} \} \in V(\boldsymbol{s}_4)$.
Let $\alpha_{14}$ be an integer such that
$\alpha_{14} \in \mathcal{I}_{15}
\setminus \{ \alpha_1, \ldots, \alpha_{13} \}$
and
\begin{eqnarray}
  \boldsymbol{s}_2 =
  \boldsymbol{h}_{\alpha_{12}} + \boldsymbol{h}_{\alpha_{14}}.
  \label{fourb14}
\end{eqnarray}
Then,
$\{ \alpha_{12}, \alpha_{14} \} \in V(\boldsymbol{s}_2)$.
From (\ref{fourb13}) and (\ref{fourb14}),
\begin{eqnarray}
  \boldsymbol{s}_3 & = &
  \boldsymbol{s}_1 + \boldsymbol{s}_2
  = (\boldsymbol{h}_{\alpha_{12}} + \boldsymbol{h}_{\alpha_{13}})
  + (\boldsymbol{h}_{\alpha_{12}} + \boldsymbol{h}_{\alpha_{14}})
  \nonumber \\
  & = & \boldsymbol{h}_{\alpha_{13}} + \boldsymbol{h}_{\alpha_{14}}.
  \label{fourb313plus14}
\end{eqnarray}
Hence,
$\{ \alpha_{13}, \alpha_{14} \} \in V(\boldsymbol{s}_3)$.
From (\ref{fourb10}) and (\ref{fourb14}),
\begin{eqnarray}
  \boldsymbol{h}_{\alpha_8} + \boldsymbol{h}_{\alpha_{10}}
  = \boldsymbol{h}_{\alpha_{12}} + \boldsymbol{h}_{\alpha_{14}}.
  \label{fourb8plus1012plus14}
\end{eqnarray}
From (\ref{fourb8plus1012plus14})
and (\ref{fourb12}),
\begin{eqnarray}
  \boldsymbol{h}_{\alpha_{10}} + \boldsymbol{h}_{\alpha_{14}}
  = \boldsymbol{h}_{\alpha_8} + \boldsymbol{h}_{\alpha_{12}}
  = \boldsymbol{s}_4.
  \label{fourb10plus144}
\end{eqnarray}
Therefore,
$\{ \alpha_{10}, \alpha_{14} \} \in V(\boldsymbol{s}_4)$.
Let $\alpha_{15}$ be an integer such that
$\alpha_{15} \in \mathcal{I}_{15}
\setminus \{ \alpha_1, \ldots, \alpha_{14} \}$
and
\begin{eqnarray}
  \boldsymbol{s}_1 =
  \boldsymbol{h}_{\alpha_{14}} + \boldsymbol{h}_{\alpha_{15}}.
  \label{fourb15}
\end{eqnarray}
Then,
$\{ \alpha_{14}, \alpha_{15} \} \in V(\boldsymbol{s}_1)$.
From (\ref{fourb13}) and (\ref{fourb15}),
\begin{eqnarray}
  \boldsymbol{h}_{\alpha_{12}} + \boldsymbol{h}_{\alpha_{13}}
  = \boldsymbol{h}_{\alpha_{14}} + \boldsymbol{h}_{\alpha_{15}}.
  \label{fourb12plus1314plus15}
\end{eqnarray}
From (\ref{fourb12plus1314plus15})
and (\ref{fourb14}),
\begin{eqnarray}
  \boldsymbol{h}_{\alpha_{13}} + \boldsymbol{h}_{\alpha_{15}}
  = \boldsymbol{h}_{\alpha_{12}} + \boldsymbol{h}_{\alpha_{14}}
  = \boldsymbol{s}_2.
  \label{fourb13plus152}
\end{eqnarray}
Hence,
$\{ \alpha_{13}, \alpha_{15} \} \in V(\boldsymbol{s}_2)$.
From (\ref{fourb12plus1314plus15}) and
(\ref{fourb313plus14}),
\begin{eqnarray}
  \boldsymbol{h}_{\alpha_{12}} + \boldsymbol{h}_{\alpha_{15}}
  = \boldsymbol{h}_{\alpha_{13}} + \boldsymbol{h}_{\alpha_{14}}
  = \boldsymbol{s}_3.
  \label{fourb12plus153}
\end{eqnarray}
Therefore,
$\{ \alpha_{12}, \alpha_{15} \} \in V(\boldsymbol{s}_3)$.
From (\ref{fourb11}) and (\ref{fourb15}),
\begin{eqnarray}
  \boldsymbol{h}_{\alpha_{10}} + \boldsymbol{h}_{\alpha_{11}}
  = \boldsymbol{h}_{\alpha_{14}} + \boldsymbol{h}_{\alpha_{15}}.
  \label{fourb10plus1114plus15}
\end{eqnarray}
From (\ref{fourb10plus1114plus15})
and (\ref{fourb10plus144}),
\begin{eqnarray}
  \boldsymbol{h}_{\alpha_{11}} + \boldsymbol{h}_{\alpha_{15}}
  = \boldsymbol{h}_{\alpha_{10}} + \boldsymbol{h}_{\alpha_{14}}
  = \boldsymbol{s}_4.
\end{eqnarray}
Hence,
$\{ \alpha_{11}, \alpha_{15} \} \in V(\boldsymbol{s}_4)$.
Let $\sigma$ be the permutation of $\mathcal{I}_{15}$
such that
$\sigma(j) = \alpha_j$ for each $j \in \mathcal{I}_{15}$.
Then,
$\sigma$ satisfies (\ref{siki:sigma4b-1}),
(\ref{siki:sigma4b-2}), (\ref{siki:sigma4b-3}),
and (\ref{siki:sigma4b-4}).

(c)
Let $\alpha_4$ be an integer such that
$\alpha_4 \in \mathcal{I}_{15}$ and
\begin{eqnarray}
  \boldsymbol{s}_3 = \boldsymbol{h}_{\alpha_4}.
  \label{fourc4}
\end{eqnarray}
Then,
$\{ \alpha_{4} \} \in V(\boldsymbol{s}_3)$.
Clearly,
$\alpha_4 \not\in \{ \alpha_1, \alpha_2 \}$.
From (\ref{fourc4}), (\ref{four3}),
and (\ref{four2}),
\begin{eqnarray}
  \boldsymbol{h}_{\alpha_4} = \boldsymbol{s}_3
  \not= \boldsymbol{s}_1 + \boldsymbol{s}_2
  = (\boldsymbol{h}_{\alpha_2} + \boldsymbol{h}_{\alpha_3})
  + \boldsymbol{h}_{\alpha_2}
  = \boldsymbol{h}_{\alpha_3}
  \label{fourc4not3}
\end{eqnarray}
Therefore,
$\alpha_4 \not= \alpha_3$.
From (\ref{four1}), (\ref{four1plus32}),
and (\ref{fourc4}),
\begin{eqnarray}
  \boldsymbol{s}_4
  & = & \boldsymbol{s}_1 + \boldsymbol{s}_2 + \boldsymbol{s}_3
  = \boldsymbol{h}_{\alpha_1} +
  (\boldsymbol{h}_{\alpha_1} + \boldsymbol{h}_{\alpha_3})
  + \boldsymbol{h}_{\alpha_4}
  \nonumber \\
  & = & \boldsymbol{h}_{\alpha_3} + \boldsymbol{h}_{\alpha_4}.
  \label{fourc43plus4}
\end{eqnarray}
Hence,
$\{ \alpha_{3}, \alpha_{4} \} \in V(\boldsymbol{s}_4)$.
Let $\alpha_5$ be an integer such that
$\alpha_5 \in
\mathcal{I}_{15} \setminus \{ \alpha_1, \ldots, \alpha_4 \}$
and
\begin{eqnarray}
  \boldsymbol{s}_1 =
  \boldsymbol{h}_{\alpha_4} + \boldsymbol{h}_{\alpha_5}.
  \label{fourc5}
\end{eqnarray}
Then,
$\{ \alpha_{4}, \alpha_{5} \} \in V(\boldsymbol{s}_1)$.
From (\ref{four1}) and (\ref{fourc5}),
\begin{eqnarray}
  \boldsymbol{h}_{\alpha_1}
  = \boldsymbol{h}_{\alpha_4} + \boldsymbol{h}_{\alpha_5}.
  \label{fourc14plus5}
\end{eqnarray}
From (\ref{fourc14plus5}) and (\ref{fourc4}),
\begin{eqnarray}
  \boldsymbol{h}_{\alpha_1} + \boldsymbol{h}_{\alpha_5}
  = \boldsymbol{h}_{\alpha_4}
  = \boldsymbol{s}_3.
  \label{fourc1plus53}
\end{eqnarray}
Therefore,
$\{ \alpha_{1}, \alpha_{5} \} \in V(\boldsymbol{s}_3)$.
From (\ref{four3}) and (\ref{fourc5}),
\begin{eqnarray}
  \boldsymbol{h}_{\alpha_2} + \boldsymbol{h}_{\alpha_3}
  = \boldsymbol{h}_{\alpha_4} + \boldsymbol{h}_{\alpha_5}.
  \label{fourc2plus34plus5}
\end{eqnarray}
From (\ref{fourc2plus34plus5})
and (\ref{fourc43plus4}),
\begin{eqnarray}
  \boldsymbol{h}_{\alpha_2} + \boldsymbol{h}_{\alpha_5}
  = \boldsymbol{h}_{\alpha_3} + \boldsymbol{h}_{\alpha_4}
  = \boldsymbol{s}_4.
  \label{fourc2plus54}
\end{eqnarray}
Hence,
$\{ \alpha_{2}, \alpha_{5} \} \in V(\boldsymbol{s}_4)$.
Let $\alpha_6$
be an integer such that
$\alpha_6 \in
\mathcal{I}_{15} \setminus \{ \alpha_1, \ldots, \alpha_5 \}$
and
\begin{eqnarray}
  \boldsymbol{s}_2 =
  \boldsymbol{h}_{\alpha_4} + \boldsymbol{h}_{\alpha_6}.
  \label{fourc6}
\end{eqnarray}
Then,
$\{ \alpha_{4}, \alpha_{6} \} \in V(\boldsymbol{s}_2)$.
From (\ref{four2}) and (\ref{fourc6}),
\begin{eqnarray}
  \boldsymbol{h}_{\alpha_2}
  = \boldsymbol{h}_{\alpha_4} + \boldsymbol{h}_{\alpha_6}.
  \label{fourc24plus6}
\end{eqnarray}
From (\ref{fourc24plus6})
and (\ref{fourc4}),
\begin{eqnarray}
  \boldsymbol{h}_{\alpha_2} + \boldsymbol{h}_{\alpha_6}
  = \boldsymbol{h}_{\alpha_4}
  = \boldsymbol{s}_3.
  \label{fourc2plus63}
\end{eqnarray}
Therefore,
$\{ \alpha_{2}, \alpha_{6} \} \in V(\boldsymbol{s}_3)$.
From (\ref{four1plus32}) and (\ref{fourc6}),
\begin{eqnarray}
  \boldsymbol{h}_{\alpha_1} + \boldsymbol{h}_{\alpha_3}
  = \boldsymbol{h}_{\alpha_4} + \boldsymbol{h}_{\alpha_6}.
  \label{fourc1plus34plus6}
\end{eqnarray}
From (\ref{fourc1plus34plus6}) and
(\ref{fourc43plus4}),
\begin{eqnarray}
  \boldsymbol{h}_{\alpha_1} + \boldsymbol{h}_{\alpha_6}
  = \boldsymbol{h}_{\alpha_3} + \boldsymbol{h}_{\alpha_4}
  = \boldsymbol{s}_4.
  \label{fourc1plus64}
\end{eqnarray}
Hence,
$\{ \alpha_{1}, \alpha_{6} \} \in V(\boldsymbol{s}_4)$.
Let $\alpha_7$ be an integer such that
$\alpha_7 \in
\mathcal{I}_{15} \setminus \{ \alpha_1, \ldots, \alpha_6 \}$
and
\begin{eqnarray}
  \boldsymbol{s}_1 =
  \boldsymbol{h}_{\alpha_6} + \boldsymbol{h}_{\alpha_7}.
  \label{fourc7}
\end{eqnarray}
Then,
$\{ \alpha_{6}, \alpha_{7} \} \in V(\boldsymbol{s}_1)$.
From (\ref{fourc5}) and (\ref{fourc7}),
\begin{eqnarray}
  \boldsymbol{h}_{\alpha_4} + \boldsymbol{h}_{\alpha_5}
  = \boldsymbol{h}_{\alpha_6} + \boldsymbol{h}_{\alpha_7}.
  \label{fourc4plus56plus7}
\end{eqnarray}
From (\ref{fourc4plus56plus7})
and (\ref{fourc6}),
\begin{eqnarray}
  \boldsymbol{h}_{\alpha_5} + \boldsymbol{h}_{\alpha_7}
  = \boldsymbol{h}_{\alpha_4} + \boldsymbol{h}_{\alpha_6}
  = \boldsymbol{s}_2.
  \label{fourc5plus72}
\end{eqnarray}
Therefore,
$\{ \alpha_{5}, \alpha_{7} \} \in V(\boldsymbol{s}_2)$.
From (\ref{four3}) and (\ref{fourc7}),
\begin{eqnarray}
  \boldsymbol{h}_{\alpha_2} + \boldsymbol{h}_{\alpha_3}
  = \boldsymbol{h}_{\alpha_6} + \boldsymbol{h}_{\alpha_7}.
  \label{fourc2plus36plus7}
\end{eqnarray}
From (\ref{fourc2plus36plus7}) and
(\ref{fourc2plus63}),
\begin{eqnarray}
  \boldsymbol{h}_{\alpha_3} + \boldsymbol{h}_{\alpha_7}
  = \boldsymbol{h}_{\alpha_2} + \boldsymbol{h}_{\alpha_6}
  = \boldsymbol{s}_3.
  \label{fourc3plus73}
\end{eqnarray}
Hence,
$\{ \alpha_{3}, \alpha_{7} \} \in V(\boldsymbol{s}_3)$.
From (\ref{fourc7}), (\ref{fourc6}),
and (\ref{fourc4}),
\begin{eqnarray}
  \boldsymbol{s}_4 & = &
  \boldsymbol{s}_1 + \boldsymbol{s}_2 + \boldsymbol{s}_3
  = (\boldsymbol{h}_{\alpha_6} + \boldsymbol{h}_{\alpha_7})
  + (\boldsymbol{h}_{\alpha_4} + \boldsymbol{h}_{\alpha_6})
  + \boldsymbol{h}_{\alpha_4}
  \nonumber \\
  & = & \boldsymbol{h}_{\alpha_7}.
  \label{fourc47}
\end{eqnarray}
Therefore,
$\{ \alpha_{7} \} \in V(\boldsymbol{s}_4)$.
Let $\alpha_{8}$ be an integer such that
$\alpha_8 \in
\mathcal{I}_{15} \setminus \{ \alpha_1, \ldots, \alpha_7 \}$,
and
let $\alpha_9$ be an integer such that
$\alpha_9 \in
\mathcal{I}_{15} \setminus \{ \alpha_1, \ldots, \alpha_8 \}$
and
\begin{eqnarray}
  \boldsymbol{s}_1 =
  \boldsymbol{h}_{\alpha_8} + \boldsymbol{h}_{\alpha_9}.
  \label{fourc9}
\end{eqnarray}
Then,
$\{ \alpha_{8}, \alpha_{9} \} \in V(\boldsymbol{s}_1)$.
Let $\alpha_{10}$ be an integer such that
$\alpha_{10} \in
\mathcal{I}_{15} \setminus \{ \alpha_1, \ldots, \alpha_9 \}$
and
\begin{eqnarray}
  \boldsymbol{s}_2 =
  \boldsymbol{h}_{\alpha_8} + \boldsymbol{h}_{\alpha_{10}}.
  \label{fourc10}
\end{eqnarray}
Then,
$\{ \alpha_{8}, \alpha_{10} \} \in V(\boldsymbol{s}_2)$.
Let $\alpha_{11}$ be an integer such that
$\alpha_{11} \in
\mathcal{I}_{15} \setminus \{ \alpha_1, \ldots, \alpha_{10} \}$
and
\begin{eqnarray}
  \boldsymbol{s}_1 =
  \boldsymbol{h}_{\alpha_{10}} + \boldsymbol{h}_{\alpha_{11}}.
  \label{fourc11}
\end{eqnarray}
Then,
$\{ \alpha_{10}, \alpha_{11} \} \in V(\boldsymbol{s}_1)$.
From (\ref{fourc9}) and (\ref{fourc11}),
\begin{eqnarray}
  \boldsymbol{h}_{\alpha_8} + \boldsymbol{h}_{\alpha_9}
  = \boldsymbol{h}_{\alpha_{10}} + \boldsymbol{h}_{\alpha_{11}}.
  \label{fourc8plus910plus11}
\end{eqnarray}
From (\ref{fourc8plus910plus11})
and (\ref{fourc10}),
\begin{eqnarray}
  \boldsymbol{h}_{\alpha_9} + \boldsymbol{h}_{\alpha_{11}}
  = \boldsymbol{h}_{\alpha_8} + \boldsymbol{h}_{\alpha_{10}}
  = \boldsymbol{s}_2.
  \label{fourc9plus112}
\end{eqnarray}
Hence,
$\{ \alpha_{9}, \alpha_{11} \} \in V(\boldsymbol{s}_2)$.
Let $\alpha_{12}$ be an integer such that
$\alpha_{12} \in
\mathcal{I}_{15} \setminus \{ \alpha_1, \ldots, \alpha_{10} \}$
and
\begin{eqnarray}
  \boldsymbol{s}_3 =
  \boldsymbol{h}_{\alpha_{8}} + \boldsymbol{h}_{\alpha_{12}}.
  \label{fourc12}
\end{eqnarray}
Then,
$\{ \alpha_{8}, \alpha_{12} \} \in V(\boldsymbol{s}_3)$.
From (\ref{fourc12}), (\ref{fourc9}), and (\ref{fourc9plus112}),
\begin{eqnarray}
  & & \boldsymbol{h}_{\alpha_{8}} + \boldsymbol{h}_{\alpha_{12}}
  = \boldsymbol{s}_3
  \nonumber \\
  & \not= &
  \boldsymbol{s}_1 + \boldsymbol{s}_2
  = (\boldsymbol{h}_{\alpha_8} + \boldsymbol{h}_{\alpha_9})
  + (\boldsymbol{h}_{\alpha_9} + \boldsymbol{h}_{\alpha_{11}})
  \nonumber \\
  & & = \boldsymbol{h}_{\alpha_8} + \boldsymbol{h}_{\alpha_{11}}.
  \label{fourc8plus12not8plus11}
\end{eqnarray}
Therefore,
$\alpha_{12} \not= \alpha_{11}$.
From (\ref{fourc9}), (\ref{fourc9plus112}),
and (\ref{fourc12}),
\begin{eqnarray}
  \boldsymbol{s}_4
  & = & \boldsymbol{s}_1 + \boldsymbol{s}_2 + \boldsymbol{s}_3
  \nonumber \\
  & = & (\boldsymbol{h}_{\alpha_8} + \boldsymbol{h}_{\alpha_9})
  + (\boldsymbol{h}_{\alpha_9} + \boldsymbol{h}_{\alpha_{11}})
  + (\boldsymbol{h}_{\alpha_{8}} + \boldsymbol{h}_{\alpha_{12}})
  \nonumber \\
  & = & \boldsymbol{h}_{\alpha_{11}} + \boldsymbol{h}_{\alpha_{12}}.
  \label{fourc411plus12}
\end{eqnarray}
Hence,
$\{ \alpha_{11}, \alpha_{12} \} \in V(\boldsymbol{s}_4)$.
Let $\alpha_{13}$ be an integer such that
$\alpha_{13} \in
\mathcal{I}_{15} \setminus \{ \alpha_1, \ldots, \alpha_{12} \}$
and
\begin{eqnarray}
  \boldsymbol{s}_1 =
  \boldsymbol{h}_{\alpha_{12}} + \boldsymbol{h}_{\alpha_{13}}.
  \label{fourc13}
\end{eqnarray}
Then,
$\{ \alpha_{12}, \alpha_{13} \} \in V(\boldsymbol{s}_1)$.
From (\ref{fourc9}) and (\ref{fourc13}),
\begin{eqnarray}
  \boldsymbol{h}_{\alpha_8} + \boldsymbol{h}_{\alpha_9}
  = \boldsymbol{h}_{\alpha_{12}} + \boldsymbol{h}_{\alpha_{13}}.
  \label{fourc8plus912plus13}
\end{eqnarray}
From (\ref{fourc8plus912plus13})
and (\ref{fourc12}),
\begin{eqnarray}
  \boldsymbol{h}_{\alpha_9} + \boldsymbol{h}_{\alpha_{13}}
  = \boldsymbol{h}_{\alpha_8} + \boldsymbol{h}_{\alpha_{12}}
  = \boldsymbol{s}_3.
  \label{fourc9plus133}
\end{eqnarray}
Therefore,
$\{ \alpha_{9}, \alpha_{13} \} \in V(\boldsymbol{s}_3)$.
From (\ref{fourc11}) and (\ref{fourc13}),
\begin{eqnarray}
  \boldsymbol{h}_{\alpha_{10}} + \boldsymbol{h}_{\alpha_{11}}
  = \boldsymbol{h}_{\alpha_{12}} + \boldsymbol{h}_{\alpha_{13}}.
  \label{fourc10plus1112plus13}
\end{eqnarray}
From (\ref{fourc10plus1112plus13})
and (\ref{fourc411plus12}),
\begin{eqnarray}
  \boldsymbol{h}_{\alpha_{10}} + \boldsymbol{h}_{\alpha_{13}}
  = \boldsymbol{h}_{\alpha_{11}} + \boldsymbol{h}_{\alpha_{12}}
  = \boldsymbol{s}_4.
  \label{fourc10plus134}
\end{eqnarray}
Hence,
$\{ \alpha_{10}, \alpha_{13} \} \in V(\boldsymbol{s}_4)$.
Let $\alpha_{14}$ be an integer such that
$\alpha_{14} \in
\mathcal{I}_{15} \setminus \{ \alpha_1, \ldots, \alpha_{13} \}$
and
\begin{eqnarray}
  \boldsymbol{s}_2 =
  \boldsymbol{h}_{\alpha_{12}} + \boldsymbol{h}_{\alpha_{14}}.
  \label{fourc14}
\end{eqnarray}
Then,
$\{ \alpha_{12}, \alpha_{14} \} \in V(\boldsymbol{s}_2)$.
From (\ref{fourc10}) and (\ref{fourc14}),
\begin{eqnarray}
  \boldsymbol{h}_{\alpha_8} + \boldsymbol{h}_{\alpha_{10}}
  = \boldsymbol{h}_{\alpha_{12}} + \boldsymbol{h}_{\alpha_{14}}.
  \label{fourc8plus1012plus14}
\end{eqnarray}
From (\ref{fourc8plus1012plus14}) and
(\ref{fourc12}),
\begin{eqnarray}
  \boldsymbol{h}_{\alpha_{10}} + \boldsymbol{h}_{\alpha_{14}}
  = \boldsymbol{h}_{\alpha_8} + \boldsymbol{h}_{\alpha_{12}}
  = \boldsymbol{s}_3.
  \label{fourc10plus143}
\end{eqnarray}
Therefore,
$\{ \alpha_{10}, \alpha_{14} \} \in V(\boldsymbol{s}_3)$.
From (\ref{fourc9plus112}) and (\ref{fourc14}),
\begin{eqnarray}
  \boldsymbol{h}_{\alpha_9} + \boldsymbol{h}_{\alpha_{11}}
  = \boldsymbol{h}_{\alpha_{12}} + \boldsymbol{h}_{\alpha_{14}}.
  \label{fourc9plus1112plus14}
\end{eqnarray}
From (\ref{fourc9plus1112plus14})
and (\ref{fourc411plus12}),
\begin{eqnarray}
  \boldsymbol{h}_{\alpha_9} + \boldsymbol{h}_{\alpha_{14}}
  = \boldsymbol{h}_{\alpha_{11}} + \boldsymbol{h}_{\alpha_{12}}
  = \boldsymbol{s}_4.
  \label{fourc9plus144}
\end{eqnarray}
Hence,
$\{ \alpha_{9}, \alpha_{14} \} \in V(\boldsymbol{s}_4)$.
Let $\alpha_{15}$ be an integer such that
$\alpha_{15} \in
\mathcal{I}_{15} \setminus \{ \alpha_1, \ldots, \alpha_{14} \}$
and
\begin{eqnarray}
  \boldsymbol{s}_1 =
  \boldsymbol{h}_{\alpha_{14}} + \boldsymbol{h}_{\alpha_{15}}.
  \label{fourc15}
\end{eqnarray}
Then, $\{ \alpha_{14}, \alpha_{15} \} \in V(\boldsymbol{s}_1)$.
From (\ref{fourc13}) and (\ref{fourc15}),
\begin{eqnarray}
  \boldsymbol{h}_{\alpha_{12}} + \boldsymbol{h}_{\alpha_{13}}
  = \boldsymbol{h}_{\alpha_{14}} + \boldsymbol{h}_{\alpha_{15}}.
  \label{fourc12plus1314plus15}
\end{eqnarray}
From (\ref{fourc12plus1314plus15})
and (\ref{fourc14}),
\begin{eqnarray}
  \boldsymbol{h}_{\alpha_{13}} + \boldsymbol{h}_{\alpha_{15}}
  = \boldsymbol{h}_{\alpha_{12}} + \boldsymbol{h}_{\alpha_{14}}
  = \boldsymbol{s}_2.
  \label{fourc13plus152}
\end{eqnarray}
Therefore,
$\{ \alpha_{13}, \alpha_{15} \} \in V(\boldsymbol{s}_2)$.
From (\ref{fourc11}) and (\ref{fourc15}),
\begin{eqnarray}
  \boldsymbol{h}_{\alpha_{10}} + \boldsymbol{h}_{\alpha_{11}}
  = \boldsymbol{h}_{\alpha_{14}} + \boldsymbol{h}_{\alpha_{15}}.
  \label{fourc10plus1114plus15}
\end{eqnarray}
From (\ref{fourc10plus1114plus15})
and (\ref{fourc10plus143}),
\begin{eqnarray}
  \boldsymbol{h}_{\alpha_{11}} + \boldsymbol{h}_{\alpha_{15}}
  = \boldsymbol{h}_{\alpha_{10}} + \boldsymbol{h}_{\alpha_{14}}
  = \boldsymbol{s}_3.
  \label{fourc11plus153}
\end{eqnarray}
Hence,
$\{ \alpha_{11}, \alpha_{15} \} \in V(\boldsymbol{s}_3)$.
From (\ref{fourc9}) and (\ref{fourc15}),
\begin{eqnarray}
  \boldsymbol{h}_{\alpha_8} + \boldsymbol{h}_{\alpha_9}
  = \boldsymbol{h}_{\alpha_{14}} + \boldsymbol{h}_{\alpha_{15}}.
  \label{fourc8plus914plus15}
\end{eqnarray}
From (\ref{fourc8plus914plus15})
and (\ref{fourc9plus144}),
\begin{eqnarray}
  \boldsymbol{h}_{\alpha_8} + \boldsymbol{h}_{\alpha_{15}}
  = \boldsymbol{h}_{\alpha_9} + \boldsymbol{h}_{\alpha_{14}}
  = \boldsymbol{s}_4.
  \label{fourc8plus154}
\end{eqnarray}
Therefore,
$\{ \alpha_{8}, \alpha_{15} \} \in V(\boldsymbol{s}_4)$.
Let $\sigma$ be the permutation of $\mathcal{I}_{15}$
such that
$\sigma(j) = \alpha_j$ for each $j \in \mathcal{I}_{15}$.
Then, $\sigma$ satisfies
(\ref{siki:sigma4c-1}),
(\ref{siki:sigma4c-2}),
(\ref{siki:sigma4c-3}),
and
(\ref{siki:sigma4c-4}).
\end{IEEEproof}

\subsection{Construction of $[7,3,4]$ P-RIO Code}

Let $C$ be a $(7,4)$ Hamming code.
\begin{rei}
The parity check matrix $H$ of $C$ is as follows.
\begin{equation}
H = \left(
 \begin{array}{ccccccc}
   1 & 0 & 1 & 0 & 1 & 0 & 1 \\
   0 & 1 & 1 & 0 & 0 & 1 & 1 \\
   0 & 0 & 0 & 1 & 1 & 1 & 1
 \end{array}
 \right). \label{siki:matrix}
\end{equation}
For each $\boldsymbol{s} \in \{0,1\}^3
\setminus \{000\}$,
$V(\boldsymbol{s})$ is as follows \cite{floatingcode}.
\begin{eqnarray}
  V(100) & = & 
\{ \{ 1 \}, \{ 2, 3 \},
\{ 4, 5 \} ,
\{ 6, 7 \} \}, \nonumber \\
V(010) & = &
\{ \{ 2 \}, \{ 1, 3 \},
\{ 4, 6 \} ,
\{ 5, 7 \} \}, \nonumber \\
V(110) & = &
\{ \{ 3 \}, \{ 1, 2 \},
\{ 4, 7 \} ,
\{ 5, 6 \} \}, \nonumber \\
V(001) & = &
\{ \{ 4 \}, \{ 1, 5 \},
\{ 2, 6 \} ,
\{ 3, 7 \} \}, \nonumber \\
V(101) & = &
\{ \{ 5 \}, \{ 1, 4 \},
\{ 2, 7 \} ,
\{ 3, 6 \} \}, \nonumber \\
V(011) & = &
\{ \{ 6 \}, \{ 1, 7 \},
\{ 2, 4 \} ,
\{ 3, 5 \} \}, \nonumber \\
V(111) & = &
\{ \{ 7 \}, \{ 1, 6 \},
\{ 2, 5 \} ,
\{ 3, 4 \} \}. \nonumber
\end{eqnarray}
For some $\boldsymbol{s}_1, \ldots,
\boldsymbol{s}_4 \in \{0,1\}^3$,
we obtain $\boldsymbol{x}_1, \ldots,
\boldsymbol{x}_4 \in \{0,1\}^7$
that
satisfy the conditions of
Theorem \ref{teiri:sufficientcondition}.

\textit{Case :}
$\boldsymbol{s}_1 = 111, \boldsymbol{s}_2 = 100,
\boldsymbol{s}_3 = 110,$ and $\boldsymbol{s}_4 = 101$.

Let $I(\boldsymbol{x}_1) = \{ 7 \},
I(\boldsymbol{x}_2) = \{ 1 \},
I(\boldsymbol{x}_3) = \{ 3 \},$
and
$I(\boldsymbol{x}_4) = \{ 5 \}$.
Therefore,
$\boldsymbol{x}_1 = 0000001,
\boldsymbol{x}_2 = 1000000,
\boldsymbol{x}_3 = 0010000,$
and
$\boldsymbol{x}_4 = 0000100$.

\textit{Case :}
$\boldsymbol{s}_1 = 001, \boldsymbol{s}_2 = 110,
\boldsymbol{s}_3 = 100,$ and $\boldsymbol{s}_4 = 001$.

Let $I(\boldsymbol{x}_1) = \{ 4 \},
I(\boldsymbol{x}_2) = \{ 3 \},
I(\boldsymbol{x}_3) = \{ 1 \},$
and
$I(\boldsymbol{x}_4) = \{ 2, 6 \}$.
Therefore,
$\boldsymbol{x}_1 = 0001000,
\boldsymbol{x}_2 = 0010000,
\boldsymbol{x}_3 = 1000000,$
and
$\boldsymbol{x}_4 = 0100010$.

\textit{Case :}
$\boldsymbol{s}_1 = 010, \boldsymbol{s}_2 = 101,
\boldsymbol{s}_3 = 010,$ and $\boldsymbol{s}_4 = 101$.

Let $I(\boldsymbol{x}_1) = \{ 2 \},
I(\boldsymbol{x}_2) = \{ 1, 4 \},
I(\boldsymbol{x}_3) = \{ 5, 7 \},$
and
$I(\boldsymbol{x}_4) = \{ 3, 6 \}$.
Therefore,
$\boldsymbol{x}_1 = 0100000,
\boldsymbol{x}_2 = 1001000,
\boldsymbol{x}_3 = 0000101,$
and
$\boldsymbol{x}_4 = 0010010$.
\end{rei}
For any
$\boldsymbol{s}_1, \boldsymbol{s}_2,
\boldsymbol{s}_3, \boldsymbol{s}_4 \in \{0,1\}^3$,
we have
$\boldsymbol{x}_1, \boldsymbol{x}_2,
\boldsymbol{x}_3, \boldsymbol{x}_4 \in \{0,1\}^7$,
which satisfy the conditions of
Theorem \ref{teiri:sufficientcondition} as described below.

We define
$I_0 = \{ i \mid i \in \mathcal{I}_4, \boldsymbol{s}_i = 000 \}$.
For any $\boldsymbol{s}_i \not= 000$,
let $j_i$ be an integer such that
$j_i \in \mathcal{I}_7$
and
$\boldsymbol{s}_i = \boldsymbol{h}_{j_i}$.

\noindent
\textit{Case 1:} $I_0 \not= \emptyset$.

Let $t' = 4 - |I_0|$.
For any $\boldsymbol{s}_{k_1}, \ldots, \boldsymbol{s}_{k_{t'}}$
($\{ k_1, \ldots, k_{t'} \} = \mathcal{I}_4 \setminus I_0$),
we use one of the following cases to
obtain $\boldsymbol{x}_{k_1}, \ldots, \boldsymbol{x}_{k_{t'}}$
that satisfy the conditions of 
Theorem \ref{teiri:sufficientcondition}.
Further,
for every $i \in I_0$,
let $\boldsymbol{x}_i = 0000000$.

In the following,
we assume $I_0 = \emptyset$.

\noindent
\textit{Case 2:}
$\boldsymbol{s}_i \not= \boldsymbol{s}_{i'}$
for any $i, i' \in \mathcal{I}_4, i \not= i'$.

Let $I(\boldsymbol{x}_i) = \{ j_i \}$
for each $i \in \mathcal{I}_4$.

\noindent
\textit{Case 3:}
There exist
$2 \leq m \leq 4$ and the permutation $\pi$ of $\mathcal{I}_4$
such that
$\boldsymbol{s}_{\pi(1)} = \cdots = \boldsymbol{s}_{\pi(m)}$
and $\boldsymbol{s}_{\pi(i)} \not= \boldsymbol{s}_{\pi(i')}$
for any $i, i' \in \{1, m+1, \ldots, 4\}, i \not= i'$.

Let
$I(\boldsymbol{x}_{\pi(1)}) = \{j_{\pi(1)}\}$,
$I(\boldsymbol{x}_{\pi(m+1)}) = \{j_{\pi(m+1)}\}$,
$\ldots$,
$I(\boldsymbol{x}_{\pi(4)}) = \{j_{\pi(4)}\}$.
From Theorem \ref{teiri:kouho1},
we have the permutation $\sigma$ of
$\mathcal{I}_7$ such that
\begin{eqnarray}
V(\boldsymbol{s}_{\pi(2)}) & = &
\{ \{ \sigma(1) \}, \{ \sigma(2), \sigma(3) \},
\nonumber \\
& & \{ \sigma(4), \sigma(5) \} ,
\{ \sigma(6), \sigma(7) \} \}. \nonumber
\end{eqnarray}
Clearly, $\sigma(1) = j_{\pi(1)}$.
Hence,
there are a minimum of $(m-1)$ elements $\alpha$
in $\{ 2, 4, 6 \}$ such that
$\{j_{\pi(m+1)}, \ldots, j_{\pi(4)}\} \cap
\{ \sigma(\alpha), \sigma(\alpha+1) \} = \emptyset$.
We denote these $\alpha$ by
$\alpha_1, \ldots, \alpha_{m-1}$.
Then, let
$I(\boldsymbol{x}_{\pi(2)}) =
\{ \sigma(\alpha_1), \sigma(\alpha_1+1) \}$,
$\ldots$,
$I(\boldsymbol{x}_{\pi(m)}) =
\{ \sigma(\alpha_{m-1}), \sigma(\alpha_{m-1}+1) \}$.

\noindent
\textit{Case 4:}
There exists the permutation $\pi$ of $\mathcal{I}_4$
such that
$\boldsymbol{s}_{\pi(1)} = \boldsymbol{s}_{\pi(2)}$,
$\boldsymbol{s}_{\pi(3)} = \boldsymbol{s}_{\pi(4)}$,
and
$\boldsymbol{s}_{\pi(1)} \not= \boldsymbol{s}_{\pi(3)}$.

From Theorem \ref{teiri:kouho2},
we have the permutation $\sigma$ of
$\mathcal{I}_7$ such that
\begin{eqnarray}
V(\boldsymbol{s}_{\pi(1)}) & = &
\{ \{ \sigma(1) \}, \{ \sigma(2), \sigma(3) \},
\nonumber \\
& & \{ \sigma(4), \sigma(5) \} ,
\{ \sigma(6), \sigma(7) \} \}, \nonumber \\
V(\boldsymbol{s}_{\pi(3)}) & = &
\{ \{ \sigma(2) \}, \{ \sigma(1), \sigma(3) \},
\nonumber \\
& & \{ \sigma(4), \sigma(6) \} ,
\{ \sigma(5), \sigma(7) \} \}. \nonumber
\end{eqnarray}
Hence,
let $I(\boldsymbol{x}_{\pi(1)}) = \{ \sigma(1) \}$,
$I(\boldsymbol{x}_{\pi(2)}) = \{ \sigma(2), \sigma(3) \}$,
$I(\boldsymbol{x}_{\pi(3)}) = \{ \sigma(4), \sigma(6) \}$,
and
$I(\boldsymbol{x}_{\pi(4)}) = \{ \sigma(5), \sigma(7) \}$.

Therefore,
from Theorem \ref{teiri:sufficientcondition},
the $[7,3,4]$ P-RIO code can be constructed.
This code stores more pages
than the $[7,3,3]$ RIO code,
which 
is constructed using
the same $(7,4)$ Hamming code.

\subsection{Construction of $[15,4,8]$ P-RIO Code}

Let $C$ be a $(15,11)$ Hamming code.
For any
$\boldsymbol{s}_1, \ldots, \boldsymbol{s}_8 \in \{0,1\}^4$,
we have
$\boldsymbol{x}_1, \ldots, \boldsymbol{x}_8 \in \{0,1\}^{15}$,
which
satisfy the conditions
of Theorem \ref{teiri:sufficientcondition}
as follows.

We define
$I_0 = \{ i \mid i \in \mathcal{I}_8, \boldsymbol{s}_i = 0000 \}$.
For any $\boldsymbol{s}_i \not= 0000$,
let $j_i$ be an integer such that
$j_i \in \mathcal{I}_{15}$
and
$\boldsymbol{s}_i = \boldsymbol{h}_{j_i}$.

\noindent
\textit{Case 1:}
$I_0 \not= \emptyset$.

We have $\boldsymbol{x}_1, \ldots, \boldsymbol{x}_8$
as for the case of the
$[7,3,4]$ P-RIO code.

\noindent
\textit{Case 2:}
$\boldsymbol{s}_i \not= \boldsymbol{s}_{i'}$
for any
$i, i' \in \mathcal{I}_8, i \not= i'$.

Let $I(\boldsymbol{x}_i) = \{ j_i \}$
for each $i \in \mathcal{I}_8$.

\noindent
\textit{Case 3:} There exist
$2 \leq m \leq 8$ and the permutation $\pi$ of $\mathcal{I}_8$
such that
$\boldsymbol{s}_{\pi(1)} = \cdots = \boldsymbol{s}_{\pi(m)}$
and $\boldsymbol{s}_{\pi(i)} \not= \boldsymbol{s}_{\pi(i')}$
for any $i, i' \in \{1, m+1, \ldots, 8\}, i \not= i'$.

Let $I(\boldsymbol{x}_{\pi(1)}) = \{j_{\pi(1)}\}$,
$I(\boldsymbol{x}_{\pi(m+1)}) = \{j_{\pi(m+1)}\}$,
$\ldots$,
$I(\boldsymbol{x}_{\pi(8)}) = \{j_{\pi(8)}\}$.
From Theorem \ref{teiri:kouho1},
we have the permutation $\sigma$ of
$\mathcal{I}_{15}$ such that
\begin{eqnarray}
V(\boldsymbol{s}_{\pi(2)})
& = &
\{ \{ \sigma(1) \}, \{ \sigma(2), \sigma(3) \},
\{ \sigma(4), \sigma(5) \} ,
\nonumber \\
& & \{ \sigma(6), \sigma(7) \},
\{ \sigma(8), \sigma(9) \}, \{ \sigma(10), \sigma(11) \} ,
\nonumber \\
& & \{ \sigma(12), \sigma(13) \}, \{ \sigma(14), \sigma(15) \}\}.
\nonumber
\end{eqnarray}
Clearly,
$\sigma(1) = j_{\pi(1)}$.
Hence,
there are a minimum of $(m-1)$ elements $\alpha$
in $\{ 2, 4, 6, 8, 10, 12, 14 \}$, such that
$\{j_{\pi(m+1)}, \ldots, j_{\pi(8)}\} \cap
\{ \sigma(\alpha), \sigma(\alpha+1) \} = \emptyset$.
We denote these $\alpha$ by
$\alpha_1, \ldots, \alpha_{m-1}$.
Then, let
$I(\boldsymbol{x}_{\pi(2)}) =
\{ \sigma(\alpha_1), \sigma(\alpha_1+1) \}$,
$\ldots$,
$I(\boldsymbol{x}_{\pi(m)}) =
\{ \sigma(\alpha_{m-1}), \sigma(\alpha_{m-1}+1) \}$.

\noindent
\textit{Case 4:} There exist
$2 \leq m_1, m_2 \leq 6$,
where $m_1 \geq m_2$,
and the permutation $\pi$ of
$\mathcal{I}_8$
such that
$\boldsymbol{s}_{\pi(1)} = \cdots = \boldsymbol{s}_{\pi(m_1)}$,
$\boldsymbol{s}_{\pi(m_1+1)} = \cdots = \boldsymbol{s}_{\pi(m_1+m_2)}$,
and $\boldsymbol{s}_{\pi(i)} \not= \boldsymbol{s}_{\pi(i')}$
for any
$i, i' \in \{1, m_1+1, m_1+m_2+1, \ldots, 8\}, i \not= i'$.

Let $I(\boldsymbol{x}_{\pi(m_1+m_2+1)}) = \{j_{\pi(m_1+m_2+1)}\}$,
$\ldots$,
$I(\boldsymbol{x}_{\pi(8)}) = \{j_{\pi(8)}\}$.
From Theorem \ref{teiri:kouho2},
we have the permutation $\sigma$ of $\mathcal{I}_{15}$
such that
\begin{eqnarray}
& & V(\boldsymbol{s}_{\pi(1)}) \nonumber \\
& = &
\{ \{ \sigma(1) \}, \{ \sigma(2), \sigma(3) \},
\{ \sigma(4), \sigma(5) \},
\nonumber \\
& & \{ \sigma(6), \sigma(7) \},
\{ \sigma(8), \sigma(9) \}, \{ \sigma(10), \sigma(11) \},
\nonumber \\
& & \{ \sigma(12), \sigma(13) \}, \{ \sigma(14), \sigma(15) \}\},
\nonumber \\
& & V(\boldsymbol{s}_{\pi(m_1+1)}) \nonumber \\
& = &
\{ \{ \sigma(2) \}, \{ \sigma(1), \sigma(3) \},
\{ \sigma(4), \sigma(6) \},
\nonumber \\
& & \{ \sigma(5), \sigma(7) \},
\{ \sigma(8), \sigma(10) \}, \{ \sigma(9), \sigma(11) \},
\nonumber \\
& & \{ \sigma(12), \sigma(14) \}, \{ \sigma(13), \sigma(15) \}\}.
\nonumber
\end{eqnarray}
Clearly,
$\{ j_{\pi(m_1+m_2+1)}, \ldots, j_{\pi(8)} \} \cap
\{ \sigma(1), \sigma(2) \}
= \emptyset$.
We define
$A_1 = \{ j_{\pi(i)} \mid i \in \{ m_1+m_2+1, \ldots, 8 \},
j_{\pi(i)} \in \{ \sigma(3), \ldots, \sigma(7) \} \}$,
$A_2 = \{ j_{\pi(i)} \mid i \in \{ m_1+m_2+1, \ldots, 8 \},
j_{\pi(i)} \in \{ \sigma(8), \ldots, \sigma(15) \} \}$,
$a_1 = | A_1 |$,
$a_2 = | A_2 |$.
Then, $a_1 + a_2 = 8-m_1-m_2$.

\noindent
\textit{Case 4-1:}
$m_1 = m_2 = 2$.

\noindent
\textit{Case 4-1-1:} $a_1 = 0$ and $a_2 = 4$.

Let
$I(\boldsymbol{x}_{\pi(1)}) = \{ \sigma(1) \}$,
$I(\boldsymbol{x}_{\pi(2)}) = \{ \sigma(2), \sigma(3) \}$,
$I(\boldsymbol{x}_{\pi(3)}) = \{ \sigma(4), \sigma(6) \}$,
$I(\boldsymbol{x}_{\pi(4)}) = \{ \sigma(5), \sigma(7) \}$.

\noindent
\textit{Case 4-1-2:} $a_1 = 1$ and $a_2 = 3$.

Let
$I(\boldsymbol{x}_{\pi(1)}) = \{ \sigma(1) \}$
and
$I(\boldsymbol{x}_{\pi(2)}) = \{ \sigma(\alpha), \sigma(\alpha+1) \}$
such that
$\alpha \in \{4,6\}$
and
$A_1 \cap \{ \sigma(\alpha), \sigma(\alpha+1) \} = \emptyset$.
Let $I(\boldsymbol{x}_{\pi(3)}) = \{ \sigma(2) \}$
and
$I(\boldsymbol{x}_{\pi(4)}) = \{ \sigma(\beta), \sigma(\beta+2) \}$
such that
$\beta \in \{8,9,12,13\}$
and
$A_2 \cap \{ \sigma(\beta), \sigma(\beta+2) \} = \emptyset$.

\noindent
\textit{Case 4-1-3:} $a_1 = 2$ and $a_2 = 2$.

Let
$I(\boldsymbol{x}_{\pi(1)}) = \{ \sigma(1) \}$
and
$I(\boldsymbol{x}_{\pi(2)}) = \{ \sigma(\alpha), \sigma(\alpha+1) \}$
such that
$\alpha \in \{2,4,6\}$
and
$A_1 \cap \{ \sigma(\alpha), \sigma(\alpha+1) \} = \emptyset$.
Let
$I(\boldsymbol{x}_{\pi(3)}) = \{ \sigma(\beta_1), \sigma(\beta_1+2) \}$
and
$I(\boldsymbol{x}_{\pi(4)}) = \{ \sigma(\beta_2), \sigma(\beta_2+2) \}$
such that
$\beta_1, \beta_2 \in \{8,9,12,13\}, \beta_1 \not= \beta_2$,
and
$A_2 \cap
\{ \sigma(\beta_1), \sigma(\beta_1+2), \sigma(\beta_2), \sigma(\beta_2+2) \}
= \emptyset$.

\noindent
\textit{Case 4-1-4:} $a_1 = 3$ and $a_2 = 1$.

Let
$I(\boldsymbol{x}_{\pi(1)}) = \{ \sigma(1) \}$ and
$I(\boldsymbol{x}_{\pi(2)}) = \{ \sigma(\alpha), \sigma(\alpha+1) \}$
such that
$\alpha \in \{8,10\}$
and
$A_2 \cap \{ \sigma(\alpha), \sigma(\alpha+1) \} = \emptyset$.
Let
$I(\boldsymbol{x}_{\pi(3)}) = \{ \sigma(2) \}$
and
$I(\boldsymbol{x}_{\pi(4)}) = \{ \sigma(\beta), \sigma(\beta+2) \}$
such that
$\beta \in \{12,13\}$
and
$A_2 \cap \{ \sigma(\beta), \sigma(\beta+2) \} = \emptyset$.

\noindent
\textit{Case 4-1-5:} $a_1 = 4$ and $a_2 = 0$.

Let
$I(\boldsymbol{x}_{\pi(1)}) = \{ \sigma(1) \}$,
$I(\boldsymbol{x}_{\pi(2)}) = \{ \sigma(8), \sigma(9) \}$,
$I(\boldsymbol{x}_{\pi(3)}) = \{ \sigma(2) \}$,
and
$I(\boldsymbol{x}_{\pi(4)}) = \{ \sigma(12), \sigma(14) \}$.

\noindent
\textit{Case 4-2:} $m_1 = 3$ and $m_2 = 2$.

\noindent
\textit{Case 4-2-1:} $a_1 = 0$ and $a_2 = 3$.

Let
$I(\boldsymbol{x}_{\pi(1)}) = \{ \sigma(1) \}$,
$I(\boldsymbol{x}_{\pi(2)}) = \{ \sigma(4), \sigma(5) \}$,
and
$I(\boldsymbol{x}_{\pi(3)}) = \{ \sigma(6), \sigma(7) \}$.
Let
$I(\boldsymbol{x}_{\pi(4)}) = \{ \sigma(2) \}$
and
$I(\boldsymbol{x}_{\pi(5)}) = \{ \sigma(\alpha), \sigma(\alpha+2) \}$
such that
$\alpha \in \{8,9,12,13\}$
and
$A_2 \cap \{ \sigma(\alpha), \sigma(\alpha+2) \} = \emptyset$.

\noindent
\textit{Case 4-2-2:} $a_1 = 1$ and $a_2 = 2$.

Let
$I(\boldsymbol{x}_{\pi(1)}) = \{ \sigma(1) \}$,
$I(\boldsymbol{x}_{\pi(2)}) = \{ \sigma(\alpha_1), \sigma(\alpha_1+1) \}$,
and
$I(\boldsymbol{x}_{\pi(3)}) = \{ \sigma(\alpha_2), \sigma(\alpha_2+1) \}$
such that
$\alpha_1, \alpha_2 \in \{2,4,6\}, \alpha_1 \not= \alpha_2$,
and
$A_1 \cap
\{ \sigma(\alpha_1), \sigma(\alpha_1+1), \sigma(\alpha_2), \sigma(\alpha_2+1) \}
= \emptyset$.
Let
$I(\boldsymbol{x}_{\pi(4)}) = \{ \sigma(\beta_1), \sigma(\beta_1+2) \}$
and
$I(\boldsymbol{x}_{\pi(5)}) = \{ \sigma(\beta_2), \sigma(\beta_2+2) \}$
such that
$\beta_1, \beta_2 \in \{8,9,12,13\}, \beta_1 \not= \beta_2$,
and
$A_2 \cap
\{ \sigma(\beta_1), \sigma(\beta_1+2), \sigma(\beta_2), \sigma(\beta_2+2) \}
= \emptyset$.

\noindent
\textit{Case 4-2-3:} $a_1 = 2$ and $a_2 = 1$.

Let
$I(\boldsymbol{x}_{\pi(1)}) = \{ \sigma(1) \}$,
$I(\boldsymbol{x}_{\pi(2)}) = \{ \sigma(\alpha), \sigma(\alpha+1) \}$,
and
$I(\boldsymbol{x}_{\pi(3)}) = \{ \sigma(\alpha+2), \sigma(\alpha+3) \}$
such that
$\alpha \in \{8,12\}$
and
$A_2 \cap
\{ \sigma(\alpha), \sigma(\alpha+1), \sigma(\alpha+2), \sigma(\alpha+3) \}
= \emptyset$.
Let
$I(\boldsymbol{x}_{\pi(4)}) = \{ \sigma(2) \}$
and
$I(\boldsymbol{x}_{\pi(5)}) = \{ \sigma(\beta), \sigma(\beta+2) \}$
such that
$\beta \in \{8,9,12,13\}$
and
$(A_2 \cup
\{ \sigma(\alpha), \sigma(\alpha+1), \sigma(\alpha+2), \sigma(\alpha+3) \})
\cap \{ \sigma(\beta), \sigma(\beta+2) \} = \emptyset$.

\noindent
\textit{Case 4-2-4:} $a_1 = 3$ and $a_2 = 0$.

Let
$I(\boldsymbol{x}_{\pi(1)}) = \{ \sigma(1) \}$,
$I(\boldsymbol{x}_{\pi(2)}) = \{ \sigma(8), \sigma(9) \}$,
$I(\boldsymbol{x}_{\pi(3)}) = \{ \sigma(10), \sigma(11) \}$,
$I(\boldsymbol{x}_{\pi(4)}) = \{ \sigma(2) \}$,
and
$I(\boldsymbol{x}_{\pi(5)}) = \{ \sigma(12), \sigma(14) \}$.

\noindent
\textit{Case 4-3:} $m_1 = m_2 = 3$.

\noindent
\textit{Case 4-3-1:} $a_1 = 0$ and $a_2 = 2$.

Let
$I(\boldsymbol{x}_{\pi(1)}) = \{ \sigma(1) \}$,
$I(\boldsymbol{x}_{\pi(2)}) = \{ \sigma(4), \sigma(5) \}$,
and
$I(\boldsymbol{x}_{\pi(3)}) = \{ \sigma(6), \sigma(7) \}$.
Let
$I(\boldsymbol{x}_{\pi(4)}) = \{ \sigma(2) \}$,
$I(\boldsymbol{x}_{\pi(5)}) = \{ \sigma(\alpha_1), \sigma(\alpha_1+2) \}$,
and
$I(\boldsymbol{x}_{\pi(6)}) = \{ \sigma(\alpha_2), \sigma(\alpha_2+2) \}$
such that
$\alpha_1, \alpha_2 \in \{8,9,12,13\}, \alpha_1 \not= \alpha_2$,
and
$A_2 \cap
\{ \sigma(\alpha_1), \sigma(\alpha_1+2), \sigma(\alpha_2), \sigma(\alpha_2+2) \}
= \emptyset$.

\noindent
\textit{Case 4-3-2:} $a_1 = a_2 = 1$.

Let
$I(\boldsymbol{x}_{\pi(1)}) = \{ \sigma(1) \}$,
$I(\boldsymbol{x}_{\pi(2)}) = \{ \sigma(\alpha_1), \sigma(\alpha_1+1) \}$,
and
$I(\boldsymbol{x}_{\pi(3)}) = \{ \sigma(\alpha_2), \sigma(\alpha_2+1) \}$
such that
$\alpha_1, \alpha_2 \in \{2,4,6\}, \alpha_1 \not= \alpha_2$,
and
$A_1 \cap
\{ \sigma(\alpha_1), \sigma(\alpha_1+1), \sigma(\alpha_2), \sigma(\alpha_2+1) \}
= \emptyset$.
Let
$I(\boldsymbol{x}_{\pi(4)}) = \{ \sigma(\beta_1), \sigma(\beta_1+2) \}$,
$I(\boldsymbol{x}_{\pi(5)}) = \{ \sigma(\beta_2), \sigma(\beta_2+2) \}$,
and
$I(\boldsymbol{x}_{\pi(6)}) = \{ \sigma(\beta_3), \sigma(\beta_3+2) \}$
such that
$\beta_1, \beta_2, \beta_3 \in \{8,9,12,13\}, \beta_1 \not= \beta_2,
\beta_1 \not= \beta_3, \beta_2 \not= \beta_3$,
and
$A_2 \cap
\{ \sigma(\beta_1), \sigma(\beta_1+2), \sigma(\beta_2), \sigma(\beta_2+2),
\sigma(\beta_3), \sigma(\beta_3+2) \}
= \emptyset$.

\noindent
\textit{Case 4-3-3:} $a_1 = 2$ and $a_2 = 0$.

Let
$I(\boldsymbol{x}_{\pi(1)}) = \{ \sigma(1) \}$,
$I(\boldsymbol{x}_{\pi(2)}) = \{ \sigma(8), \sigma(9) \}$,
$I(\boldsymbol{x}_{\pi(3)}) = \{ \sigma(10), \sigma(11) \}$,
$I(\boldsymbol{x}_{\pi(4)}) = \{ \sigma(2) \}$,
$I(\boldsymbol{x}_{\pi(5)}) = \{ \sigma(12), \sigma(14) \}$,
and
$I(\boldsymbol{x}_{\pi(6)}) = \{ \sigma(13), \sigma(15) \}$.

\noindent
\textit{Case 4-4:} $m_1 = 4$ and $m_2 = 2$.

\noindent
\textit{Case 4-4-1:} $a_1 = 0$ and $a_2 = 2$.

Let
$I(\boldsymbol{x}_{\pi(1)}) = \{ \sigma(1) \}$,
$I(\boldsymbol{x}_{\pi(2)}) = \{ \sigma(2), \sigma(3) \}$,
$I(\boldsymbol{x}_{\pi(3)}) = \{ \sigma(4), \sigma(5) \}$,
and
$I(\boldsymbol{x}_{\pi(4)}) = \{ \sigma(6), \sigma(7) \}$.
Let
$I(\boldsymbol{x}_{\pi(5)}) = \{ \sigma(\alpha_1), \sigma(\alpha_1+2) \}$
and
$I(\boldsymbol{x}_{\pi(6)}) = \{ \sigma(\alpha_2), \sigma(\alpha_2+2) \}$
such that
$\alpha_1, \alpha_2 \in \{8,9,12,13\}, \alpha_1 \not= \alpha_2$,
and
$A_2 \cap
\{ \sigma(\alpha_1), \sigma(\alpha_1+2), \sigma(\alpha_2), \sigma(\alpha_2+2) \}
= \emptyset$.

\noindent
\textit{Case 4-4-2:} $a_1 = a_2 = 1$.

Let
$I(\boldsymbol{x}_{\pi(1)}) = \{ \sigma(1) \}$
and
$I(\boldsymbol{x}_{\pi(2)}) = \{ \sigma(\alpha), \sigma(\alpha+1) \}$
such that
$\alpha \in \{4,6\}$ and
$A_1 \cap
\{ \sigma(\alpha), \sigma(\alpha+1) \}
= \emptyset$.
Let
$I(\boldsymbol{x}_{\pi(3)}) = \{ \sigma(\beta), \sigma(\beta+1) \}$
and
$I(\boldsymbol{x}_{\pi(4)}) = \{ \sigma(\beta+2), \sigma(\beta+3) \}$
such that
$\beta \in \{8,12\}$
and
$A_2 \cap
\{ \sigma(\beta), \sigma(\beta+1), \sigma(\beta+2), \sigma(\beta+3) \}
= \emptyset$.
Let
$I(\boldsymbol{x}_{\pi(5)}) = \{ \sigma(2) \}$
and
$I(\boldsymbol{x}_{\pi(6)}) = \{ \sigma(\gamma), \sigma(\gamma+2) \}$
such that
$\gamma \in \{8,9,12,13\}$
and
$(A_2 \cup
\{ \sigma(\beta), \sigma(\beta+1), \sigma(\beta+2), \sigma(\beta+3) \})
\cap \{ \sigma(\gamma), \sigma(\gamma+2) \} = \emptyset$.

\noindent
\textit{Case 4-4-3:} $a_1 = 2$ and $a_2 = 0$.

Let
$I(\boldsymbol{x}_{\pi(1)}) = \{ \sigma(1) \}$,
$I(\boldsymbol{x}_{\pi(2)}) = \{ \sigma(8), \sigma(9) \}$,
$I(\boldsymbol{x}_{\pi(3)}) = \{ \sigma(10), \sigma(11) \}$,
and
$I(\boldsymbol{x}_{\pi(4)}) = \{ \sigma(\alpha), \sigma(\alpha+1) \}$
such that
$\alpha \in \{2,4,6\}$
and
$A_1 \cap
\{ \sigma(\alpha), \sigma(\alpha+1) \}
= \emptyset$.
Let
$I(\boldsymbol{x}_{\pi(5)}) = \{ \sigma(12), \sigma(14) \}$
and
$I(\boldsymbol{x}_{\pi(6)}) = \{ \sigma(13), \sigma(15) \}$.

\noindent
\textit{Case 4-5:} $m_1 = 4$ and $m_2 = 3$.

\noindent
\textit{Case 4-5-1:} $a_1 = 0$ and $a_2 = 1$.

Let
$I(\boldsymbol{x}_{\pi(1)}) = \{ \sigma(1) \}$,
$I(\boldsymbol{x}_{\pi(2)}) = \{ \sigma(2), \sigma(3) \}$,
$I(\boldsymbol{x}_{\pi(3)}) = \{ \sigma(4), \sigma(5) \}$,
and
$I(\boldsymbol{x}_{\pi(4)}) = \{ \sigma(6), \sigma(7) \}$.
Let
$I(\boldsymbol{x}_{\pi(5)}) = \{ \sigma(\alpha_1), \sigma(\alpha_1+2) \}$,
$I(\boldsymbol{x}_{\pi(6)}) = \{ \sigma(\alpha_2), \sigma(\alpha_2+2) \}$,
and
$I(\boldsymbol{x}_{\pi(7)}) = \{ \sigma(\alpha_3), \sigma(\alpha_3+2) \}$
such that
$\alpha_1, \alpha_2, \alpha_3 \in \{8,9,12,13\}, \alpha_1 \not= \alpha_2,
\alpha_1 \not= \alpha_3, \alpha_2 \not= \alpha_3$,
and
$A_2 \cap
\{ \sigma(\alpha_1), \sigma(\alpha_1+2), \sigma(\alpha_2), \sigma(\alpha_2+2),
\sigma(\alpha_3), \sigma(\alpha_3+2) \}
= \emptyset$.

\noindent
\textit{Case 4-5-2:} $a_1 = 1$ and $a_2 = 0$.

Let
$I(\boldsymbol{x}_{\pi(1)}) = \{ \sigma(8), \sigma(9) \}$,
$I(\boldsymbol{x}_{\pi(2)}) = \{ \sigma(10), \sigma(11) \}$,
$I(\boldsymbol{x}_{\pi(3)}) = \{ \sigma(12), \sigma(13) \}$,
and
$I(\boldsymbol{x}_{\pi(4)}) = \{ \sigma(14), \sigma(15) \}$.
Let
$I(\boldsymbol{x}_{\pi(5)}) = \{ \sigma(2) \}$,
$I(\boldsymbol{x}_{\pi(6)}) = \{ \sigma(\alpha_1), \sigma(\alpha_1+2) \}$,
and
$I(\boldsymbol{x}_{\pi(7)}) = \{ \sigma(\alpha_2), \sigma(\alpha_2+2) \}$
such that
$\alpha_1, \alpha_2 \in \{1,4,5\}, \alpha_1 \not= \alpha_2$,
and
$A_1 \cap
\{ \sigma(\alpha_1), \sigma(\alpha_1+2), \sigma(\alpha_2), \sigma(\alpha_2+2) \}
= \emptyset$.

\noindent
\textit{Case 4-6:} $m_1 = m_2 = 4$.

Let
$I(\boldsymbol{x}_{\pi(1)}) = \{ \sigma(1) \}$,
$I(\boldsymbol{x}_{\pi(2)}) = \{ \sigma(2), \sigma(3) \}$,
$I(\boldsymbol{x}_{\pi(3)}) = \{ \sigma(4), \sigma(5) \}$,
$I(\boldsymbol{x}_{\pi(4)}) = \{ \sigma(6), \sigma(7) \}$,
$I(\boldsymbol{x}_{\pi(5)}) = \{ \sigma(8), \sigma(10) \}$,
$I(\boldsymbol{x}_{\pi(6)}) = \{ \sigma(9), \sigma(11) \}$,
$I(\boldsymbol{x}_{\pi(7)}) = \{ \sigma(12), \sigma(14) \}$,
and
$I(\boldsymbol{x}_{\pi(8)}) = \{ \sigma(13), \sigma(15) \}$.

\noindent
\textit{Case 4-7:} $m_1 = 5$ and $m_2 = 2$.

\noindent
\textit{Case 4-7-1:} $a_1 = 0$ and $a_2 = 1$.

Let
$I(\boldsymbol{x}_{\pi(1)}) = \{ \sigma(1) \}$,
$I(\boldsymbol{x}_{\pi(2)}) = \{ \sigma(4), \sigma(5) \}$,
$I(\boldsymbol{x}_{\pi(3)}) = \{ \sigma(6), \sigma(7) \}$,
$I(\boldsymbol{x}_{\pi(4)}) = \{ \sigma(\alpha), \sigma(\alpha+1) \}$,
and
$I(\boldsymbol{x}_{\pi(5)}) = \{ \sigma(\alpha+2), \sigma(\alpha+3) \}$
such that
$\alpha \in \{8,12\}$
and
$A_2 \cap
\{ \sigma(\alpha), \sigma(\alpha+1), \sigma(\alpha+2), \sigma(\alpha+3) \}
= \emptyset$.
Let
$I(\boldsymbol{x}_{\pi(6)}) = \{ \sigma(2) \}$
and
$I(\boldsymbol{x}_{\pi(7)}) = \{ \sigma(\beta), \sigma(\beta+2) \}$
such that
$\beta \in \{8,9,12,13\}$
and
$(A_2 \cup
\{ \sigma(\alpha), \sigma(\alpha+1), \sigma(\alpha+2), \sigma(\alpha+3) \})
\cap \{ \sigma(\beta), \sigma(\beta+2) \} = \emptyset$.

\noindent
\textit{Case 4-7-2:} $a_1 = 1$ and $a_2 = 0$.

Let
$I(\boldsymbol{x}_{\pi(1)}) = \{ \sigma(1) \}$,
$I(\boldsymbol{x}_{\pi(2)}) = \{ \sigma(8), \sigma(9) \}$,
$I(\boldsymbol{x}_{\pi(3)}) = \{ \sigma(10), \sigma(11) \}$,
$I(\boldsymbol{x}_{\pi(4)}) = \{ \sigma(12), \sigma(13) \}$,
and
$I(\boldsymbol{x}_{\pi(5)}) = \{ \sigma(14), \sigma(15) \}$.
Let
$I(\boldsymbol{x}_{\pi(6)}) = \{ \sigma(2) \}$
and
$I(\boldsymbol{x}_{\pi(7)}) = \{ \sigma(\alpha), \sigma(\alpha+2) \}$
such that
$\alpha \in \{4,5\}$
and
$A_1 \cap
\{ \sigma(\alpha), \sigma(\alpha+2) \}
= \emptyset$.

\noindent
\textit{Case 4-8:} $m_1 = 5$ and $m_2 = 3$.

Let
$I(\boldsymbol{x}_{\pi(1)}) = \{ \sigma(1) \}$,
$I(\boldsymbol{x}_{\pi(2)}) = \{ \sigma(4), \sigma(5) \}$,
$I(\boldsymbol{x}_{\pi(3)}) = \{ \sigma(6), \sigma(7) \}$,
$I(\boldsymbol{x}_{\pi(4)}) = \{ \sigma(8), \sigma(9) \}$,
$I(\boldsymbol{x}_{\pi(5)}) = \{ \sigma(10), \sigma(11) \}$,
$I(\boldsymbol{x}_{\pi(6)}) = \{ \sigma(2) \}$,
$I(\boldsymbol{x}_{\pi(7)}) = \{ \sigma(12), \sigma(14) \}$,
and
$I(\boldsymbol{x}_{\pi(8)}) = \{ \sigma(13), \sigma(15) \}$.

\noindent
\textit{Case 4-9:} $m_1 = 6$ and $m_2 = 2$.

Let
$I(\boldsymbol{x}_{\pi(1)}) = \{ \sigma(1) \}$,
$I(\boldsymbol{x}_{\pi(2)}) = \{ \sigma(2), \sigma(3) \}$,
$I(\boldsymbol{x}_{\pi(3)}) = \{ \sigma(4), \sigma(5) \}$,
$I(\boldsymbol{x}_{\pi(4)}) = \{ \sigma(6), \sigma(7) \}$,
$I(\boldsymbol{x}_{\pi(5)}) = \{ \sigma(8), \sigma(9) \}$,
$I(\boldsymbol{x}_{\pi(6)}) = \{ \sigma(10), \sigma(11) \}$,
$I(\boldsymbol{x}_{\pi(7)}) = \{ \sigma(12), \sigma(14) \}$,
and
$I(\boldsymbol{x}_{\pi(8)}) = \{ \sigma(13), \sigma(15) \}$.

\noindent
\textit{Case 5:}
There exist
$2 \leq m_1, m_2, m_3 \leq 4$,
where $m_1 \geq m_2 \geq m_3$,
and the permutation $\pi$ of $\mathcal{I}_8$
such that
$\boldsymbol{s}_{\pi(1)} = \cdots = \boldsymbol{s}_{\pi(m_1)}$,
$\boldsymbol{s}_{\pi(m_1+1)} = \cdots = \boldsymbol{s}_{\pi(m_1+m_2)}$,
$\boldsymbol{s}_{\pi(m_1+m_2+1)} = \cdots = \boldsymbol{s}_{\pi(m_1+m_2+m_3)}$,
and $\boldsymbol{s}_{\pi(i)} \not= \boldsymbol{s}_{\pi(i')}$
for any
$i, i' \in \{1, m_1+1, m_1+m_2+1, m_1+m_2+m_3+1, \ldots, 8\}, i \not= i'$.

Let
$I(\boldsymbol{x}_{\pi(m_1+m_2+m_3+1)}) = \{j_{\pi(m_1+m_2+m_3+1)}\}$,
$\ldots$,
$I(\boldsymbol{x}_{\pi(8)}) = \{j_{\pi(8)}\}$.

\noindent
\textit{Case 5-1:}
$\boldsymbol{s}_{\pi(1)} + \boldsymbol{s}_{\pi(m_1+1)}
\not= \boldsymbol{s}_{\pi(m_1+m_2+1)}$.

From Theorem \ref{teiri:kouho3},
we have the permutation $\sigma$ of $\mathcal{I}_{15}$
such that
\begin{eqnarray}
  & & V(\boldsymbol{s}_{\pi(1)})
  \nonumber \\
& = &
  \{ \{ \sigma(1) \}, \{ \sigma(2), \sigma(3) \}, \{ \sigma(4), \sigma(5) \},
  \nonumber \\
& & \{ \sigma(6), \sigma(7) \},
  \{ \sigma(8), \sigma(9) \}, \{ \sigma(10), \sigma(11) \},
  \nonumber \\
  & & \{ \sigma(12), \sigma(13) \}, \{ \sigma(14), \sigma(15) \}\},
  \nonumber \\
  & & V(\boldsymbol{s}_{\pi(m_1+1)})
  \nonumber \\
& = &
  \{ \{ \sigma(2) \}, \{ \sigma(1), \sigma(3) \}, \{ \sigma(4), \sigma(6) \},
  \nonumber \\
& & \{ \sigma(5), \sigma(7) \},
  \{ \sigma(8), \sigma(10) \}, \{ \sigma(9), \sigma(11) \},
  \nonumber \\
  & & \{ \sigma(12), \sigma(14) \}, \{ \sigma(13), \sigma(15) \}\},
  \nonumber \\
  & & V(\boldsymbol{s}_{\pi(m_1+m_2+1)})
  \nonumber \\
& = &
  \{ \{ \sigma(4) \}, \{ \sigma(1), \sigma(5) \}, \{ \sigma(2), \sigma(6) \},
  \nonumber \\
& & \{ \sigma(3), \sigma(7) \},
  \{ \sigma(8), \sigma(12) \}, \{ \sigma(9), \sigma(13) \},
  \nonumber \\
& & \{ \sigma(10), \sigma(14) \}, \{ \sigma(11), \sigma(15) \}\}. \nonumber
\end{eqnarray}
Clearly,
$\{ j_{\pi(m_1+m_2+m_3+1)}, \ldots, j_{\pi(8)} \} \cap
\{ \sigma(1), \sigma(2), \sigma(4) \}
= \emptyset$.
We define
$A_1 = \{ j_{\pi(i)} \mid i \in \{ m_1+m_2+m_3+1, \ldots, 8 \},
j_{\pi(i)} \in \{ \sigma(3), \sigma(5), \sigma(6), \sigma(7) \} \}$,
$A_2 = \{ j_{\pi(i)} \mid i \in \{ m_1+m_2+m_3+1, \ldots, 8 \},
j_{\pi(i)} \in \{ \sigma(8), \ldots, \sigma(15) \} \}$,
$a_1 = | A_1 |$,
$a_2 = | A_2 |$.
Then,
$a_1 + a_2 = 8-m_1-m_2-m_3$.

\noindent
\textit{Case 5-1-1:}
$m_1 = m_2 = m_3 = 2$.

\noindent
\textit{Case 5-1-1-1:}
$a_1 = 0$ and $a_2 = 2$.

Let
$I(\boldsymbol{x}_{\pi(1)}) = \{ \sigma(1) \}$,
$I(\boldsymbol{x}_{\pi(2)}) = \{ \sigma(2), \sigma(3) \}$,
$I(\boldsymbol{x}_{\pi(3)}) = \{ \sigma(4), \sigma(6) \}$,
and
$I(\boldsymbol{x}_{\pi(4)}) = \{ \sigma(5), \sigma(7) \}$.
Let
$I(\boldsymbol{x}_{\pi(5)}) = \{ \sigma(\alpha_1), \sigma(\alpha_1+4) \}$
and
$I(\boldsymbol{x}_{\pi(6)}) = \{ \sigma(\alpha_2), \sigma(\alpha_2+4) \}$
such that
$\alpha_1, \alpha_2 \in \{8,9,10,11\}, \alpha_1 \not= \alpha_2$,
and
$A_2 \cap
\{ \sigma(\alpha_1), \sigma(\alpha_1+4), \sigma(\alpha_2), \sigma(\alpha_2+4) \}
= \emptyset$.

\noindent
\textit{Case 5-1-1-2:}
$a_1 = a_2 = 1$.

Let
$I(\boldsymbol{x}_{\pi(1)}) = \{ \sigma(1) \}$,
$I(\boldsymbol{x}_{\pi(3)}) = \{ \sigma(2) \}$,
and
$I(\boldsymbol{x}_{\pi(5)}) = \{ \sigma(4) \}$.

\noindent
\textit{Case 5-1-1-2-1:}
$A_2 = \{ \sigma(8) \}$ or $A_2 = \{ \sigma(15) \}$.

Let
$I(\boldsymbol{x}_{\pi(2)}) = \{ \sigma(10), \sigma(11) \}$,
$I(\boldsymbol{x}_{\pi(4)}) = \{ \sigma(12), \sigma(14) \}$,
and
$I(\boldsymbol{x}_{\pi(6)}) = \{ \sigma(9), \sigma(13) \}$.

\noindent
\textit{Case 5-1-1-2-2:}
$A_2 = \{ \sigma(9) \}$ or $A_2 = \{ \sigma(14) \}$.

Let
$I(\boldsymbol{x}_{\pi(2)}) = \{ \sigma(12), \sigma(13) \}$,
$I(\boldsymbol{x}_{\pi(4)}) = \{ \sigma(8), \sigma(10) \}$,
and
$I(\boldsymbol{x}_{\pi(6)}) = \{ \sigma(11), \sigma(15) \}$.

\noindent
\textit{Case 5-1-1-2-3:}
$A_2 = \{ \sigma(10) \}$ or $A_2 = \{ \sigma(13) \}$.

Let
$I(\boldsymbol{x}_{\pi(2)}) = \{ \sigma(8), \sigma(9) \}$,
$I(\boldsymbol{x}_{\pi(4)}) = \{ \sigma(12), \sigma(14) \}$,
and
$I(\boldsymbol{x}_{\pi(6)}) = \{ \sigma(11), \sigma(15) \}$.

\noindent
\textit{Case 5-1-1-2-4:}
$A_2 = \{ \sigma(11) \}$ or $A_2 = \{ \sigma(12) \}$.

Let
$I(\boldsymbol{x}_{\pi(2)}) = \{ \sigma(14), \sigma(15) \}$,
$I(\boldsymbol{x}_{\pi(4)}) = \{ \sigma(8), \sigma(10) \}$,
and
$I(\boldsymbol{x}_{\pi(6)}) = \{ \sigma(9), \sigma(13) \}$.

\noindent
\textit{Case 5-1-1-3:}
$a_1 = 2$ and $a_2 = 0$.

Let
$I(\boldsymbol{x}_{\pi(1)}) = \{ \sigma(1) \}$,
$I(\boldsymbol{x}_{\pi(2)}) = \{ \sigma(8), \sigma(9) \}$,
$I(\boldsymbol{x}_{\pi(3)}) = \{ \sigma(2) \}$,
$I(\boldsymbol{x}_{\pi(4)}) = \{ \sigma(12), \sigma(14) \}$,
$I(\boldsymbol{x}_{\pi(5)}) = \{ \sigma(4) \}$,
and
$I(\boldsymbol{x}_{\pi(6)}) = \{ \sigma(11), \sigma(15) \}$.

\noindent
\textit{Case 5-1-2:}
$m_1 = 3$ and $m_2 = m_3 = 2$.

\noindent
\textit{Case 5-1-2-1:}
$a_1 = 0$ and $a_2 = 1$.

Let
$I(\boldsymbol{x}_{\pi(1)}) = \{ \sigma(1) \}$,
$I(\boldsymbol{x}_{\pi(2)}) = \{ \sigma(4), \sigma(5) \}$,
$I(\boldsymbol{x}_{\pi(3)}) = \{ \sigma(6), \sigma(7) \}$,
and
$I(\boldsymbol{x}_{\pi(4)}) = \{ \sigma(2) \}$.

\noindent
\textit{Case 5-1-2-1-1:}
$A_2 = \{ \sigma(8) \}$ or $A_2 = \{ \sigma(10) \}$.

Let
$I(\boldsymbol{x}_{\pi(5)}) = \{ \sigma(12), \sigma(14) \}$,
$I(\boldsymbol{x}_{\pi(6)}) = \{ \sigma(9), \sigma(13) \}$,
and
$I(\boldsymbol{x}_{\pi(7)}) = \{ \sigma(11), \sigma(15) \}$.

\noindent
\textit{Case 5-1-2-1-2:}
$A_2 = \{ \sigma(9) \}$ or $A_2 = \{ \sigma(11) \}$.

Let
$I(\boldsymbol{x}_{\pi(5)}) = \{ \sigma(13), \sigma(15) \}$,
$I(\boldsymbol{x}_{\pi(6)}) = \{ \sigma(8), \sigma(12) \}$,
and
$I(\boldsymbol{x}_{\pi(7)}) = \{ \sigma(10), \sigma(14) \}$.

\noindent
\textit{Case 5-1-2-1-3:}
$A_2 = \{ \sigma(12) \}$ or $A_2 = \{ \sigma(14) \}$.

Let
$I(\boldsymbol{x}_{\pi(5)}) = \{ \sigma(8), \sigma(10) \}$,
$I(\boldsymbol{x}_{\pi(6)}) = \{ \sigma(9), \sigma(13) \}$,
and
$I(\boldsymbol{x}_{\pi(7)}) = \{ \sigma(11), \sigma(15) \}$.

\noindent
\textit{Case 5-1-2-1-4:}
$A_2 = \{ \sigma(13) \}$ or $A_2 = \{ \sigma(15) \}$.

Let
$I(\boldsymbol{x}_{\pi(5)}) = \{ \sigma(9), \sigma(11) \}$,
$I(\boldsymbol{x}_{\pi(6)}) = \{ \sigma(8), \sigma(12) \}$,
and
$I(\boldsymbol{x}_{\pi(7)}) = \{ \sigma(10), \sigma(14) \}$.

\noindent
\textit{Case 5-1-2-2:}
$a_1 = 1$ and $a_2 = 0$.

Let
$I(\boldsymbol{x}_{\pi(1)}) = \{ \sigma(1) \}$,
$I(\boldsymbol{x}_{\pi(2)}) = \{ \sigma(\alpha_1), \sigma(\alpha_1+1) \}$,
and
$I(\boldsymbol{x}_{\pi(3)}) = \{ \sigma(\alpha_2), \sigma(\alpha_2+1) \}$
such that
$\alpha_1, \alpha_2 \in \{2,4,6\}, \alpha_1 \not= \alpha_2$,
and
$A_1 \cap
\{ \sigma(\alpha_1), \sigma(\alpha_1+1), \sigma(\alpha_2), \sigma(\alpha_2+1) \}
= \emptyset$.
Let
$I(\boldsymbol{x}_{\pi(4)}) = \{ \sigma(8), \sigma(10) \}$,
$I(\boldsymbol{x}_{\pi(5)}) = \{ \sigma(12), \sigma(14) \}$,
$I(\boldsymbol{x}_{\pi(6)}) = \{ \sigma(9), \sigma(13) \}$,
and
$I(\boldsymbol{x}_{\pi(7)}) = \{ \sigma(11), \sigma(15) \}$.

\noindent
\textit{Case 5-1-3:}
$m_1 = m_2 = 3$ and $m_3 = 2$.

Let
$I(\boldsymbol{x}_{\pi(1)}) = \{ \sigma(1) \}$,
$I(\boldsymbol{x}_{\pi(2)}) = \{ \sigma(4), \sigma(5) \}$,
$I(\boldsymbol{x}_{\pi(3)}) = \{ \sigma(6), \sigma(7) \}$,
$I(\boldsymbol{x}_{\pi(4)}) = \{ \sigma(2) \}$,
$I(\boldsymbol{x}_{\pi(5)}) = \{ \sigma(8), \sigma(10) \}$,
$I(\boldsymbol{x}_{\pi(6)}) = \{ \sigma(12), \sigma(14) \}$,
$I(\boldsymbol{x}_{\pi(7)}) = \{ \sigma(9), \sigma(13) \}$,
and
$I(\boldsymbol{x}_{\pi(8)}) = \{ \sigma(11), \sigma(15) \}$.

\noindent
\textit{Case 5-1-4:}
$m_1 = 4$ and $m_2 = m_3 = 2$.

Let
$I(\boldsymbol{x}_{\pi(1)}) = \{ \sigma(1) \}$,
$I(\boldsymbol{x}_{\pi(2)}) = \{ \sigma(2), \sigma(3) \}$,
$I(\boldsymbol{x}_{\pi(3)}) = \{ \sigma(4), \sigma(5) \}$,
$I(\boldsymbol{x}_{\pi(4)}) = \{ \sigma(6), \sigma(7) \}$,
$I(\boldsymbol{x}_{\pi(5)}) = \{ \sigma(8), \sigma(10) \}$,
$I(\boldsymbol{x}_{\pi(6)}) = \{ \sigma(12), \sigma(14) \}$,
$I(\boldsymbol{x}_{\pi(7)}) = \{ \sigma(9), \sigma(13) \}$,
and
$I(\boldsymbol{x}_{\pi(8)}) = \{ \sigma(11), \sigma(15) \}$.

\noindent
\textit{Case 5-2:}
$\boldsymbol{s}_{\pi(1)} + \boldsymbol{s}_{\pi(m_1+1)}
= \boldsymbol{s}_{\pi(m_1+m_2+1)}$.

From Theorem \ref{teiri:kouho3},
we have the permutation $\sigma$ of $\mathcal{I}_{15}$
such that
\begin{eqnarray}
  & & V(\boldsymbol{s}_{\pi(1)})
  \nonumber \\
& = &
  \{ \{ \sigma(1) \}, \{ \sigma(2), \sigma(3) \}, \{ \sigma(4), \sigma(5) \},
  \nonumber \\
& & \{ \sigma(6), \sigma(7) \},
  \{ \sigma(8), \sigma(9) \}, \{ \sigma(10), \sigma(11) \},
  \nonumber \\
  & & \{ \sigma(12), \sigma(13) \}, \{ \sigma(14), \sigma(15) \}\},
  \nonumber \\
  & & V(\boldsymbol{s}_{\pi(m_1+1)})
  \nonumber \\
& = &
  \{ \{ \sigma(2) \}, \{ \sigma(1), \sigma(3) \}, \{ \sigma(4), \sigma(6) \},
  \nonumber \\
& & \{ \sigma(5), \sigma(7) \},
  \{ \sigma(8), \sigma(10) \}, \{ \sigma(9), \sigma(11) \},
  \nonumber \\
  & & \{ \sigma(12), \sigma(14) \}, \{ \sigma(13), \sigma(15) \}\},
  \nonumber \\
  & & V(\boldsymbol{s}_{\pi(m_1+m_2+1)})
  \nonumber \\
& = &
  \{ \{ \sigma(3) \}, \{ \sigma(1), \sigma(2) \}, \{ \sigma(4), \sigma(7) \},
  \nonumber \\
& & \{ \sigma(5), \sigma(6) \},
  \{ \sigma(8), \sigma(11) \}, \{ \sigma(9), \sigma(10) \},
  \nonumber \\
  & & \{ \sigma(12), \sigma(15) \}, \{ \sigma(13), \sigma(14) \}\}.
  \nonumber
\end{eqnarray}
Clearly,
$\{ j_{\pi(m_1+m_2+m_3+1)}, \ldots, j_{\pi(8)} \} \cap
\{ \sigma(1), \sigma(2), \sigma(3) \}
= \emptyset$.
We define
$A_1 = \{ j_{\pi(i)} \mid i \in \{ m_1+m_2+m_3+1, \ldots, 8 \},
j_{\pi(i)} \in \{ \sigma(4), \sigma(5), \sigma(6), \sigma(7) \} \}$,
$A_2 = \{ j_{\pi(i)} \mid i \in \{ m_1+m_2+m_3+1, \ldots, 8 \},
j_{\pi(i)} \in \{ \sigma(8), \ldots, \sigma(15) \} \}$,
$a_1 = | A_1 |$,
$a_2 = | A_2 |$.
Then,
$a_1 + a_2 = 8-m_1-m_2-m_3$.

\noindent
\textit{Case 5-2-1:}
$m_1 = m_2 = m_3 = 2$.

\noindent
\textit{Case 5-2-1-1:}
$a_1 = 0$ and $a_2 = 2$.

Let
$I(\boldsymbol{x}_{\pi(1)}) = \{ \sigma(1) \}$,
$I(\boldsymbol{x}_{\pi(2)}) = \{ \sigma(2), \sigma(3) \}$,
$I(\boldsymbol{x}_{\pi(3)}) = \{ \sigma(4), \sigma(6) \}$,
and
$I(\boldsymbol{x}_{\pi(4)}) = \{ \sigma(5), \sigma(7) \}$.
Let
$I(\boldsymbol{x}_{\pi(5)}) = \{ \sigma(\alpha_1), \sigma(\alpha_1') \}$
and
$I(\boldsymbol{x}_{\pi(6)}) = \{ \sigma(\alpha_2), \sigma(\alpha_2') \}$
such that
$\alpha_1, \alpha_2 \in \{8,9,12,13\}, 
\alpha_1', \alpha_2' \in \{10,11,14,15\}, \alpha_1 \not= \alpha_2$,
$\{ \sigma(\alpha_1), \sigma(\alpha_1') \} , \{ \sigma(\alpha_2), \sigma(\alpha_2') \}
\in V(\boldsymbol{s}_{\pi(5)})$,
and
$A_2 \cap
\{ \sigma(\alpha_1), \sigma(\alpha_1'), \sigma(\alpha_2), \sigma(\alpha_2') \}
= \emptyset$.

\noindent
\textit{Case 5-2-1-2:}
$a_1 = a_2 = 1$.

Let
$I(\boldsymbol{x}_{\pi(1)}) = \{ \sigma(1) \}$
and
$I(\boldsymbol{x}_{\pi(2)}) = \{ \sigma(\alpha), \sigma(\alpha+1) \}$
such that
$\alpha \in \{4,6\}$
and
$A_1 \cap
\{ \sigma(\alpha), \sigma(\alpha+1) \}
= \emptyset$.
Let
$I(\boldsymbol{x}_{\pi(3)}) = \{ \sigma(2) \}$
and
$I(\boldsymbol{x}_{\pi(4)}) = \{ \sigma(\beta), \sigma(\beta+2) \}$
such that
$\beta \in \{8,9,12,13\}$
and
$A_2 \cap
\{ \sigma(\beta), \sigma(\beta+2) \}
= \emptyset$.
Let
$I(\boldsymbol{x}_{\pi(5)}) = \{ \sigma(3) \}$
and
$I(\boldsymbol{x}_{\pi(6)}) = \{ \sigma(\gamma), \sigma(\gamma') \}$
such that
$\gamma \in \{8,9,12,13\}, 
\gamma' \in \{10,11,14,15\}$,
$\{ \sigma(\gamma), \sigma(\gamma') \}
\in V(\boldsymbol{s}_{\pi(5)})$,
and
$(A_2 \cup
\{ \sigma(\beta), \sigma(\beta+2) \})
\cap \{ \sigma(\gamma), \sigma(\gamma') \} = \emptyset$.

\noindent
\textit{Case 5-2-1-3:}
$a_1 = 2$ and $a_2 = 0$.

Let
$I(\boldsymbol{x}_{\pi(1)}) = \{ \sigma(1) \}$,
$I(\boldsymbol{x}_{\pi(2)}) = \{ \sigma(2), \sigma(3) \}$,
$I(\boldsymbol{x}_{\pi(3)}) = \{ \sigma(8), \sigma(10) \}$,
$I(\boldsymbol{x}_{\pi(4)}) = \{ \sigma(9), \sigma(11) \}$,
$I(\boldsymbol{x}_{\pi(5)}) = \{ \sigma(12), \sigma(15) \}$,
and
$I(\boldsymbol{x}_{\pi(6)}) = \{ \sigma(13), \sigma(14) \}$.

\noindent
\textit{Case 5-2-2:}
$m_1 = 3$ and $m_2 = m_3 = 2$.

\noindent
\textit{Case 5-2-2-1:}
$a_1 = 0$ and $a_2 = 1$.

Let
$I(\boldsymbol{x}_{\pi(1)}) = \{ \sigma(1) \}$,
$I(\boldsymbol{x}_{\pi(2)}) = \{ \sigma(4), \sigma(5) \}$,
$I(\boldsymbol{x}_{\pi(3)}) = \{ \sigma(6), \sigma(7) \}$,
$I(\boldsymbol{x}_{\pi(4)}) = \{ \sigma(2) \}$,
and
$I(\boldsymbol{x}_{\pi(5)}) = \{ \sigma(\alpha), \sigma(\alpha+2) \}$
such that
$\alpha \in \{8,9\}$
and
$A_2 \cap
\{ \sigma(\alpha), \sigma(\alpha+2) \}
= \emptyset$.
Let
$I(\boldsymbol{x}_{\pi(6)}) = \{ \sigma(3) \}$
and
$I(\boldsymbol{x}_{\pi(7)}) = \{ \sigma(\beta), \sigma(\beta') \}$
such that
$\beta \in \{12,13\}, 
\beta' \in \{14,15\}$,
$\{ \sigma(\beta), \sigma(\beta') \}
\in V(\boldsymbol{s}_{\pi(6)})$,
and
$A_2 \cap
\{ \sigma(\beta), \sigma(\beta') \}
= \emptyset$.

\noindent
\textit{Case 5-2-2-2:}
$a_1 = 1$ and $a_2 = 0$.

Let
$I(\boldsymbol{x}_{\pi(1)}) = \{ \sigma(1) \}$,
$I(\boldsymbol{x}_{\pi(2)}) = \{ \sigma(2), \sigma(3) \}$,
and
$I(\boldsymbol{x}_{\pi(3)}) = \{ \sigma(\alpha), \sigma(\alpha+1) \}$
such that
$\alpha \in \{4,6\}$
and
$A_1 \cap
\{ \sigma(\alpha), \sigma(\alpha+1) \}
= \emptyset$.
Let
$I(\boldsymbol{x}_{\pi(4)}) = \{ \sigma(8), \sigma(10) \}$,
$I(\boldsymbol{x}_{\pi(5)}) = \{ \sigma(9), \sigma(11) \}$,
$I(\boldsymbol{x}_{\pi(6)}) = \{ \sigma(12), \sigma(15) \}$,
and
$I(\boldsymbol{x}_{\pi(7)}) = \{ \sigma(13), \sigma(14) \}$.

\noindent
\textit{Case 5-2-3:}
$m_1 = m_2 = 3$ and $m_3 = 2$.

Let
$I(\boldsymbol{x}_{\pi(1)}) = \{ \sigma(1) \}$,
$I(\boldsymbol{x}_{\pi(2)}) = \{ \sigma(4), \sigma(5) \}$,
$I(\boldsymbol{x}_{\pi(3)}) = \{ \sigma(6), \sigma(7) \}$,
$I(\boldsymbol{x}_{\pi(4)}) = \{ \sigma(2) \}$,
$I(\boldsymbol{x}_{\pi(5)}) = \{ \sigma(8), \sigma(10) \}$,
$I(\boldsymbol{x}_{\pi(6)}) = \{ \sigma(9), \sigma(11) \}$,
$I(\boldsymbol{x}_{\pi(7)}) = \{ \sigma(3) \}$,
and
$I(\boldsymbol{x}_{\pi(8)}) = \{ \sigma(12), \sigma(15) \}$.

\noindent
\textit{Case 5-2-4:}
$m_1 = 4$ and $m_2 = m_3 = 2$.

Let
$I(\boldsymbol{x}_{\pi(1)}) = \{ \sigma(1) \}$,
$I(\boldsymbol{x}_{\pi(2)}) = \{ \sigma(2), \sigma(3) \}$,
$I(\boldsymbol{x}_{\pi(3)}) = \{ \sigma(4), \sigma(5) \}$,
$I(\boldsymbol{x}_{\pi(4)}) = \{ \sigma(6), \sigma(7) \}$,
$I(\boldsymbol{x}_{\pi(5)}) = \{ \sigma(8), \sigma(10) \}$,
$I(\boldsymbol{x}_{\pi(6)}) = \{ \sigma(9), \sigma(11) \}$,
$I(\boldsymbol{x}_{\pi(7)}) = \{ \sigma(12), \sigma(15) \}$,
and
$I(\boldsymbol{x}_{\pi(8)}) = \{ \sigma(13), \sigma(14) \}$.

\noindent
\textit{Case 6:}
There exists
the permutation $\pi$ of $\mathcal{I}_8$
such that
$\boldsymbol{s}_{\pi(1)} = \boldsymbol{s}_{\pi(2)}$,
$\boldsymbol{s}_{\pi(3)} = \boldsymbol{s}_{\pi(4)}$,
$\boldsymbol{s}_{\pi(5)} = \boldsymbol{s}_{\pi(6)}$,
$\boldsymbol{s}_{\pi(7)} = \boldsymbol{s}_{\pi(8)}$,
and
$\boldsymbol{s}_{\pi(i)} \not= \boldsymbol{s}_{\pi(i')}$
for any
$i, i' \in \{1, 3, 5, 7\}, i \not= i'$.

\noindent
\textit{Case 6-1:}
$\boldsymbol{s}_{\pi(1)}, \boldsymbol{s}_{\pi(3)},
\boldsymbol{s}_{\pi(5)},$ and $\boldsymbol{s}_{\pi(7)}$
are linearly independent.

From Theorem \ref{teiri:kouho4},
we have the permutation $\sigma$ of $\mathcal{I}_{15}$
such that
\begin{eqnarray}
  & & V(\boldsymbol{s}_{\pi(1)})
  \nonumber \\
& = &
  \{ \{ \sigma(1) \}, \{ \sigma(2), \sigma(3) \}, \{ \sigma(4), \sigma(5) \},
  \nonumber \\
& & \{ \sigma(6), \sigma(7) \},
  \{ \sigma(8), \sigma(9) \}, \{ \sigma(10), \sigma(11) \},
  \nonumber \\
  & & \{ \sigma(12), \sigma(13) \}, \{ \sigma(14), \sigma(15) \}\},
  \nonumber \\
  & & V(\boldsymbol{s}_{\pi(3)})
  \nonumber \\
& = &
  \{ \{ \sigma(2) \}, \{ \sigma(1), \sigma(3) \}, \{ \sigma(4), \sigma(6) \},
  \nonumber \\
& & \{ \sigma(5), \sigma(7) \},
  \{ \sigma(8), \sigma(10) \}, \{ \sigma(9), \sigma(11) \},
  \nonumber \\
  & & \{ \sigma(12), \sigma(14) \}, \{ \sigma(13), \sigma(15) \}\},
  \nonumber \\
  & & V(\boldsymbol{s}_{\pi(5)})
  \nonumber \\
& = &
  \{ \{ \sigma(4) \}, \{ \sigma(1), \sigma(5) \}, \{ \sigma(2), \sigma(6) \},
  \nonumber \\
& & \{ \sigma(3), \sigma(7) \},
  \{ \sigma(8), \sigma(12) \}, \{ \sigma(9), \sigma(13) \},
  \nonumber \\
  & & \{ \sigma(10), \sigma(14) \}, \{ \sigma(11), \sigma(15) \}\},
  \nonumber \\
  & & V(\boldsymbol{s}_{\pi(7)})
  \nonumber \\
& = &
  \{ \{ \sigma(8) \}, \{ \sigma(1), \sigma(9) \}, \{ \sigma(2), \sigma(10) \},
  \nonumber \\
& & \{ \sigma(3), \sigma(11) \},
  \{ \sigma(4), \sigma(12) \}, \{ \sigma(5), \sigma(13) \},
  \nonumber \\
& & \{ \sigma(6), \sigma(14) \}, \{ \sigma(7), \sigma(15) \}\}. \nonumber
\end{eqnarray}
Let
$I(\boldsymbol{x}_{\pi(1)}) = \{ \sigma(1) \}$,
$I(\boldsymbol{x}_{\pi(2)}) = \{ \sigma(2), \sigma(3) \}$,
$I(\boldsymbol{x}_{\pi(3)}) = \{ \sigma(4), \sigma(6) \}$,
$I(\boldsymbol{x}_{\pi(4)}) = \{ \sigma(9), \sigma(11) \}$,
$I(\boldsymbol{x}_{\pi(5)}) = \{ \sigma(8), \sigma(12) \}$,
$I(\boldsymbol{x}_{\pi(6)}) = \{ \sigma(10), \sigma(14) \}$,
$I(\boldsymbol{x}_{\pi(7)}) = \{ \sigma(5), \sigma(13) \}$,
and
$I(\boldsymbol{x}_{\pi(8)}) = \{ \sigma(7), \sigma(15) \}$.

\noindent
\textit{Case 6-2:}
There exists
the permutation $\tau$ of
$\{\pi(1), \pi(3), \pi(5), \pi(7) \}$
such that
$\boldsymbol{s}_{\tau(\pi(1))} + \boldsymbol{s}_{\tau(\pi(3))}
= \boldsymbol{s}_{\tau(\pi(5))}$.

For simplicity,
suppose
$\boldsymbol{s}_{\pi(1)} + \boldsymbol{s}_{\pi(3)} =
\boldsymbol{s}_{\pi(5)}$.
From Theorem \ref{teiri:kouho4},
we have the permutation $\sigma$ of $\mathcal{I}_{15}$
such that
\begin{eqnarray}
  & & V(\boldsymbol{s}_{\pi(1)})
  \nonumber \\
& = &
  \{ \{ \sigma(1) \}, \{ \sigma(2), \sigma(3) \}, \{ \sigma(4), \sigma(5) \},
  \nonumber \\
& & \{ \sigma(6), \sigma(7) \},
  \{ \sigma(8), \sigma(9) \}, \{ \sigma(10), \sigma(11) \},
  \nonumber \\
  & & \{ \sigma(12), \sigma(13) \}, \{ \sigma(14), \sigma(15) \}\},
  \nonumber \\
  & & V(\boldsymbol{s}_{\pi(3)})
  \nonumber \\
& = &
  \{ \{ \sigma(2) \}, \{ \sigma(1), \sigma(3) \}, \{ \sigma(4), \sigma(6) \},
  \nonumber \\
& & \{ \sigma(5), \sigma(7) \},
  \{ \sigma(8), \sigma(10) \}, \{ \sigma(9), \sigma(11) \},
  \nonumber \\
  & & \{ \sigma(12), \sigma(14) \}, \{ \sigma(13), \sigma(15) \}\},
  \nonumber \\
  & & V(\boldsymbol{s}_{\pi(5)})
  \nonumber \\
& = &
  \{ \{ \sigma(3) \}, \{ \sigma(1), \sigma(2) \}, \{ \sigma(4), \sigma(7) \},
  \nonumber \\
& & \{ \sigma(5), \sigma(6) \},
  \{ \sigma(8), \sigma(11) \}, \{ \sigma(9), \sigma(10) \},
  \nonumber \\
  & & \{ \sigma(12), \sigma(15) \}, \{ \sigma(13), \sigma(14) \}\},
  \nonumber \\
  & & V(\boldsymbol{s}_{\pi(7)})
  \nonumber \\
& = &
  \{ \{ \sigma(4) \}, \{ \sigma(1), \sigma(5) \}, \{ \sigma(2), \sigma(6) \},
  \nonumber \\
& & \{ \sigma(3), \sigma(7) \},
  \{ \sigma(8), \sigma(12) \}, \{ \sigma(9), \sigma(13) \},
  \nonumber \\
  & & \{ \sigma(10), \sigma(14) \}, \{ \sigma(11), \sigma(15) \}\}.
  \nonumber
\end{eqnarray}
Let
$I(\boldsymbol{x}_{\pi(1)}) = \{ \sigma(1) \}$,
$I(\boldsymbol{x}_{\pi(2)}) = \{ \sigma(2), \sigma(3) \}$,
$I(\boldsymbol{x}_{\pi(3)}) = \{ \sigma(4), \sigma(6) \}$,
$I(\boldsymbol{x}_{\pi(4)}) = \{ \sigma(5), \sigma(7) \}$,
$I(\boldsymbol{x}_{\pi(5)}) = \{ \sigma(8), \sigma(11) \}$,
$I(\boldsymbol{x}_{\pi(6)}) = \{ \sigma(12), \sigma(15) \}$,
$I(\boldsymbol{x}_{\pi(7)}) = \{ \sigma(9), \sigma(13) \}$,
and
$I(\boldsymbol{x}_{\pi(8)}) = \{ \sigma(10), \sigma(14) \}$.

\noindent
\textit{Case 6-3:}
$\boldsymbol{s}_{\pi(1)} + \boldsymbol{s}_{\pi(3)}
= \boldsymbol{s}_{\pi(5)} + \boldsymbol{s}_{\pi(7)}$.

From Theorem \ref{teiri:kouho4},
we have the permutation $\sigma$ of $\mathcal{I}_{15}$
such that
\begin{eqnarray}
  & & V(\boldsymbol{s}_{\pi(1)})
  \nonumber \\
& = &
  \{ \{ \sigma(1) \}, \{ \sigma(2), \sigma(3) \}, \{ \sigma(4), \sigma(5) \},
  \nonumber \\
& & \{ \sigma(6), \sigma(7) \},
  \{ \sigma(8), \sigma(9) \}, \{ \sigma(10), \sigma(11) \},
  \nonumber \\
  & & \{ \sigma(12), \sigma(13) \}, \{ \sigma(14), \sigma(15) \}\},
  \nonumber \\
  & & V(\boldsymbol{s}_{\pi(3)})
  \nonumber \\
& = &
  \{ \{ \sigma(2) \}, \{ \sigma(1), \sigma(3) \}, \{ \sigma(4), \sigma(6) \},
  \nonumber \\
& & \{ \sigma(5), \sigma(7) \},
  \{ \sigma(8), \sigma(10) \}, \{ \sigma(9), \sigma(11) \},
  \nonumber \\
  & & \{ \sigma(12), \sigma(14) \}, \{ \sigma(13), \sigma(15) \}\},
  \nonumber \\
  & & V(\boldsymbol{s}_{\pi(5)})
  \nonumber \\
& = &
  \{ \{ \sigma(4) \}, \{ \sigma(1), \sigma(5) \}, \{ \sigma(2), \sigma(6) \},
  \nonumber \\
& & \{ \sigma(3), \sigma(7) \},
\{ \sigma(8), \sigma(12) \}, \{ \sigma(9), \sigma(13) \},
\nonumber \\
& & \{ \sigma(10), \sigma(14) \}, \{ \sigma(11), \sigma(15) \}\},
\nonumber \\
& & V(\boldsymbol{s}_{\pi(7)})
\nonumber \\
& = &
\{ \{ \sigma(7) \}, \{ \sigma(1), \sigma(6) \}, \{ \sigma(2), \sigma(5) \},
\nonumber \\
& & \{ \sigma(3), \sigma(4) \},
\{ \sigma(8), \sigma(15) \}, \{ \sigma(9), \sigma(14) \},
\nonumber \\
& & \{ \sigma(10), \sigma(13) \}, \{ \sigma(11), \sigma(12) \}\}.
\nonumber
\end{eqnarray}
Let
$I(\boldsymbol{x}_{\pi(1)}) = \{ \sigma(1) \}$,
$I(\boldsymbol{x}_{\pi(2)}) = \{ \sigma(2), \sigma(3) \}$,
$I(\boldsymbol{x}_{\pi(3)}) = \{ \sigma(4), \sigma(6) \}$,
$I(\boldsymbol{x}_{\pi(4)}) = \{ \sigma(5), \sigma(7) \}$,
$I(\boldsymbol{x}_{\pi(5)}) = \{ \sigma(8), \sigma(12) \}$,
$I(\boldsymbol{x}_{\pi(6)}) = \{ \sigma(11), \sigma(15) \}$,
$I(\boldsymbol{x}_{\pi(7)}) = \{ \sigma(9), \sigma(14) \}$,
and
$I(\boldsymbol{x}_{\pi(8)}) = \{ \sigma(10), \sigma(13) \}$.

Therefore,
from Theorem \ref{teiri:sufficientcondition},
the $[15,4,8]$ P-RIO code can be constructed
using the coset coding.
The number of pages
stored by
this code
is greater
than that of the $[15,4,6]$ RIO code
constructed using
the same $(15,11)$ Hamming code.

An upper bound on the sum-rate of RIO codes
and P-RIO codes that store $t$ pages
is $\log (t+1)$ \cite{parallel}.
Table \ref{table:capacity} shows the sum-rates
and the upper bounds on that of 
the RIO codes and the P-RIO codes
constructed via coset coding with Hamming codes.
\begin{table}[tb]
\caption{Comparison with capacity}
\label{table:capacity}
\begin{center}
\begin{tabular}{|c|c|c|c|c|} \hline
  Code length & \multicolumn{2}{|c|}{RIO code} &
  \multicolumn{2}{|c|}{P-RIO code}
  \\ \cline{2-5}
   & Sum-rate & Upper bound & Sum-rate & Upper bound \\ \hline
7 & 1.2857 & 2 & 1.7142 & 2.3219 \\ \hline
15 & 1.6 & 2.8073 & 2.1333 & 3.1699 \\ \hline
\end{tabular}%
\end{center}
\end{table}

\section{Conclusion}
\label{section:conclusion}

In this paper,
we have constructed P-RIO codes
using coset coding with Hamming codes.
When each page is encoded,
the information on the data of the other pages
is leveraged to
increase the number of stored pages
compared to that for
RIO codes.
Our P-RIO codes are constructive,
whereas the approach in \cite{parallel}
is based on exhaustive search.
However,
in our approach,
the number of cases required for the encoding
increases with the number of pages.
Therefore,
we should consider another approach
to deriving
the number of pages of P-RIO codes
that can be constructed using Hamming codes
of length $(2^r-1)$
for $r \geq 5$.


\section*{Acknowledgment}

This work was partially supported by
the Advanced Storage Research Consortium
and JSPS KAKENHI Grant Number
15K00010.



%

\end{document}